\documentclass[12pt, reqno]{amsart}

\usepackage{xr}
\usepackage{amsmath,amssymb}
\usepackage[authoryear,longnamesfirst]{natbib}
\usepackage{graphicx,epsfig,epstopdf}
\usepackage{amssymb}
\usepackage{amsmath}
\usepackage{amsmath}
\usepackage{amsthm}
\usepackage{amsfonts}
\usepackage{graphicx}
\usepackage{bbm}
\usepackage{mathrsfs}
\usepackage{enumerate}
\usepackage{dsfont}
\usepackage{natbib}
\usepackage{multirow}
\usepackage{dsfont}
\usepackage{color}
\usepackage{caption}
\usepackage{subcaption}
\newtheorem{theorem}{Theorem}[section]
\newtheorem{assumption}{Assumption}[section]
\newtheorem{algorithm}{Algorithm}[section]

\newtheorem{corollary}{Corollary}[section]
\newtheorem{lemma}{Lemma}[section]

\theoremstyle{definition}

\newtheorem{remark}{Remark}[section]
\numberwithin{remark}{section}

\newcommand{\Real}{\mathbb R}

\newcommand{\be}{\begin{eqnarray}}
\newcommand{\ee}{\end{eqnarray}}

\newcommand{\Ss}{\mathcal{S}}
\newcommand{\NN}{\mathbbm{N}}

\newcommand{\ba}{\begin{array}}
\newcommand{\ea}{\end{array}}
\newcommand{\bs}{\begin{align}\begin{split}\nonumber}
\newcommand{\bsnumber}{\begin{align}\begin{split}}
\newcommand{\es}{\end{split}\end{align}}

\renewcommand{\hat}{\widehat}
\renewcommand{\H}{\mathcal H}

\newcommand{\calF}{\mathcal{F}}

\newcommand{\Ep}{{\mathrm{E}}}

\newcommand{\EG}{\frac{1}{G}\sum_{g=1}^G}

\newcommand{\EP}{{\mathrm{E}}_\mathrm{P}}

\newcommand{\Op}{O_\Pr}
\newcommand{\op}{o_\Pr}
\renewcommand{\Pr}{{\mathrm{P}}}

\def\RR{ {\mathbb{R}}}

\def\supp{{\rm support}}

\newcommand{\semin}[1]{\phi_{{\rm min}}(#1)}
\newcommand{\semax}[1]{\phi_{{\rm max}}(#1)}
\renewcommand{\hat}{\widehat}

\newcommand{\diag}{{\rm diag}}

\newcommand{\argmin}{\rm argmin}

\newcommand{\sumg}{\sum_{g=1}^G}
\newcommand{\sumi}{\sum_{i=1}^{n_g}}
\newcommand{\1}{\mathbbm 1}
\newcommand{\ktc}{\bar{\kappa}_{2\widetilde c}}

\newcommand{\hPsi}{\hat\Psi_{k0}}
\newcommand{\calC}{\mathcal{C}}
\newcommand{\calG}{\mathcal{G}}

\newcommand{\sg}{\text{sg}}
\begin{document}

\title[Many APE$\text{s}$]{Many average partial effects:\\ with an application to text regression}

\author[Chiang]{Harold D. Chiang
}\address[Harold D. Chiang]{Department of Economics, University of Wisconsin-Madison, United States}

\thanks{Based on my job market paper. I am indebted to Alex Belloni, Otilia Boldea, Irene Botosaru, Mehmet Caner, Bruce Hansen, Atsushi Inoue, Hugo Jales, Kengo Kato, Frank Kleibergen,  
Tong Li, Pedro Sant'Anna, Yuya Sasaki, Takuya Ura, Matt Webb, and the participants at AMES 2019, Bristol Econometrics Study Group 2019, Canadian Econometric Study Group 2019, ESEM 2019, International PhD Conference at Erasmus University Rotterdam, NY Camp Econometrics XIV, TER 2019, and seminal participants at Cornell, Illinois, Iowa, McGill, Michigan, UQAM, Rutgers, Syracuse, UCL, UNC Chapel-Hill, and UW-Madison for their insightful comments and discussions. I also benefited from the programming discussions with Terry Kim and Bin Yang Tan. }

\date{
This version is of  \today.  
 }


\maketitle
\allowdisplaybreaks

\begin{abstract}
We study estimation, pointwise and simultaneous inference, and confidence intervals for many average partial effects of lasso Logit. Focusing on high-dimensional cluster-sampling environments, we propose a new average partial effect estimator and explore its asymptotic properties. Practical penalty choices compatible with our asymptotic theory are also provided. The estimator allow for valid inference without requiring oracle property. We provide easy-to-implement algorithms for cluster-robust high-dimensional hypothesis testing and construction of simultaneously valid confidence intervals using a multiplier cluster bootstrap. We apply the proposed algorithms to the text regression model of \cite{Wu18} to examine the presence of gendered language on the internet.
\end{abstract}

\section{Introduction}

Binary response models are some of the most commonly used nonlinear econometric models. When studying such models, the average partial effect, henceforth APE, is a popular target parameter of interest.
Under big data environments, as often happens in text analysis, dimension reduction via lasso, or other type of machine learning algorithms, is often unavoidable. Failure to account for the model selection step often leads to severely biased estimates, which invalidate the usual inference procedures (see Figure 1 for an illustration). Few results are available for valid post-selection inference for a single nonlinear functional of high-dimensional nuisance parameters, such as APE, let alone simultaneous inference for potentially many of such parameters. To fill this void, this paper considers simultaneous inference and confidence intervals for lasso Logit APEs. In addition, as we illustrate in simulation studies in Section \ref{sec:simulation_studies} that ignoring cluster sampling can lead to severely distorted testing results, even when cluster sizes are small. The size distortion is further aggravated in testing multiple hypotheses. Importantly, all our theoretical results stay valid for cluster-sampled data with heterogeneous cluster sizes.

\begin{center}
\begin{figure}\label{fig:1}
\begin{subfigure}[b]{0.5\textwidth}
\includegraphics[width=80mm]{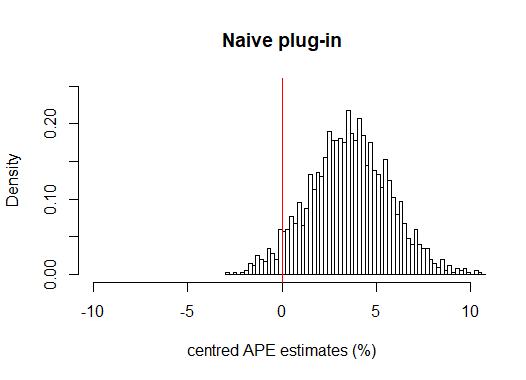}
\caption{APE estimates based on direct plug-in of lasso Logit coefficient estimates.}
\end{subfigure}
\quad
~
\begin{subfigure}[b]{0.5\textwidth}
\includegraphics[width=80mm]{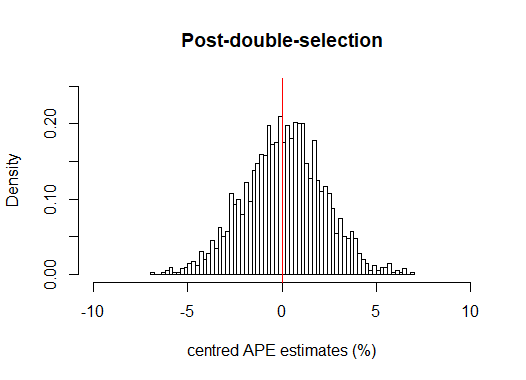}
\caption{APE estimates based on the proposed post-double-selection estimator.}
\end{subfigure}
\caption{\small Simulations for low-dimensional lasso Logit APE estimation based on $2,000$ iterations. Each iteration has sample size $n=200$. The dimensionality of covariates is set to be $p=10$. We set true parameter vector as $\beta^0=[.1,-1,1,0,...,0]$.  Covariates $X$ are generated as i.i.d. zero-mean multivariate normal random vectors with Toeplitz covariance matrix $\Sigma$ with $\Sigma_{ij}=0.5^{|i-j|}$. Outcome variables are generated following $Y=\1\{X'\beta^0+U\}$ with i.i.d. $U$ following standard logistic distribution. The lasso estimations are implemented using R package \textbf{glmnet} with penalty selection algorithms discussed in Section \ref{sec:nuisance_parameters}.}
\end{figure}
\end{center}

To our knowledge, this is the first paper handling multiple testing and simultaneous confidence interval problems for more than a single APE under high-dimensional or big-data environments. In addition, cluster sampling with heterogeneous cluster sizes is allowed. Using the Neyman orthogonalization technique, we propose a new lasso-based post-double selection APE estimator. To accompany the main theoretical results, we propose valid nuisance parameter estimators as well as their practical tuning parameter selection algorithms that are compatible with our theory. 
To address the multiple-testing problem, we develop a new, simple-to-implement, multiplier cluster bootstrap. We provide
simple algorithms for testing high-dimensional hypotheses and constructing simultaneously valid confidence intervals.
Simulation studies suggest the proposed methods have favorable finite-sample performance. We illustrate the applicability of our theoretical results through examining a claim of \cite{Wu18} on the presence of genderally biased use of language following Wu's text regression model using internet forum textual data from Economics Job Market Rumors (EJMR) forum - see the following section.

\section{Motivation: Text Analysis and Gendered Language on the Internet}\label{sec:motivating_application}
Text analysis using machine learning algorithms has become a useful alternative to the more traditional data analysis used in economics and other social sciences.  Popular categories of text analysis models include text regression models, generative models, dictionary-based methods and word embeddings. 
The first two categories
link attributes and word counts through conditional probabilities\footnote{Roughly speaking, given attributes $v_i$ and word counts $c_i$, a text regression model considers $\Pr(v_i|c_i)$ and a generative model considers $\Pr(c_i|v_i)$. } 
 and, therefore, naturally relate to common econometric models. Notable examples of applications using text regression include stock prices prediction (e.g. \cite{JW13}) and the Google Flu Trends, which is summarized in \cite{G_etal09}, among others. \cite{GST19} is a representative recent example for generative models applied to economics. For more details and applications, see \cite{GKT19} for an up-to-date review.

\begin{table}
\caption{Top 10 most predictive words for female/male from \cite{Wu18}}
\qquad\\
\begin{tabular}{cccc}
\hline 
\hline 
 \multicolumn{2}{c}{\it Female }& \multicolumn{2}{c}{\it Male }\\
Word & APE & Word & APE  \\
\hline 
\hline
Pregnancy&$0.292$&Knocking&$-0.329$\\
Hotter&$0.289$&Testosterone&$-0.204$\\
Pregnant&$0.258$&Blog&$-0.183$\\
Hp&$0.238$&Hateukbro&$-0.176$\\
Vagina&$0.228$&Adviser&$-0.175$\\
Breast&$0.220$&Hero&$-0.174$\\
Plow&$0.219$& Cuny&$-0.173$\\
Shopping & $0.207$ & Handsome & $-0.166$\\
Marry&$0.207$& Mod & $-0.166$\\
Gorgeous&$0.201$ &Homo& $-0.160$\\
\hline
\end{tabular}\label{table:Wu_18}
\\
(pronoun sample; a replication of Table 2 in \cite{Wu18})
\end{table}

Using a text regression model,
 \cite{Wu18} examines how women and men are discussed and depicted in the anonymous 
Economics Job Market Rumors forum. The author first extracted a list of female/male classifier vocabularies. According to Wu, a post is considered to be female if it contains any female classifier and male if it contains any male classifier\footnote{Wu makes use of a classification procedure to decide the posts that contains both female and male classifiers. See Section II A of \cite{Wu18} for more details}. 
Let $Female_i$ be an indicator of whether post $i$ is female. $X_i$ denotes a vector
of counts for each of the top 10,000 most common words\footnote{It is also possible to use frequency and $n$-grams in place of word count and words, respectively, as suggested in \cite{GKT19}.
}
(excluding all gender classifiers) 
that are present
in gendered post $i$. Wu considers the text regression model with the logistic\footnote{For text regression models with binary attributes, a penalized logistic model is recommended by \cite{GKT19}; see their Section 3.1.1 for more details.} specification,
\begin{align*}
\Pr(Female_i|X_i)=\Lambda(X_i'\beta)
\end{align*}
where $\Lambda$ is the logistic function, using a lasso Logit procedure. The $Male$ counterpart is estimated analogously.
For interpretability, Wu computes estimates for the APE for each of the $9,540$ words, where the APE for the word count of the $k$-th word is defined as
\begin{align*}
\text{APE}_k=\Ep[\beta_k\Lambda'(X_i'\beta)].
\end{align*}
Some top estimates of Wu's are listed in Table \ref{table:Wu_18}.  Based on these results, Wu concludes that the words that predict a post about a woman are typically about physical appearance or personal information, whereas those most predictive of a post about a man tend to focus on academic or professional characteristics. 


\cite{Wu18} focuses on estimation. To further examine the magnitude and statistical significance of these estimates, the researcher may be interested in conducting hypotheses testing or constructing confidence intervals. To do so, several issues need to be carefully accounted for. First, as posts in EJMR data of \cite{Wu18} are sampled from different threads of various discussion topics, it is likely that posts coming from the same thread are highly correlated. Therefore, statistical testing should be conducted using a cluster robust inference method. Secondly, 
\cite{Wu18} highlights that females are often described with words about appearance or personal information. 
To formally quantify such statements, one may want to conduct multiple testing for APEs of a (potentially large) set of vocabularies related to appearance or personal information. 
Furthermore, in many cases, words with the same or close meaning are double-counted in this dataset, e.g. ``attractive" and ``attractiveness" or ``homo", ``homosexual," and ``gay."\footnote{If the researcher is only concerned about joint testing, an easy alternative is to combine these words. However, this is not desirable when one wants to obtain separate estimates.}  Thus, the researcher may want to consider a joint test that controls family-wise error rates for APEs of these words. This results in a multiple testing problem. Therefore, the testing procedure needs to be able to control the family-wise error rate while testing potentially many variables. To our best knowledge, no method in the literature is capable of addressing all these issues simultaneously. This paper attempts to provide a useful and easy-to-implement method that can be applied to such problems.


\section{Background and Literature Review}

\subsection{Contributions}\label{sec:contributions} Our main contribution is to provide a theory for high-dimensional multiple-testing and simultaneous confidence intervals for APEs of binomial and fractional response regression models under clustered data. To our best knowledge, no results were previously available for this purpose. As a by-product, this paper also complements existing papers by proposing a practical method for studying low-dimensional APEs of interest under high-dimensional settings. Furthermore, cluster sizes are allowed to be heterogeneous - this is essential to our application as number of posts varies from thread to thread. Inference and construction of confidence intervals for such models are practically challenging; despite that methods are proposed in the literature, no simulation evidence for inference of even a single APE under lasso-regularization with these methods is available. In addition, we present practical and theoretically justified penalty choices for all the lasso estimators. Furthermore, easy-to-implement bootstrap procedures are also provided for inference/confidence intervals that hold valid, regardless of whether the researcher is interested in one or multiple APEs. 

\subsection{Relations to the Literature}\label{sec:literature}
The past decade has seen an explosive development in the literature of post-selection inference for lasso-based high-dimensional methods. This includes \cite{BCCH12} for instrumental variable models, \cite{BCH14}, \cite{JM14}, \cite{ZZ14}, \cite{Farrell15}, \cite{CK18} and \cite{AIW18} for linear regression/treatment effects models. Post-selection inference for generalized linear models such as Logit has been studied by \cite{vdGBRD14}, \cite{BCK15}, \cite{BCW16}, \cite{BCFH17} 
and \cite{BCCW18}, to list a few. This line of research predominately focuses on regression coefficients of the generalized linear models rather than nonlinear functionals such as an APE. Recently, \cite{CNS18} study $L_2$-continuous functionals using lasso and Dantzig selector. While focusing on affine-functionals, they provide an extension of their method to nonlinear functionals. Their method makes use of a linear Riesz representer to approximate the linearization of a nonlinear functional, which differs from our approach. In addition, all of the aforementioned papers are based on i.i.d. or independent sampling assumptions. On the other hand, there are some results available for high-dimensional linear panel data. This includes \cite{BCHK16}, \cite{Kock16} and \cite{KT18}.

Cluster-robust inference under various fixed-dimensional parametric settings has been well-studied and widely applied in the literature. See \cite{Wooldridge10} and \cite{CM15} for textbook treatment and comprehensive reviews. 
There has been recent literary focus on cluster-robust bootstrap inference. This includes \cite{KS12}, \cite{Hagemann17},
\cite{MW17} and \cite{DMN18}, among others. These results cannot be generalized in a straightforward manner to high-dimensional settings as the delta-method does not, in general, hold in an asymptotic framework with increasing dimensionality, see \cite{Caner17} for more details.

APE for binomial/fractional regression models has been discussed extensively in the literature (cf \cite{chamberlain84}, \cite{wooldridge05} and \cite{wooldridge18}, etc).
Inference for APEs of lasso-based binomial regression models are first studied by \cite{WZ17} under a short (balanced) panel data setting. They make use of a single-selection step with a lasso Probit estimator and propose a de-biased estimator for a single APE and obtain asymptotic normality. More recently, \cite{HW18} highlight the estimator of \cite{WZ17} for its requirement of a ``soft" beta-min assumption that rules out regularization bias asymptotically\footnote{Such post model selection inference issues are widely discussed in the literature (see e.g. \cite{PL09} and the reference within).}. 
For i.i.d. data, \cite{HW18} provide an alternative augmented minimax estimator based on the novel framework for linear functionals developed in \cite{HS17}. However, no variance estimator for this approach is proposed. Also, the aforementioned results are available only for a single APE; multiple testing and simultaneous confidence intervals for more than one APE remain unavailable. In addition, implementing inference for even a single APE under such settings presents practical challenges; to our best knowledge, there has been no simulation evidence presented for the proposed estimators in the aforementioned papers.

This paper aims to address all the aforementioned issues simultaneously. To do so, we extend the general framework for i.i.d. data developed in the important works of \cite{BCK15} and \cite{BCCW18} to allow for cluster sampling and adapt it to the studies of APEs.  The pointwise/simultaneous inference and confidence intervals are based on a multiplier cluster bootstrap which is built upon the high-dimensional central limit theorem of \cite{CCK13}.


\subsection{Notations}\label{sec:notation}
Denote $(\Omega,\mathcal A)$ the underlying measurable space and for each $G\in \mathbb N$, $\mathcal P_G$ is a set of probability measures $\Pr \in \mathcal P_G$ defined on $\mathcal A$. Consider triangular array data $\{W^G_{g}: g=1,...,G, G=1,2,3,...\}$
defined on probability space $(\Omega, \mathcal{A}, \Pr)$,
where $\Pr$ depends on $G$ through $\mathcal P_G$.  Each $W^G_{g}= \{ W^G_{ig}: 1\le i \le n_g \}$,
is a random vector that is independent across $g$, but not necessarily identically distributed. All parameters that characterize the distribution of  $\{W^G_g; g=1,...,G\}$ are
implicitly indexed by $\Pr_G$ and thus by $G$.  This dependence is henceforth omitted for simplicity.
$W_{ig}=(Y_{ig},X_{ig}')'$ takes values in $\Real^{p+1}$. For each $g\le G$, $G\in\mathds N$, the deterministic size of cluster $n_g$ satisfies $1\le n_g\le\bar n$ for a constant $\bar n$ independent of $G$. Therefore, for $i$ such that $n_g<i \le \bar n$, we can set $W_{ig}=0$ and thus each $W_g$ can be represented as a $\bar n(p+1)$-dimensional random vector. 
 Let $\EP$ be the expectation with respect to law $\Pr$. 

For a vector $\beta$, the $k$-th component is denoted as $\beta_k$. For vectors, denote the $\ell_1$-norm as $\|\cdot\|_1$, ${l}_2$-norm as $\|\cdot\|$, $\ell_\infty$-norm as $\|\cdot\|_\infty$, and the ``$\ell_0$-norm" as $\|\cdot\|_0$ to denote the number of non-zero components. For a matrix $A$, let $A'$ be the transpose of $A$. For $1\le q < \infty$, $\|A\|_{q}$ denotes the induced ${l}_q$-norm and $\|A\|_\infty=\max_{1\le j,k\le p}|A_{j,k}|$.  For a vector $\delta \in \RR^{p}$ and given data, $\| X_{ig}'\delta \|_{G} = \sqrt{ \frac{1}{G} \sumg \sumi(X_{ig}'\delta)^{2}}$ denotes the prediction norm of $\delta$. Let $e_j$ be the $j$-th vector of the standard basis for $\Real^p$. 
Given a vector $\delta \in \RR^p$, and a set of
indices $T \subseteq \{1,\ldots,p\}$, denote $\delta_T \in \RR^p$ the vector such that $(\delta_{T})_{j} = \delta_j$ if $j\in T$ and $(\delta_{T})_{j}=0$ if $j \notin T$. The support of $\delta$ is defined as $\supp(\delta) = \{ j \in \{1,...,p\}: \delta_j \neq 0\} $.
We denote $a \vee b = \max\{ a, b\}$, and $a \wedge b = \min\{ a , b \}$. The notaion $[k]=\{1,...,k\}$ is used for $k\in\mathbbm N$. We use $c$, $C$ to denote strictly positive constants that is independent of $G$ and $\Pr\in \mathcal P_G$. Their values may change at each presence.
The notation $a_G \lesssim b_G$ denotes $a_G \le C b_G$ for all $G$ and some $C>0$ that does not depend on $G$. $a_G=o(1)$ means that there exists a sequence $(b_G)_{G\ge 1}$ of positive numbers that do not depend on $\Pr\in \mathcal P_G$ for all $G$ such that $|a_G|\le b_G$ for all $G$ and $b_G  = o(1) $ as $G $ converges to zero. $a_G\lesssim_P b_G$ means that for any $\epsilon>0$, there exists $C$ such that $\Pr_\Pr(a_G > C b_G)\le \epsilon$ for all $G$. 
Throughout the paper we assume $G \ge 3$.
 
\subsection{Outline}\label{sec:outline}
The rest of the paper is structured as follows. In Section \ref{sec:overview}, an overview of the method and algorithms are given. Section \ref{sec:main} contains the main asymptotic results. Section \ref{sec:nuisance_parameters} covers algorithms for penalty choices and the auxiliary results for theoretical performance of nuisance parameters. Results of simulation studies are demonstrated in Section \ref{sec:simulation_studies}. In Section \ref{sec:empirical_illustration}, we apply the proposed method to conduct simultaneous testing to verify a statement about gendered language in \cite{Wu18}. We concludes in Section \ref{sec:conclusion}. All the mathematical proofs and additional details are delegated to the appendix.

\section{An Overview}\label{sec:overview} 
Recall that $W_{ig}=(Y_{ig},X'_{ig})'$. Suppose that the researcher observes data sampled from $G$ clusters, $\{W_{ig}:i=1,...,n_g,\:g=1,...,G\}$. Each cluster size $n_g$ is considered non-random, 
and $1\le n_g\le \bar n<\infty$ for a constant $\bar n$ that does not depend on $G$. Denote $n=\sumg n_g$. 
Throughout the paper, we assume that the conditional expectation of $Y$ given $X$ follows the following single-index structure
\begin{align*}
\EP(Y_{ig}|X_{ig})=\Lambda( X'_{ig}\beta^0).
\end{align*}
for each cluster $g$.
Any $W_{i_1 g},W_{i_2 g}$ can be arbitrarily correlated while any $W_{i_1 g_1}, W_{i_2 g_2}$ are independent if $g_1 \ne g_2$.
The dimensionality of $\beta^0$ is allowed to increase with $G$. 
This is the population-averaged approach as $\beta^0$ represents an averaged parameter after integrating out heterogeneity. 
 The target parameter is the APE with respect to the $k$-th continuous covariate of interest,
\begin{align*}
\text{APE}_k=\EP \left[\frac{1}{n}\sumg\sumi\beta^0_k \Lambda'(X'_{ig}\beta^0)\right]
\end{align*}
where $\Lambda'$ stands for the derivative of $\Lambda$. As $n_g \le \bar n  $, it suffices to consider $\alpha_k$, the rescaled APE\footnote{The original APE can be simply recovered by $\text{APE}_k=(G/n)\cdot\alpha_k $.} defined as
\begin{align*}
\alpha_k=&\EP\left[\frac{1}{G}\sumg\sumi \beta^0_k \Lambda'(X'_{ig}\beta^0)\right].
\end{align*}

\subsection{Estimation and Inference Procedures}\label{sec:estimation_and_inference_procedures}
We now summarize the estimation, inference and construction of simultaneous confidence intervals procedures based on the theoretical results to be presented in Section \ref{sec:main} and \ref{sec:nuisance_parameters} ahead. First, we describe the procedures for computing the proposed APE estimators. Set $\alpha_k$ as the parameter of interest. The post-double-selection estimator for $\alpha_k$ is defined as
\begin{align}
\widetilde \alpha_k= \frac{1}{G}\sumg\sumi \check \beta^k_k \Lambda'(X_{ig}'\check\beta^k) \label{eq:estimator_DS}
\end{align}
where  $\check \beta^k$ is the pooled Logit estimate with its support restricted to the set of covariates
\begin{align}
\widetilde T_k=
\{k\}\cup\supp(\hat \beta)\cup \supp(\hat \zeta^k)\cup \supp(\hat \gamma^k),\label{eq:T_support}
\end{align}
and $\hat \beta$, $\hat \zeta^k$ and $\hat \gamma^k$ are nuisance parameter estimators to be defined below. 
Therefore, once $\widetilde T_k$ is obtained, estimation of $\widetilde \alpha_k$ becomes a standard pooled Logit problem. 

Suppose that we have some generic penalty tuning parameters $\lambda$, $\lambda^\gamma_k$ and $\lambda_k^\zeta$ and, in addition, $\hat \Psi$, $\hat\Psi^\gamma_k$, $\hat\Psi_k^\zeta$, diagonal normalization matrices of dimensions $p$, $p-1$ and $p$, respectively. Formal and theoretically justified choices of these objects are delayed to Section \ref{sec:nuisance_parameters}.

First, $\hat \beta$ and its two post-lasso counterparts are defined as
\begin{align}
&\hat \beta \in\underset{\beta \in \Real^p}{\text{argmin}}\:\frac{1}{G} \sumg\sumi \{-Y_{ig} X_{ig}'\beta + \log(1+\exp(X'_{ig}\beta)\,)\}+ \frac{\lambda}{G}\|\hat \Psi \beta \|_1,\label{eq:hat_beta}\\
&\widetilde \beta \in \underset{\beta\in \Real^p:\supp(\beta)\subset\supp(\hat\beta)}{\text{argmin}}\:\frac{1}{G} \sumg\sumi 
\{-Y_{ig} X_{ig}'\beta + \log(1+\exp(X'_{ig}\beta)\,)\},\label{eq:estimator_PL_beta_0}\\
&\widetilde \beta^k \in \underset{\beta\in \Real^p:\supp(\beta)\subset\supp(\hat\beta_{-k})}{\text{argmin}}\:\frac{1}{G} \sumg\sumi \{-Y_{ig} X_{ig}'\beta + \log(1+\exp(X'_{ig}\beta)\,)\}.\label{eq:estimator_PL_beta_k}
\end{align}
Using the above post-lasso estimates, we compute $\hat f^2_{ig}=\Lambda'(X_{ig}'\widetilde\beta)$ and $\hat S_{ig}^k= \widetilde\beta^k_k \cdot \{1-2\Lambda(X_{ig}'\widetilde\beta)\}$. Throughout the rest of this paper, denote $D^j_{ig}=X_{ig,j}$, the $j$-th component of $X_{ig}$, and $X^j_{ig}=X_{ig,-j}'$, the remaining $p-1$ variables. Using these quantities, the remaining two nuisance parameter estimates can be obtained as
\begin{align}
&\hat \gamma^{k}=\:\underset{\gamma\in\RR^{p-1}}{\argmin}\:\frac{1}{G} \sumg\sumi \hat f^2_{ig}(D^{k}_{ig}-X^{k\prime}_{ig}\gamma)^2 + 2\frac{\lambda_k^\gamma }{G}\|\hat\Psi_k^\gamma\gamma\|_1,\label{eq:hat_gamma}\\
&\hat \zeta^k=\:\underset{\zeta\in\RR^{p}}{\argmin}\:\frac{1}{G} \sumg\sumi \hat f^2_{ig}(\hat S^{k}_{ig}-X'_{ig}\zeta)^2 + 2\frac{\lambda_k^\zeta }{G}\|\hat\Psi^\zeta_k\zeta\|_1.\label{eq:hat_zeta}
\end{align}
Now $\widetilde T_k$ can be calculated following (\ref{eq:T_support}) and thus
\begin{align}
 \check \beta^k =& \underset{\beta\in \Real^p:\beta_j=0 \text{ for all } j\in \widetilde T^c_k}{\argmin}\;\frac{1}{G} \sumg \sumi \{-Y_{ig} X_{ig}'\beta + \log(1+\exp(X'_{ig}\beta)\,)\},\label{eq:beta_check}
\end{align}
and $\widetilde \alpha_k$ can be obtained following equation (\ref{eq:estimator_DS}).

Suppose that the researcher is interested in $\alpha_k$ for a set of continuous covariates with $k\in A$ for an index set $A \subset [p]$\footnote{There is no restriction on the cardinality of $A$. $A=[p]$ is also allowed.}. We present a concrete estimation procedure as the following algorithm.
\begin{algorithm}[Post-double-selection estimator]\label{alg:DS_APE}
For each $k \in A$,
 \begin{enumerate}[(1)]
 \item Run lasso and post-lasso Logit to compute $\widetilde \beta$ following (\ref{eq:hat_beta}) and (\ref{eq:estimator_PL_beta_0}).
 \item Define generated weights $\hat f^2_{ig}=\Lambda'(X_{ig}'\widetilde\beta)$.
  \item Run lasso to compute $\hat \gamma^k$ following (\ref{eq:hat_gamma}).
 \item Run lasso to compute $\hat\zeta^k$ following (\ref{eq:hat_zeta}).

 \item Let $\widetilde T_k=\{k\}\cup\supp(\hat \beta)\cup \supp(\hat \zeta^k)\cup \supp(\hat \gamma^k)$ and compute $\check \beta^k$ following (\ref{eq:beta_check}).
 \item Compute plug-in estimator $\widetilde \alpha_k$ following (\ref{eq:estimator_DS}).
 \end{enumerate}
 \end{algorithm} 
\begin{remark}\label{remark:double-selection}
The post-double-selection estimator is theoretically related to the post-double-selection estimators for linear models in \cite{BCH14} and for Logit regression coefficients in \cite{BCW16} and \cite{BCCW18}. However, because our target parameters of interest are APEs, the nonlinear transformations of high-dimensional nuisance parameters, rather than regression coefficients themselves, the structure of our nuisance parameters are fundamentally different. Estimation of these nuisance parameters requires different strategies and therefore presents extra challenges. We discuss the theory of nuisance parameters estimation in Section \ref{sec:nuisance_parameters}.
\end{remark} 
For inference, let us define the post-lasso counterparts of $\hat \gamma^{k}$ and $\hat \zeta^{k}$
\begin{align}
\widetilde \gamma^{k}=&\underset{\supp(\gamma)\subset \supp(\hat \gamma^k)}{\argmin}\frac{1}{G} \sumg\sumi \hat f^2_{ig}(D^{k}_{ig}-X^k_{ig}\gamma)^2, \label{eq:estimator_PL_gamma_j}\\
\widetilde \zeta^k=&\underset{\supp(\zeta)\subset \supp(\hat \zeta^k)}{\argmin}\frac{1}{G} \sumg\sumi \hat f^2_{ig} (\hat S^k_{ig}-X'_{ig}\zeta)^2,\label{eq:estimator_PL_zeta_k}
\end{align}
and the nuisance parameter estimate
\begin{align}
\widetilde \theta^k=[-\widetilde \gamma^k_1,...,-\widetilde \gamma^k_{k-1},1,-\widetilde \gamma^k_k,...,-\widetilde \gamma^k_{p-1}]'\cdot\left\{\frac{1}{G\hat \tau_k^2} \sumg \sumi \hat f_{ig}^2\right\}, \label{eq:estimator_theta_k}
\end{align}
where each $\hat \tau^2_k$ is calculated using
\begin{align}
\hat \tau_{k}^2:=\frac{1}{G} \sumg\sumi \hat f_{ig}^2(D^k_{ig} - X^k_{ig}\widetilde \gamma^k )^2. \label{eq:tau_2}
\end{align}
Define the additional nuisance parameter estimate
\begin{align}
\widetilde \mu^k = \widetilde \zeta^k + \widetilde \theta^k.  \label{eq:post_mu}
\end{align}
Finally, define the variance estimate as
 \begin{align}
\widetilde \sigma^2_k
=&\frac{1}{G} \sumg \left\{\sumi \left(\widetilde \alpha_k  \left(\frac{G}{n}\right)- \widetilde\beta_k \Lambda'(X_{ig}'\widetilde\beta)+\widetilde\mu^{k\prime}X_{ig} 
  \{Y_{ig}-\Lambda(X_{ig}'\widetilde\beta)\} \right)\right\}^2. \label{eq:estimator_variance}
\end{align}
We are now ready to introduce a procedure for simultaneous inference. Suppose that the null hypothesis of interest is 
\begin{align*}
\text{H}_0: \alpha_k=\alpha_k^0 \text{ for all $k\in A$}
\end{align*}
for some values $(\alpha^0_k)_{k\in A}$. We present a concrete simultaneous inference procedure as the following algorithm.
 \begin{algorithm}[Simultaneous inference via nultiplier cluster bootstrap]\label{alg:simultaneous_inference}
For each $k\in A$,
 \begin{enumerate}[(1)]
 \item Compute $\widetilde\sigma_k$ for $k\in A$ following (\ref{eq:estimator_variance}).
 \item Compute the test statistic
$
 T=\max_{k\in [p]}\sqrt{G}\widetilde \sigma_k^{-1} |\widetilde \alpha_k -  \alpha_k^0|.
$
\item For each $k\in A$, compute $\widetilde\mu^k$ following (\ref{eq:post_mu}).
\item Set the number of bootstrap iterations to $B$. For each $b\in[B]$, generate i.i.d. standard normal random variables $\{\xi^b_g\}_{g=1}^G$ independently from data.
\item For each $k\in A$ and $b\in[B]$, compute 
\begin{align}
W^b=\max_{k\in A}\Big|\frac{1}{\sqrt{G}\widetilde\sigma_k}\sumg \xi_g^b\sumi \left(\widetilde\alpha_k \left(\frac{G}{n}\right)-\widetilde \beta^k_k \Lambda'(X_{ig}'\widetilde\beta)+\widetilde\mu^{k\prime}X_{ig} 
  \{Y_{ig}-\Lambda(X_{ig}'\widetilde\beta)\}\right)\Big| \label{eq:multiplier_process_Wb}
\end{align} 
and $c_a$, the $(1-a)$-th quantile of $\{W^b\}_{b=1}^B$.
\item If $T>c_a$, reject $\text{H}_0$. Otherwise do not reject $\text{H}_0$.
 \end{enumerate}
 \end{algorithm}
 Finally, we illustrate the procedure for constructing simultaneously valid confidence intervals with $(1-a)$ coverage probability for $\alpha_k$, $k\in A$.
 \begin{algorithm}[Simultaneous confidence intervals via multiplier cluster bootstrap]\label{alg:simultaneous_confidence_intervals}
  For each $k\in A$,
  \begin{enumerate}[(1)]
  \item Compute $\widetilde \sigma^2_k$ for $k\in A$ following (\ref{eq:estimator_variance}).
\item Set the number of bootstrap iterations to $B$. For each $b\in[B]$, generate i.i.d. standard normal random variables $\{\xi^b_g\}_{g=1}^G$ independently from data.
\item For each $k\in A$ and $b\in[B]$, compute $W^b$ following (\ref{eq:multiplier_process_Wb})
and $c_a$, the $(1-a)$-th quantile of $\{W^b\}_{b=1}^B$.
\item Compute simultaneous confidence intervals $I=\times_{k\in A} I_k $, where $I_k=\widetilde \alpha_k\pm \widetilde \sigma_k \cdot c_a/\sqrt{G}$.
 \end{enumerate}
 \end{algorithm}
 \begin{remark}\label{rem:normalization_inference}
 Note that it is also possible to conduct multiple testing and simultaneous confidence intervals without normalization (studentization). To do so, one simply follows every step in Algorithms \ref{alg:simultaneous_inference} and \ref{alg:simultaneous_confidence_intervals} with $1$ in place of $\widehat \sigma_k$ for all $k$. 
 \end{remark}
\section{Main Theoretical Results}\label{sec:main}
In this section, we present our main theoretical results for simultaneous inference and construction of confidence intervals. These results justify the validity of the algorithms proposed in Section \ref{sec:overview}.
First, we introduce some notations.
Recall
\begin{align*}
\EP(Y_{ig}|X_{g})=\EP(Y_{ig}|X_{ig})=\Lambda( X'_{ig}\beta^0).
\end{align*}
Define the Neyman orthogonal score for $\alpha_k$ by
\begin{align}
\bar\psi_k (W_{ig},\alpha,\eta)
 =&\alpha \cdot\frac{G}{n}- \beta_k \Lambda'(X_{ig}'\beta)+\mu'X_{ig} 
  \{Y_{ig}-\Lambda(X_{ig}'\beta)\}\label{eq:neyman_orthogonal_score_APE}\\
=& \alpha \cdot\frac{G}{n} -\psi_k (W_{ig},\eta), \nonumber
\end{align}
where $\psi_k (W_{ig},\eta)= \beta_k \Lambda'(X_{ig}'\beta)-\mu'X_{ig} \{Y_{ig}-\Lambda(X_{ig}'\beta)\}$. In addition, let the ``ideal" population nuisance parameters\footnote{See Section \ref{sec:orthogonalization} in the Appendix for derivation of this moment condition.} for $\alpha_k$ be $\eta^k=(\beta^{0\prime},\mu^{k\prime})'\in \Real^{2p}$ with
\begin{align}
\mu^k=&\zeta^k + \theta^k,\label{eq:pop_mu_k}\\
 \zeta^k =&\left\{\EP\left[\frac{1}{G}\sum_{g=1}^{G}  \sumi f^2_{ig} X_{ig} X_{ig}'\right]\right\}^{-1}  \EP\left[\frac{1}{G}\sum_{g=1}^{G} \sumi f^2_{ig} X_{ig} S_{ig}^k\right] \label{eq:pop_zeta_k},\\
\theta^k=&\left\{\EP\left[\frac{1}{G}\sum_{g=1}^{G}  \sumi f^2_{ig} X_{ig} X_{ig}'\right]\right\}^{-1} \EP\left[\frac{1}{G}\sum_{g=1}^{G} \sumi f^2_{ig} e_k\right], \label{eq:pop_theta_k}
\end{align}
where
$S^k_{ig}=\beta_k^0 (1-2\Lambda(X_{ig}'\beta^0))$ is an auxiliary regressor and $f^2_{ig}=\Lambda'(X_{ig}'\beta^0)$ is a regression weight.
Also denote the population nodewise regression coefficients for the $j$-th covariate as $\gamma^j$. We can also rewrite the population nuisance parameter regression coefficients $\zeta^j$ as a weighted projection of $S^j_{ig}$ on $X_{ig}$. Thus, we have the following
\begin{align}
\gamma^{j}=\underset{\gamma \in \Real^{p-1}}{\argmin} \EP\left[\frac{1}{G}\sum_{g=1}^{G} \sumi f^2_{ig} (D^{j}_{ig}-X^j_{ig}\gamma)^2\right], \label{eq:population_gamma}\\
\zeta^{j}=\underset{\gamma \in \Real^{p}}{\argmin} \EP\left[\frac{1}{G}\sum_{g=1}^{G} \sumi f^2_{ig} (S^{j}_{ig}-X_{ig}\zeta)^2\right]. \label{eq:population_zeta}
\end{align}
Denote the projection errors by $Z^j_{ig}=D^j_{ig}-X^j_{ig}\gamma^j$ and $\varepsilon^j_{ig}=S^j_{ig}-X_{ig}\zeta^j$. Let $q>4$ be a constant independent of $G$. Let $c_1$ and $C_1$ be some strictly positive constants independent of $G$. Furthermore, let $a_G=p \vee G$ and $\check \delta_G $ be a sequence of positive constants that converge to zero. $M_{G,1}\ge 1$ and $M_{G,2}\ge 1$ be some sequence of positive constants possibly diverging to infinity. $s=s_G$ is a non-decreasing sequence of constants. We make the follow assumptions.
\begin{assumption}[Parameters]\label{a:parameters}
	The true parameters satisfy that
\begin{align*}
 \|\beta^0\|_2 + \max_{j \in [p]} \|\gamma^j\|_2 + \max_{k \in [p]} \|\zeta^k\|_2\le C_1.
\end{align*} 
\end{assumption}
\begin{assumption}[Sparsity]\label{a:sparsity}
There exist vectors $\bar\gamma^j\in \Real^{p-1}$ and $\bar \zeta^k\in \Real^{p}$ for all $j,k \in [p]$ such that
\begin{align*}
\|\beta^0\|_0 + \max_{j \in [p]}\|\bar \gamma^j\|_0  +  \max_{k \in [p]}\|\bar \zeta^k\|_0  \le s
\end{align*}
and 
\begin{align*}
\max_{j,k \in [p]} (\|\bar \gamma^j - \gamma^j \|_2\vee \|\bar \theta^k - \theta^k \|_2 + s^{-1/2}\|\bar \gamma^j - \gamma^j \|_1\vee \|\bar \zeta^k - \zeta^k \|_1)\le C_1 (s \log a_G/G)^{1/2}.
\end{align*}
\end{assumption}
\begin{remark}\label{rem:approximately_sparse_nuisance_para}
Assumption \ref{a:parameters} requires bounded $\ell_2$ norm of nuisance parameters, which is mild and standard in the lasso literature. The $\ell_1$ norm of the nuisance parameters are allowed to be growing with $G$. 
Note that we do not require exact sparsity of $\gamma^j$ and $\zeta^k$ in Assumption \ref{a:sparsity} 
since the exact sparsity of nodewise lasso coefficients could be more difficult to justify in many applications.
Also, note that for each $j\in[p]$, we can without loss of generality assume $\bar \gamma^j= \gamma_T^j$, where $T=\supp(\gamma^j)$. The same applies to $\bar\zeta^k$ and $\zeta^k$.
\end{remark}

For the following assumption, define $U_{gk}= \bar n\cdot \max_{i \in [n_g]} |X_{ig,k}|$, $U_g=[U_{gk}]_{k \in [p]}$ and $V^j_g=\max_{i \in [n_g]} (|Z^j_{ig}|\vee |\varepsilon^j_{ig}|)$.
\begin{assumption}[Covariates]\label{a:covariates}
Suppose that there exist finite positive constants $c_1$, $C_1$ and $\check \delta_G=o(1)$ such that the following moment conditions hold for all $G$,
\begin{enumerate}
\item $\inf_{\|\xi\|_2=1}  \left(\EP[G^{-1}\sum_{g=1}^{G} \sumi (f_{ig} X_{ig}'\xi)^2]\bigwedge  \EP[G^{-1}\sum_{g=1}^{G} (\sumi \{Y_{ig}-\Lambda(X_{ig}'\beta^0)\} X_{ig}'\xi)^2]\right)\ge c_1$.
\item $\min_{j,k} \left(\EP[G^{-1}\sum_{g=1}^{G} (\sumi f_{ig}^2 Z^j_{ig} X_{ig,k})^2]\bigwedge\EP[G^{-1}\sum_{g=1}^{G} (\sumi f_{ig}^2 X_{ig,j} X_{ig,k})^2]\right)\ge c_1$. 
\item $\max_{j,k}\{ \EP[G^{-1}\sum_{g=1}^{G}  | V^j_{g} U_{gk}|^3]\}^{1/3}\log^{1/2} a_G\le \check \delta_G G^{1/6}$.
\item $\sup_{\|\xi\|_2=1} \EP[G^{-1}\sum_{g=1}^{G}  (U_g'\xi)^4]+ \max_{j\in[p]} \EP[G^{-1}\sum_{g=1}^{G}  (V^j_g)^4]\le C_1$.
\item $M_{G,1}\ge \{ \EP[G^{-1}\sum_{g=1}^{G} \max_{j\in [p]}|V^j_{g}|^{2q}]\}^{1/2q}$.
\item $M_{G,1}^4 s \log a_G \le \check \delta_G G^{1/2-1/q}$.
\item $M_{G,2}\ge \{ \EP[G^{-1}\sum_{g=1}^{G} \|U_g\|_\infty^{2q}]\}^{1/2q}$.
\item $M_{G,2}^4 s \log a_G \le \check \delta_G G^{1/2-1/q}$. 
\item $(M_{G,1}^2 \vee s \log^2 a_G)M_{G,2}^4 s \le \check\delta_G G^{1-3/q}$.
\end{enumerate}
\end{assumption}

\begin{assumption}[Sparse eigenvalues]\label{a:sparse_eigenvalues}
Let $\Delta(m)=\{\delta\in \Real^p:\|\delta\|_0\le m, \|\delta\|_2=1\}$. With probability at least $1-C(\log G)^{-1}$, we have
\begin{align*}
&1\lesssim \min_{j\in [p]}\min_{\delta \in \Delta(Cs)}\frac{1}{G} \sumg\sumi( Z^j_{ig} X_{ig}'\delta)^2 \le \max_{j\in [p]}\max_{\delta \in \Delta(Cs)}\frac{1}{G} \sumg\sumi( Z^j_{ig} X_{ig}'\delta)^2 \lesssim 1,\\
&1\lesssim \min_{\delta \in \Delta(Cs)}\frac{1}{G} \sumg\sumi( X_{ig}'\delta)^2 \le \max_{\delta \in \Delta(Cs)}\frac{1}{G} \sumg\sumi( X_{ig}'\delta)^2 \lesssim 1.
\end{align*}
\end{assumption}

\begin{remark}\label{rem:sufficient_conditions}
Assumptions \ref{a:parameters}, \ref{a:sparsity}, \ref{a:covariates} are the cluster sampling counterpart of the Assumptions 3.1, 3.2, 3.4 and 3.5 of \cite{BCCW18}.  To deal with APEs, however, we do need extra conditions on the growth of some moments that are listed below in the statement of Theorem \ref{theorem:main_sufficient}. These growth conditions are satisfied when, for example, the covariates are sub-gaussian and/or uniformly bounded. When regressors are uniformly bounded, which is assumed in both \cite{WZ17} and \cite{HW18}, the rate requirement would be $s\log p/G^{1/2}=o(1)$ ($s^{3/2}\log p/G^{1/2}=o(1)$ is required by \cite{WZ17}). Assumption \ref{a:sparse_eigenvalues} is analogous to condition SE in \cite{BCHK16} for the linear panel data model. 
\end{remark}

\begin{theorem}[Main result]\label{theorem:main_sufficient}
Suppose that Assumptions \ref{a:parameters}, \ref{a:sparsity}, \ref{a:covariates}, \ref{a:sparse_eigenvalues} hold, then
\begin{enumerate}
\item The following uniform Bahadur representation holds with probability at least $1-C(\log G)^{-1}$
\begin{align*}
&\sup_{\Pr\in\mathcal P_G}\max_{1 \le k \le p}\Bigg|\sqrt{G}\sigma_{k}^{-1}
(\hat \alpha_k -\alpha_k)
-
\frac{1}{\sqrt{G}}\sumg \sumi \varphi_k(W_{ig},\alpha_k,\eta^k)\Bigg|\lesssim \delta_G,
\end{align*}
where $\varphi_k(W_{ig},\alpha,\eta)=- \bar\psi_k(W_{ig},\alpha,\eta)/\sigma_k$ and $\eta^k=(\beta^{0\prime},\mu^{k\prime})'$.
\item  Let $c_W(a)$ be the $a$-th quantile of $W$, we have, with probability at least $1-C(\log G)^{-1}$,
\begin{align*}
\sup_{\Pr\in \mathcal P_G}\sup_{\alpha\in (0,1)}
\Big|\Pr_\Pr\left(\max_{1 \le k \le p}|\sqrt{G}\sigma_{k}^{-1}
(\hat \alpha_k -\alpha_k)|\le c_W(a)\right) 
-
a\Big|=o(1).
\end{align*}
 That is to say, the algorithms in Section \ref{sec:overview} provide valid simultaneous inference and confidence intervals asymptotically.
\end{enumerate}

\end{theorem}
A proof can be found in Section \ref{sec:proof for theorem:main_sufficient} in the Appendix. 
Now, it remains to find a valid variance estimator. Recall the variance estimator $\widetilde \sigma_k^2 $ defined in (\ref{eq:estimator_variance}). Denote $\widetilde \sigma_k = \{\widetilde \sigma_k^2\}^{1/2} $.

\begin{lemma}[Variance estimator]\label{lemma:variance_est}
Suppose that the conditions for Theorem \ref{theorem:main_sufficient} hold.
Then
\begin{align*}
\max_{k\in [p]}|\widetilde \sigma_k - \sigma_k |\lesssim(\log a_G)^{-1}
\end{align*}
with probability at least $1-C(\log G)^{-1}$.

\end{lemma}
A proof can be found in Section \ref{sec:proof for lemma:variance_est} in the Appendix.

\section{Nuisance Parameters}\label{sec:nuisance_parameters}
\text{}\\
Recall that the ``ideal" nuisance parameter vector $\eta^k=(\beta^{0\prime},\mu^{k\prime})'$, where
\begin{align*}
\mu^k=
&\left\{\EP\left[\frac{1}{G}\sum_{g=1}^{G}  \sumi f_{ig}^2 X_{ig}X_{ig}'\right]\right\}^{-1} \cdot\EP\left[\frac{1}{G}\sum_{g=1}^{G} \sumi\left( \beta^0_k \Lambda''(X_{ig}'\beta^0)X_{ig}+  \Lambda'(X_{ig}'\beta^0)e_k\right)\right]=\zeta^k+ \theta^k.
\end{align*}
In this section, we propose estimators for these nuisance parameters as well as some theoretically justified choices of penalty tuning parameters. The choices here are based on the moderate deviation theory of self-normalized sums, which is first adapted for penalty selection of lasso by \cite{BCCH12}.   Throughout this section, we fix a positive integer $\bar m \ge 1$ as the number of iterations used in the algorithms for choosing penalty tuning parameters.
\subsection{Post-Lasso Logit and Estimation of $\beta^0$}\label{sec:LOG}
We now establish an asymptotic theory for estimation of $\beta^0$, which plays a central role in estimation of APE.
The identification of $\beta^0$ follows from quasi-maximum likelihood and the assumption of population-averaged approach $\Ep[Y_{ig}|X_{ig}]=\Lambda(X_{ig}'\beta^0)$. Define the negative partial log-likelihood function by
\begin{align}
M (Y_{ig},X_{ig},\beta)=-\{Y_{ig} X_{ig}'\beta - \log(1+\exp(X'_{ig}\beta))\}\label{eq:LOG_loss_func}.
\end{align} 
Then, one has
\begin{align*}
\beta^0=\underset{\beta \in \Real^p}{\text{argmin}}\: \EP\left[\frac{1}{G}\sum_{g=1}^{G} \sumi M (Y_{ig},X_{ig},\beta)\right].
\end{align*}
We propose the following algorithm for the choice of $\hat \Psi$.
\begin{algorithm}[Penalty choice: clustered Lasso logit $\beta^0$]\label{alg:LOG}
Define $\lambda=c\sqrt{G}\Phi^{-1} (1-\gamma/2p)$ and set $c=1.1$ and $\gamma=0.1/\log G$. For $m=0$, let
\begin{align*}
\hat l_{j,0}=\frac{1}{2}\left\{\frac{1}{G} \sumg\left(\sumi n_g X^2_{ig,j}\right)\right\}^{1/2}
\end{align*}
and for $1\le m\le \bar m$, 
\begin{align*}
\hat l_{j,m}=\left\{\frac{1}{G} \sumg\left(\sumi \{Y_{ig}-\Lambda(X_{ig}\widetilde \beta)\} X_{ig,j}\right)^2\right\}^{1/2}
\end{align*}
with $\widetilde \beta$ coming from iteration $m-1$. Let $\hat \Psi=\diag\{\hat l_{j,m}:j\in[p]\}$.
\end{algorithm}

The following result provides convergence rates of $\widetilde \beta$ and $\widetilde \beta^k$.
\begin{theorem}\label{thm:LOG_rates}
Suppose that the Assumption \ref{a:parameters}, \ref{a:sparsity}, \ref{a:covariates} and \ref{a:sparse_eigenvalues} are satisfied. If $\check \delta_G^2 \log a_G =o(1)$, then with penalty chosen according to Algorithm (\ref{alg:LOG}), with probability $1- \gamma$, $\gamma=O(1/\log G)$,
\begin{align*}
\|\widetilde \beta - \beta^0\|_1\vee\max_{k\in[p]}\|\widetilde \beta^k - \beta^0\|_1 \lesssim \sqrt{\frac{s^2\log a_G}{G}} \text{ and } \|\widetilde \beta - \beta^0\|_2 \vee \max_{k\in[p]}\|\widetilde \beta^k - \beta^0\|_2 \lesssim\sqrt{\frac{s\log a_G}{G}}.
\end{align*}

\end{theorem}
A proof can be found in Section \ref{sec:proof for theorem:LOG_rates} in the Appendix.

\subsection{Weighted Post-Lasso with Estimated Weights}\label{sec:REGNPL}
We now establish asymptotic theory for weighted post-lasso with estimated weights that will be essential for Sections \ref{sec:Nodewise_PL} and \ref{sec:REG}. 
We propose the following algorithm for the choices of penalty tuning parameters.
\begin{algorithm}[Penalty choice: weighted clustered Lasso $\gamma^j$]\label{alg:REGNPL}
Define $\lambda^\gamma=c\sqrt{G}\Phi^{-1} (1- \gamma/2p(p-1))$ and set $c=1.1$ and $\gamma=0.1/\log G$. For each $j\in[p]$, for $m=0$, set 
\begin{align*}
\hat l_{jk,0}=&2\max_{ g \in[G]}\max_{ i\in [n_g]}|\hat f_{ig}X_{ig,k}|\left\{\frac{1}{G} \sumg\left(\sumi \hat f_{ig} D^j_{ig}\right)^2\right\}^{1/2} 
\end{align*}
and $1\le m\le \bar m$,
\begin{align*}
\hat l_{jk,m}=& 2\left\{\frac{1}{G} \sumg\left(\sumi \hat f_{ig}^2 (D^j_{ig}-X^j_{ig}\widetilde \gamma^j)X^j_{ig,k}\right)^2\right\}^{1/2}
\end{align*}
and $\hat \Psi^\gamma_j=\diag\{\hat l_{jk,m}:k\in[p-1]\}$.
\end{algorithm}
The following result provides convergence rates of $\widetilde\gamma^j$, which plays an important role in Section \ref{sec:Nodewise_PL}.
\begin{theorem}\label{thm:REGNPL_rates}
Suppose that Assumption \ref{a:parameters}, \ref{a:sparsity}, \ref{a:covariates}, \ref{a:sparse_eigenvalues} are satisfied and if $\check \delta_G^2 \log a_G =o(1)$, then with penalty chosen according to Algorithm (\ref{alg:REGNPL}), with probability $1- \gamma$, $\gamma=O(1/\log G)$
\begin{align*}
\max_{j \in [p]}\|\widetilde \gamma^j - \gamma^j\|_1 \lesssim \sqrt{\frac{s^2\log a_G}{G}}
\:\text{ and }\:
\max_{j \in [p]}\|\widetilde \gamma^j - \gamma^j\|_2 \lesssim \sqrt{\frac{s\log a_G}{G}}.
\end{align*}

\end{theorem}
A proof can be found in Section \ref{sec:proof for thm:REGNPL_rates} in the Appendix.

\subsection{Nodewise Post-Lasso and Estimation of $\theta^k$}\label{sec:Nodewise_PL}
Now we provide estimators for $\theta^k$ that are built upon the method of cluster nodewise post-lasso estimator for approximately inverting a singular matrix. The theory developed here is based on applying the weighted post-lasso with estimated weights from \cite{BCCW18} to the panel nodewise regressions of \cite{Kock16}. Recall that each nuisance parameter vector
$\theta^k$ contains the matrix
\begin{align*}
\Theta:=\left\{ \EP\left[\frac{1}{G}\sum_{g=1}^{G} \sumi f^2_{ig}X_{ig}X'_{ig}\right]\right\}^{-1}.
\end{align*}
Its sample counterpart is not invertible if $p>n$ and could be very unstable if $p$ is only moderately larger than $n$. Here, we take advantage of Assumption \ref{a:sparsity} to construct a high quality approximate inverse estimate. Denote $\Theta_{j}$ for the $j$-th row written as a column vector. If we can find some reasonable estimator $\hat\Theta_k$ for $\Theta_k$, then intuitively an estimator for $\theta^k$ can be defined as 
\begin{align*}
\widetilde \theta^k = \hat \Theta_k  \cdot \frac{1}{G} \sumg \sumi \hat f_{ig}^2. 
\end{align*}
We propose a cluster nodewise post-lasso procedure to estimate $\Theta$.
Recall that the error
$Z^j_{ig}=D^{j}_{ig}- X^j_{ig}\gamma^{j}$ which satisfies $ \EP[\frac{1}{G}\sum_{g=1}^{G} \sumi f^2_{ig} X^j_{ig} Z^j_{ig}]=0$.
Define the error variance $\tau^2_{j}= \EP[\frac{1}{G}\sum_{g=1}^{G} \sumi f^2_{ig} ( Z_{ig}^{j})^2]$. Some properties of $\tau_j^2$ can be found in Section \ref{sec:properies_tau_2}. Note that $\gamma^j$ has a sparse approximation $\bar \gamma^j$ under Assumption \ref{a:parameters}.
Then, we can use post-lasso estimate $\widetilde \gamma^j$ for $\gamma^j$ from Section \ref{sec:REGNPL} and construct a $p\times p$ matrix $\hat C$ by
\begin{align*}
\hat C=
\begin{bmatrix}
1 & -\widetilde \gamma^1_{1 }& \dots & -\widetilde \gamma^1_{p-1 }\\
-\widetilde \gamma^2_{ 1 }& 1 & \dots &-\widetilde \gamma^2_{p -1}\\
\vdots &\vdots &\ddots&\vdots\\
-\widetilde \gamma^p_{ 1 }&-\widetilde \gamma^p_{ 2 }&\dots& 1
\end{bmatrix}.
\end{align*}
That is, the off-diagonal spots of the $j$-th row of $\hat C$ consist of components of $- \widetilde \gamma^j$ and the diagonal entries are set to $1$. 
Also, denote
\begin{align*}
\hat T^2=\text{diag}\{\hat \tau_{1}^2,...,\hat \tau_{p}^2\},
\end{align*}
where
$
\hat \tau_{j}^2$ is defined in (\ref{eq:tau_2}).
Now, the cluster nodewise post-lasso estimator for $\Theta$ is defined as
\begin{align*}
\hat \Theta=\hat T^{-2}\hat C,
\end{align*}
which in turn gives the expression of (\ref{eq:estimator_theta_k}).
The following results provide validity of $\hat \Theta$ and $\widetilde \theta^k$.

\begin{lemma}\label{lemma:Theta_rates}
Suppose that the Assumption \ref{a:parameters}, \ref{a:sparsity}, \ref{a:covariates}, \ref{a:sparse_eigenvalues} are satisfied. If $\check \delta_G^2 \log a_G =o(1)$, then with penalty chosen according to Algorithm \ref{alg:REGNPL}, with probability $1- \gamma$, $\gamma=O(1/\log G)$,
\begin{align*}
&\max_{j\in [p]}\|\hat \Theta_{j} -\Theta_{j}\|_1\lesssim \sqrt{\frac{s^2\log a_G}{G}} \text{ and } \max_{j\in [p]}\|\hat \Theta_{j} -\Theta_{j}\|_2\lesssim \sqrt{\frac{s\log a_G}{G}}.
\end{align*}


\end{lemma}
\begin{theorem}\label{thm:theta_k_rates}
Suppose that all assumptions required by Lemma \ref{lemma:Theta_rates} are satisfied. Then, with probability $1- \gamma$, $\gamma=O(1/\log G)$, we have
\begin{align*}
\max_{k\in[p]}\|\widetilde \theta^k - \theta^k \|_1\lesssim \sqrt{\frac{s^2\log a_G}{G}} \text{ and }\max_{k\in[p]}\|\widetilde \theta^k - \theta^k \|_2\lesssim \sqrt{\frac{s\log a_G}{G}}.
\end{align*}
\end{theorem}
Proofs for the above two results can be found in Sections \ref{sec:proof for lemma:Theta_rates} and \ref{sec:proof for thm:theta_k_rates} in the Appendix. 
\subsection{Weighted Post-Lasso and Estimation of $\zeta^k$}\label{sec:REG} 
Recall that the nuisance parameters $\zeta^k$ is identified by
\begin{align*}
\zeta^k=\underset{\zeta \in \Real^p}{\text{argmin}}\: \EP\left[\frac{1}{G}\sum_{g=1}^{G} \sumi  f_{ig}^2 ( S_{ig}^k - X'_{ig}\zeta )^2\right].
\end{align*}
We propose the following algorithm for choice of the penalty tuning parameters.
\begin{algorithm}[Penalty Choice: Weighted Clustered Lasso $\zeta^k$]\label{alg:REG}
Define $\lambda^\zeta_j=c\sqrt{G}\Phi^{-1} (1- \gamma/2p^2)$ and set $c=1.1$ and $\gamma=0.1/\log G$. For each $k\in[p]$, for $m=0$, set 
\begin{align*}
\hat l_{kj,0}=&2\max_{ g \in[G]}\max_{i \in [n_g]}| \hat f_{ig}X_{ig,j}|\left\{\frac{1}{G} \sumg\left(\sumi \hat f_{ig} \hat S^k_{ig}\right)^2\right\}^{1/2} 
\end{align*}
and $1\le m\le \bar m$,
\begin{align*}
\hat l_{kj,m}=& 2\left\{\frac{1}{G} \sumg\left(\sumi \hat f_{ig}^2 (\hat S^k_{ig}-X_{ig}'\widetilde \zeta^k)X_{ig,j}\right)^2\right\}^{1/2}
\end{align*}
and $\hat \Psi^\zeta_k=\diag\{\hat l_{kj,m}:j\in[p]\}$.
\end{algorithm}
The following result provides convergence rates of $\widetilde\zeta^k$.
\begin{corollary}\label{corollary:weighted_lasso_rates} 
Suppose that Assumptions \ref{a:parameters}, \ref{a:sparsity}, \ref{a:covariates}, \ref{a:sparse_eigenvalues} hold. If $\check \delta_G^2 \log a_G =o(1)$, then with penalty chosen according to Algorithm \ref{alg:REG}, with probability $1- \gamma$, $\gamma=O(1/\log G)$,
\begin{align*}
\max_{k\in[p]}\|\widetilde \zeta^k - \zeta^k \|_1 \lesssim \sqrt{\frac{s^2\log a_G}{G}} \text{ and } \max_{k\in[p]}\|\widetilde \zeta^k - \zeta^k \|_2 \lesssim \sqrt{\frac{s\log a_G}{G}}. 
\end{align*}
\end{corollary}
A proof can be found in Section \ref{sec:proof for corollary:weighted_lasso_rates} in the Appendix. 

\section{Simulation Studies}\label{sec:simulation_studies}
In this section, we conduct simulation studies to investigate the finite-sample performance of the proposed procedures.
Set the number of total observations to $n$, and each observation is then randomly assigned into $G_0$ clusters. The empty clusters, if they exist, are then discarded and thus $G\le G_0$. For DGP1, let the number of covariates for each observation be $p=1.5\cdot G_0$ and
\begin{align*}
&\beta^0=[1,\beta_2,1/3,1/4,1/5,..,1/19,1/20,0,...,0]' \in \Real^p.
\end{align*}
The first component of each covariate vector is set to $1$ and the rest of the subvector, $X_{ig,-1}$, can be decomposed into an idiosyncratic part $X^1_{ig}$ and a cluster-wise component $ X^2_g$ as
\begin{align*}
X_{ig,-1}= X^1_{ig} + X^2_g
\end{align*}
and both $X^1_{ig}$ and $X^2_g$ are i.i.d. following a multivariate normal distribution with mean $0$ and a Toeplitz covariance matrix:
\begin{align*}
\Sigma_{ij} (\rho):= \rho^{|i-j|}, \; \rho=0.1,\,0.3,\,0.5,\,0.7,\,0.9,\; i,j,\in[p-1].
\end{align*}
So the larger $\rho$ is, the more correlated the covariates are.
The outcome variable is generated by
\begin{align*}
\qquad Y_{ig}= \1\left\{X_{ig}'\beta^0 + U_{ig} >0\right\},
\end{align*}
where the error term can also be decomposed into an idiosyncratic term and a cluster-wise term as
\begin{align*}
U_{ig} = \Lambda\left(\Phi^{-1} ( U^1_{ig} + U^2_{g}) \right),
\end{align*}
where both $U^1_{ig} $ and $U^2_g$ are i.i.d. following the normal $N(0,1/2)$ distribution. Thus, $U_{ig}$ is a standard logistic distribution. Thus both covariates and errors are correlated within each cluster. To consider ``outliers" and
substantial skew and kurtosis in marginal distribution of independent variables, we also consider alternative DGPs inspired by \cite{KS12} by setting $X^1_{ig}$ and $X^2_g$ to follow
a mixture between two distributions, $N(0,\Sigma(\rho))$ with probability 0.9 and a $N(0,\Sigma(\rho))-1.5\times N(1,\Sigma(\rho))$ with probability $0.1$.

\begin{table}
\caption{List of DGPs in Simulation Studies.}
\qquad\\
\begin{tabular}{cc}
DGP & Descriptions\\
\hline
\hline
M1&  $X^1_{ig},\,X^2_g\sim N(0,\Sigma(\rho))$ with $\rho=0.1$\\
M2&Same as M1 except $\rho=0.3$\\

M3&Same as M1 except $\rho=0.5$\\

M4&Same as M1 except $\rho=0.7$\\

M5&Same as M1 except $\rho=0.9$\\
\hline
\hline
M6&  $X^1_{ig},\,X^2_g\sim \left(N(0,\Sigma(\rho))-1.5* B(1,0.1)*N(1,\Sigma(\rho))\right)$ with $\rho = 0.1$\\

M7&Same as M6 except $\rho=0.3$\\

M8&Same as M6 except $\rho=0.5$\\

M9&Same as M6 except $\rho=0.7$\\

M10&Same as M6 except $\rho=0.9$\\
\hline
\hline
\end{tabular}\label{table:DGPs}
\end{table} 

Note that for the DGPs with high $\rho$ values, such as M4, M5, M9 and M10, the approximate sparsity conditions in Assumption \ref{a:parameters} are violated.  We conduct three sets of simulations. First we examine one-dimensional confidence interval coverage for $\alpha_2$ with true underlying $\beta_2\in\{0,0.25,0.5,0.75,1\}$. Our second goal is to construct simultaneous confidence intervals that control the family-wise error rate for $\alpha_k$ for $k\in A$, $A$ is set to be
\begin{align*}
&A_{1}=\{2\},\, A_{2}=\{2,3\},\,
A_{3}=\{2,3,4\},
\,A_{5}=\{2,3,...,6\},\,A_{10}=\{2,3,...,10\},\,A_{20}=\{2,3,...,20\},\\
&A_{30}=\{2,3,...,31\},\,A_{40}=\{2,3,...,41\},
\,A_{50}=\{2,3,...,51\},\,A_{100}=\{2,3,...,101\},
\end{align*}
where the APE with respect to the intercept is always omitted. In this group of simulations, we set $\beta_2=0.5$. Finally, we examine the asymptotic behaviors of coverage probabilities of simultaneous intervals for $A_{10}$. 

The estimation of all lasso and lasso Logit are conducted using R package \textbf{glmnet} and the penalty choices follow Algorithms \ref{alg:LOG}, \ref{alg:REGNPL} and \ref{alg:REG} in Section \ref{sec:nuisance_parameters} with $\bar m=1 $. For each iteration of the simulation, we set the number of bootstrap iterations to $B=600$. We then simulate $1,000$ times for each DGP. The simultaneous confidence intervals are constructed following Algorithm \ref{alg:simultaneous_confidence_intervals} and without normalization by $\widetilde \sigma_k$ for simplicity. The true $\alpha_k$ are computed 
using $3,000,000$ additional observations generated independently from data following the same marginal distribution as $X_{ig}$.
The nominal coverage probability is set to be $0.95$. The results for one-dimensional confidence intervals are presented in Table \ref{table:one_CI}.
\qquad\\
\begin{table}
\caption{Coverage probability for one-dimensional $95$\% confidence intervals for $\alpha_2$ under each DGP with $G_0=200$, $n=500$ and $p=300$:}
\qquad\\
\begin{tabular}{cccccc}
DGP &$\beta_2=0$ &$\beta_2=.25$&$\beta_2=.5$&$\beta_2=.75$&$\beta_2=1$\\
\hline
\hline
M1 &$0.953$ & $0.935$&$0.943$&$0.959$&$0.972$\\
M2 &$0.949$ & $0.944$&$0.937$&$0.956$&$0.969$\\
M3 &$0.939$ & $0.938$&$0.941$&$0.951$&$0.968$\\
M4 &$0.938$ & $0.944$&$0.937$&$0.944$&$0.954$\\
M5 &$0.928$ & $0.931$&$0.928$&$0.940$&$0.923$\\
\hline
\hline
M6 &$0.928$ & $0.919$&$0.920$&$0.946$&$0.970$\\
M7 &$0.921$ & $0.920$&$0.912$&$0.926$&$0.957$\\
M8 &$0.934$ & $0.925$&$0.929$&$0.954$&$0.956$\\
M9 &$0.926$ & $0.929$&$0.933$&$0.941$&$0.958$\\
M10 &$0.938$ & $0.935$&$0.944$&$0.938$&$0.947$\\
\hline
\end{tabular} \label{table:one_CI}
\end{table}
\quad\\
We now illustrate the necessity of cluster robust method under many-small-cluster asymptotics. Consider DGP M1-M5. Suppose we implement observation-wise multiplier bootstrap (i.e. in each iternation, i.i.d. standard normal r.v.'s $\xi_i^b$ are generated for each observation) in place of the proposed multiplier cluster bootstrap. The results are presented in Table 4. One can see that in most cases, the non-cluster robust method is severely oversized while the multiplier cluster bootstrap has close to nominal coverage rates consistently. 
\qquad\\

	\begin{table}[ht]\label{table:robust}
		\caption{Comparison of cluster robust and non-cluster robust methods' coverage probability for one-dimensional $95$\% confidence intervals for $\alpha_2$ under DGP M1-M5 with $G_0=200$, $n=500$ and $p=300$:}
	\qquad\\
	\begin{tabular}{ccccccc}
		Cluster robust &$\beta_2=0$ &$\beta_2=.25$&$\beta_2=.5$&$\beta_2=.75$&$\beta_2=1$\\
		\hline
			\hline
		&&DGP:& M1\\
		\hline
		
		Yes&$0.953$ & $0.935$&$0.943$&$0.959$&$0.972$\\
		No  & $0.907$& $0.910$ &$0.900$&$0.934$&$0.962$\\
		
		\hline
		\hline
		&&DGP:& M2\\
			\hline
	
		Yes &$0.949$ & $0.944$&$0.937$&$0.956$&$0.969$\\
		No  & $0.913$& $0.899$ &$0.898$&$0.933$&$0.943$\\
		
				\hline
			&&DGP:& M3\\
			\hline
			
			Yes &$0.939$ & $0.938$&$0.941$&$0.951$&$0.968$\\
			No  & $0.884$& $0.914$ &$0.921$&$ 0.941$&$0.962$\\
				\hline
		&&DGP:& M4\\
			\hline
			
			Yes &$0.938$ & $0.944$&$0.937$&$0.944$&$0.954$\\
			No  & $0.885$& $0.924$ &$0.901$&$0.929$&$0.936$\\
			\hline
				&&DGP:& M5\\
			\hline
			
			Yes &$0.928$ & $0.931$&$0.928$&$0.940$&$0.923$\\
			No  & $ 0.922$& $0.897$ &$0.906$&$ 0.933$&$0.923$\\
			\hline
	\end{tabular} 
\end{table}
\quad\\

We now present the coverage probabilities for simultaneous confidence intervals for different sets of covariates. For this part of experiments, we focus on models M1 to M5 with $\beta_2=0.5$. The results are shown in Table \ref{table:simul_CI}.

\qquad\\

\begin{table}
\caption{Coverage probability for $95$\% simultaneous confidence intervals for $\alpha_k$, $k\in A$ under each DGP with $G_0=200$, $n=500$ and $p=300$:} 
\qquad\\
\begin{tabular}{ccccccccccc}
DGP &$A_{1}$&$A_{2}$&$A_{3}$&$A_{5}$&$A_{10}$&$A_{20}$&$A_{30}$&$A_{40}$&$A_{50}$&$A_{100}$\\
\hline
\hline
M1  &$0.943$&$0.941$&$0.933$&$0.926$&$0.920$&$0.940$&$0.951$&$0.959$&$0.957$&$0.943$\\
\hline
M2   &$0.937$&$0.945$&$0.932$&$0.930$&$0.907$&$0.927$&$0.917$&$0.920$&$0.940$&$0.950$\\
\hline
M3  &$0.941$&$0.943$&$0.956$&$0.914$&$0.882$&$0.900$&$0.901$&$0.914$& $0.920$&$0.930$\\
\hline
M4 &$0.937$&$0.928$&$0.925$&$0.876$&$0.865$&$0.863$&$0.865$& $0.891$&  $0.897$&$0.925$\\
\hline
M5  &$0.928$&$0.926$&$0.930$&$0.891$&$0.865$&$0.861$&$0.874$& $0.898$ & $0.894$&$0.904$\\
\hline
\end{tabular} \label{table:simul_CI}
\end{table}

As before, we now illustrate the need of cluster robust method in simultaneous inference. As Table 6 shows, the oversize problem of non-cluster robust method is aggravated in simultaneous inference.

\begin{table}\label{table:robust_simultaneous}
	\caption{Comparison of cluster robust and non-cluster robust methods' coverage probability for $95$\% simultaneous confidence intervals for $\alpha_k$, $k\in A$ under each DGP with $G_0=200$, $n=500$ and $p=300$:} 
	\qquad\\
	\begin{tabular}{ccccccccc}
		Cluster robust &$A_1$&$A_2$&$A_{3}$&$A_{5}$&$A_{10}$&$A_{20}$&$A_{50}$&$A_{100}$\\
		\hline
			\hline
		&&&DGP:& M1\\
		\hline
		Yes& $0.943$&$0.941$&$0.933$&$0.926$&$0.920$&$0.940$&$0.957$&$0.943$\\
		No &$0.900$&$0.896$&$0.877$&$0.852$&$ 0.842$&$0.839$&$0.785$&$0.788$\\
		\hline
		&&&DGP: &M2\\
			\hline
		Yes &$0.937$&$0.945$&$0.932$&$0.930$&$0.907$&$0.927$&$0.940$&$0.950$\\
		No &$0.898$&$0.894$&$0.890$&$0.847$&$0.819$&$0.810$&$0.773$&$0.774$\\
		\hline
		&&&DGP: &M3\\
		\hline
		Yes &$0.941$&$0.943$&$0.956$&$0.914$&$0.882$&$0.900$&$0.920$&$0.930$\\
		No &$0.921$&$0.890$&$0.861$&$0.851$&$0.796$&$0.706$&$0.730$&$0.753$\\
		\hline
		&&&DGP: &M4\\
		\hline
		Yes&$0.937$&$0.928$&$0.925$&$0.876$&$0.865$&$0.863$&$0.897$&$0.925$\\
		No &$0.901$&$0.903$&$0.854$&$0.856$&$0.765$&$0.720$&$0.732$&$0.760$\\
		\hline
		&&&DGP: &M5\\
		\hline
		Yes&$0.928$&$0.926$&$0.930$&$0.891$&$0.865$&$0.861$&$0.894$&$0.904$\\
		No &$0.906$&$0.906$&$0.869$&$0.831$&$0.764$&$0.766$&$0.775$&$0.775$\\
		\hline
	\end{tabular} 
\end{table}

Finally, we investigate the asymptotic behaviors of the case with $A_{10}$, one of the worst-performing cases in the above simulations for simultaneous confidence intervals, to examine whether the performance improves as sample size increases. In this set of simulations, set $\beta_2=0.5$ for number of nominal clusters $G_0=200$, $400$, $600$ and $800$, $p=1.5\cdot G_0$, and $n=2.5\cdot G_0$. The results are presented in Table \ref{table:asymp_CI}.

\begin{table}
\caption{ Asymptotic behaviors of coverage probability for $95$\% simultaneous confidence intervals for $\alpha_k$, $k\in A_{10}$ under each DGP:}
\qquad\\
\begin{tabular}{ccccc}
 DGP &$G_0=200$&$G_0=400$&$G_0=600$&$G_0=800$\\
\hline
\hline
M1  &$0.920$&$0.914$&$0.930$&$0.945$\\
\hline
M2   &$0.907$&$0.917$&$0.920$&$0.921$\\
\hline
M3  &$0.882$&$0.894$& $0.898$&$0.903$\\
\hline
M4 &$0.865$&$0.856$&  $0.864$&$0.876$\\
\hline
M5  &$0.865$&$0.859$& $0.847$&$0.857$\\
\hline
\end{tabular} \label{table:asymp_CI}
\end{table}
\quad\\

In all of the simulation experiments, the coverage probabilities are mostly reasonably close to the nominal coverage rate when $\rho$ is not close to one. When $\rho$ is high, the approximate sparsity of nuisance parameters in Assumption \ref{a:parameters} is potentially violated. Thus some of the coverage probabilities deviate away from the nominal rate. In addition, the coverage probabilities improve as sample size increases. In summary, the outcomes of these experiments are consistent with our theoretical results.

\section{Application: Testing Gendered Language on the Internet}\label{sec:empirical_illustration}

In this section, we apply our method of simultaneous inference for APEs in the text regression model of \cite{Wu18} introduced in Section \ref{sec:motivating_application}. We make use of the pronoun sample
(gendered posts including either female or male pronouns) from  \cite{Wu18}. Following Wu, using the EJMR dataset\footnote{The dataset is publicly available at url:https://www.aeaweb.org/articles?id=10.1257/pandp.20181101}, we exclude the same list of words from the 10,000, including all gender classifiers, plus names of non-economist celebrities.
We conduct our analysis based on the subset of non-duplicate posts that are used as the test sample for selecting optimal probability threshold in the original paper (the posts with index labelled as test0) for classification of posts that contains both female and male classifiers. We consider only pronoun sample. This leaves 46,502 posts sampled from 31,739 threads and 9541 covariates\footnote{Since the number of observations is larger than dimensionality of parameters, regular Logit and even OLS can be applied here. We have attempted to implement Logit using \textbf{glm} package in R. However, it did not finish after 70 minutes. OLS on the other hand takes 55 minutes to complete. In contrast, the proposed estimation and inference algorithms, when applied to the testing problem in this section, takes about two minutes to complete.} that consists of an intercept and the word counts of 9,540 non-excluded vocabularies. 


\cite{Wu18} highlights that posts about males include more academically and professionally oriented vocabularies, such as ``adviser," ``supervisor," and ``Nobel." To see the joint significance of these words' APE in terms of predicting female, we test
\begin{align*}
\text{H}_0: \alpha_{\text{adviser }}=\alpha_{\text{supervisor} }=\alpha_{\text{nobel} }=0.
\end{align*}
Following the penalty choices of Algorithms \ref{alg:LOG}, \ref{alg:REGNPL} and \ref{alg:REG}, the estimates of APEs of these words calculated using Algorithm \ref{alg:DS_APE} are listed in Table \ref{table:EJMR_APE_estimates}. These estimates are qualitatively similar to the corresponding estimates in \cite{Wu18}. 
Using multiplier cluster bootstrap with $10,000$ bootstrap iterations, we obtain the test results listed in Table \ref{table:EJMR_APE_tests}.
Note that under all three confidence levels, we reject the null hypothesis and the statistical evidence supports Wu's statement
\footnote{One may be concerned about the high-correlation between "supervisor" and "adviser." However, removing either one of them does not change the significance of the tests at $99$\% confidence level.}.
\begin{table}
\caption{APE estimates for ``adviser," ``supervisor," and ``Nobel."}
\qquad\\
\begin{tabular}{cccc}
\hline
\hline
\multicolumn{1}{c}{ } &\multicolumn{1}{c}{adviser } & \multicolumn{1}{c}{supervisor } &\multicolumn{1}{c}{Nobel } \\
\hline
\hline
APE estimate &$ -0.1414$ &$-0.1214$& $-0.1214$\\
\hline
\end{tabular}\label{table:EJMR_APE_estimates}
\end{table}
\qquad\\

\begin{table}
\caption{Multiple Testing Results under $1-\alpha\%$  Confidence level.}
\qquad\\
\begin{tabular}{ccc}
\hline
\hline
\multicolumn{1}{c}{$\alpha$ }&\multicolumn{1}{c}{MCB critical value }& \multicolumn{1}{c}{test statistic }\\
\hline
\hline
$10\%$&$ 16.0889$&$25.1867$\\
$5\%$&$ 18.2870$&$25.1867$\\
$1\%$&$ 22.2930$&$25.1867$\\
\hline
\end{tabular}\label{table:EJMR_APE_tests}
\end{table}
\qquad\\


\section{Conclusion}\label{sec:conclusion}

In this paper, we study logistic average partial effects with lasso regularization when data is sampled under clustering. We proposed two valid estimators along with their theoretically justified lasso penalty choices. Based on these estimators, we provide easy-to-implement algorithms for simultaneous inference and confidence intervals and establish their asymptotic validity. Simulation studies demonstrate that the proposed procedures work as predicted by the theory in finite sample. We then apply the proposed method to conduct analysis of textual data to examine the presence of gendered language on the EJMR forum following the text regression model of \cite{Wu18}. Our analysis provides further statistical evidence to support Wu's finding.



\allowdisplaybreaks
\appendix
\section{Review on Covering Numbers and Related Definitions}\label{sec:covering_numbers}
We shall first review some definitions on classes of functions that we will constantly refer to in this appendix. For more detail, see \cite{vdVW96} or \cite{GN16}.
Let $S$ be a set and $\calC$ be a nonempty class of subsets of $S$. Pick any finite set $\{x_1,...,x_n\}$ of size $n$. We say that $\calC$ picks out a subset $A\subset \{x_1,...,x_n\}$ if there exists $C\in \calC$ such that $A=\{x_1,...,x_n\}\cap C$.
Let $\Delta^\calC(x_1,...,x_n)$ be the number of subsets of $\{x_1,...,x_n\}$ picks out by $\calC$, (i.e. $
\Delta^\calC (x_1,...,x_n) = \text{Card} (\{\{x_1,...,x_n\}\cap C: C\in \calC\}).
$.
We say the class $\calC$ shatters $\{x_1,...,x_n\}$ if $\calC$ picks out all of its $2^n$ subsets ($\Delta^\calC (x_1,...,x_n) =2^n$).  
The VC index $V(\calC)$ is defined by the smallest $n$ for which no set of size $n$ is shattered by $\calC$, i.e., with $m^\calC(n):=\max_{x_1,...,x_n}\Delta^\calC(x_1,...,x_n)$,
\begin{align*}
V(\calC)=\begin{cases}
\inf\{n:m^\calC(n)<2^n\}, \text{ if such set is non-empty,}\\
+\infty, \text{ otherwise.}
\end{cases}
\end{align*}
The class $\calC$ is called a VC(
Vapnik–Chervonenkis) class if $V(\calC)<\infty$. For a real function $f$ on $S$, its subgraph is defined as 
$
\sg(f)=\{(x,t):t<f(x)\}
$.
A function class $\calF$ on $S$ is a VC subgraph class if the collection of subgraphs of $f\in\calF$, $\sg(\calF)=\{\sg(f):f\in\calF\}$, is a VC class of sets in $S\times \Real$. We can define the VC index of $\calF$, $V(\calF)$, as the VC index of $\sg(\calF)$. Let $(T,d)$ be pseudometric space (i.e. $d(x,y)=0$ doesn't imply $x=y$). For $\varepsilon>0$, an $\varepsilon$-net of $T$ is a subset $T_\varepsilon$ of $T$ such that for every $t\in T$ there exists $t_\varepsilon\in T_\varepsilon$ with $d(t,t_\varepsilon)\le \varepsilon$. The $\varepsilon$-covering number $N(T,d,\varepsilon)$ of $T$ is defined by
\begin{align*}
N(T,d,\varepsilon)=\inf \{\text{Card}(T_\varepsilon):\, T_\varepsilon \text{ is an $\varepsilon$-net of $T$}\}.
\end{align*}
For any $r\ge 1$, denote $\|f\|_{Q,r}=\left(\int |f|^r dQ\right)^{1/r}$ for $f\in L^r(S)$.
Suppose $\calF$ be a VC-subgraph class with envelope $F$, then Theorem 2.6.7 in \cite{vdVW96} suggests that for any $r\in [1,\infty)$, we have
\begin{align*}
\sup_{Q} N(\calF,\|\cdot\|_{Q,r},\varepsilon\|F\|_{Q,r}) \le KV(\calF)(16e)^{V(\calF)} \left(\frac{2}{\varepsilon}\right)^{r (V(\calF) -1)}
\end{align*}
for all $0<\varepsilon\le 1$, where $K$ is universal and the supremum is taken over all finite discrete probability measures. Despite of this desirable property, sometimes VC subgraph is to stringent and therefore fail to be useful in more complex situations. Often times we work with the alternative definition of VC type class. We say $\calF$ being a VC type class of functions with characteristics $(A,v)$ if for some positive constants $A,v$,
\begin{align}
\sup_{Q}N(\calF,\|\cdot\|_{Q,2},\varepsilon\|F\|_{Q,2})\le \left(\frac{A}{\varepsilon}\right)^v.\label{eq:VC-type}
\end{align}
The notion of VC type class combined with Lemma \ref{lemma:entropy_algebra} in Appendix \ref{sec:technical_lemmas} cover many useful scenarios. Let $\calF$ be a pointwise measurable class of measurable functions $\Ss\mapsto\Real$ with measurable envelope $F$. For $0<\delta<\infty$, define the uniform entropy integral of $\calF$ as
$
J(\calF,F,\delta):=\int_0^\delta \sup_Q \sqrt{\log 2 N(\calF,\|\cdot\|_{Q,2},\varepsilon \| F \|_{Q,2}) }d\varepsilon.
$
 
\section{Orthogonalization of the Score}\label{sec:orthogonalization}
\text{}\\
In this Section, we shall derive the Neyman orthogonal score for $\alpha_k$, as defined in (\ref{eq:neyman_orthogonal_score_APE}), following the methodology in Section 2.2 of \cite{BCCW18} (see also Section 2 of \cite{CCDDHNR18}).
The first order condition of the population quasi-maximal likelihood and definition of the $k$-th APE give $ \EP[G^{-1}\sum_{g=1}^{G} \sumi m(W_{ig},\alpha_k,\beta^0)]=0$, where
\begin{align*}
m_k(W_{ig},\alpha,\beta)
=
\begin{bmatrix}
\partial_{\alpha} \widetilde\ell(W_{ig},\alpha,\beta)\\
\partial_{\beta} \widetilde\ell(W_{ig},\alpha,\beta)
\end{bmatrix}
:=
\begin{bmatrix}
\alpha \cdot\frac{G}{n}-\beta_k \Lambda'(X_{ig}'\beta)\\
\ell'(Y_{ig},X_{ig}'\beta)X_{ig}
\end{bmatrix},
\end{align*} 
where $\ell(a,b)=a\log \Lambda(b) + (1-a)\log(1-\Lambda(b))$, $\ell'(a,b)=\frac{\partial}{\partial b} \ell(a,b)$, $\ell''(a,b)=\frac{\partial^2}{\partial b^2} \ell(a,b)$.
Note that the order of integral and derivative are interchangeable in this case. Let us define 
\begin{align*}
J=&\partial_{(\alpha, \beta')'} \EP\left[\frac{1}{G}\sum_{g=1}^{G} \sumi m_k(W_{ig},\alpha,\beta)\right] \big|_{\alpha=\alpha_k,\beta=\beta^0}\\
=&
\begin{bmatrix}
1 & - \EP[\frac{1}{G}\sum_{g=1}^{G} \sumi(\beta_k \Lambda''(X_{ig}'\beta)X_{ig}' + \Lambda'(X_{ig}'\beta)e_k')]\\
0 & \EP[\frac{1}{G}\sum_{g=1}^{G}  \sumi \ell''(Y_{ig},X_{ig}'\beta)X_{ig} X_{ig}']
\end{bmatrix}_{\alpha=\alpha_k,\beta=\beta^0}
=
\begin{bmatrix}
J_{\alpha \alpha} & J_{\alpha \beta}\\
J_{\beta \alpha} & J_{\beta \beta}
\end{bmatrix}.
\end{align*}
Now define population nuisance parameter
{\small\begin{align*}
\mu^k=& -J_{\beta\beta}^{-1}J_{\alpha \beta}'
 =\left\{ \EP\left[\frac{1}{G}\sum_{g=1}^{G}  \sumi \ell''(Y_{ig},X_{ig}'\beta^0)X_{ig} X_{ig}'\right]\right\}^{-1}  \EP\left[\frac{1}{G}\sum_{g=1}^{G} \sumi (\beta^0_k \Lambda''(X'_{ig}\beta^0)X_{ig}+ \Lambda'(X'_{ig}\beta^0)e_k)\right] \\
 =&\left\{ \EP\left[\frac{1}{G}\sum_{g=1}^{G}  \sumi f^2_{ig} X_{ig} X_{ig}'\right]\right\}^{-1}  \EP\left[\frac{1}{G}\sum_{g=1}^{G} \sumi f^2_{ig} X_{ig} S_{ig}^k\right] + \left\{ \EP\left[\frac{1}{G}\sum_{g=1}^{G}  \sumi f^2_{ig} X_{ig} X_{ig}'\right]\right\}^{-1} \EP\left[\frac{1}{G}\sum_{g=1}^{G} \sumi f^2_{ig} e_k\right]\\
 =&\zeta^k + \theta^k,
\end{align*}}
where $S^k_{ig}=\beta_k (1-2\Lambda(X_{ig}'\beta^0))$.
Here we have used the property of the logistic function $\Lambda''(X_{ig}'\beta^0)=\Lambda(X_{ig}'\beta^0)(1-\Lambda(X_{ig}'\beta^0))(1-2\Lambda(X_{ig}'\beta^0))$ and thus
$
\beta^0_k \Lambda''(X'_{ig}\beta^0)
=f_{ig }^2\beta_k^0 (1-2\Lambda(X_{ig}'\beta^0))=f_{ig}^2 S^k_{ig}.
$
Define Neyman orthogonal score for $\alpha_k$ as 
\begin{align*}
\bar\psi_k (W_{ig},\alpha,\eta)=&\partial_{\alpha} \widetilde\ell(W_{ig},\alpha,\beta)- \mu'\partial_{\beta} \widetilde\ell(W_{ig},\alpha,\beta) \\
 =&\alpha \cdot\frac{G}{n}- \beta_k \Lambda'(X_{ig}'\beta)+\mu'X_{ig} 
  \{Y_{ig}-\Lambda(X_{ig}'\beta)\},
\end{align*}
where $\beta_k$ is the $k$-th coordinate of $\beta$ and $\eta=(\beta',\mu')\in \Real^{2p}$.
It is straightforward to verify the followings,
\begin{align*}
& \EP\left[\frac{1}{G}\sum_{g=1}^{G}  \sumi\bar\psi_k (W_{ig},\alpha_k,\eta^k)\right]=0,\qquad \text{(existance condition)}\\
&\partial_{\eta} \EP\left[\frac{1}{G}\sum_{g=1}^{G}  \sumi\bar\psi_k (W_{ig},\alpha_k,\eta^k)\right]=0,\quad\text{ (Neyman orthogonality condition)}\\
&\partial_{\alpha} \EP\left[\frac{1}{G}\sum_{g=1}^{G}  \sumi\bar\psi_k (W_{ig},\alpha_k,\eta^k)\right]=1\ne 0.\:\text{ (uniqueness condition)}
\end{align*}

\section{Main Results under High-level Assumptions}\label{sec:main_high_level}
In this section we shall introduce a version of our asymptotic results under high-level conditions. It serves as a building block for results in Section \ref{sec:main}.  Suppose that we have some generic nuisance parameter estimators $\hat \eta^k$ such that $\eta^k\in \H_k$ for $G$ large enough. Denote $A_k$, a bounded interval of $\alpha_k$ shrinking with $G$, and $\H_k\subset H_k$, a sparse neighborhood of $\eta^k$ shrinking with $G$, where $H_k\subset \Real^p$ a compact and convex set that contains $\eta^k$. Let $v_G\ge 1$, $D_{G}$, $K_G$ and $B_G$ be some positive sequences of constants that can possibly grow to infinity. Let $q \ge 2$ be some constant. Further, let $\tau_G$, $\delta_G$ and $\Delta_G$ be some positive sequences of constants that converge to zero and $\Delta_G<1$. 

\begin{assumption}\label{a:score_moment}
For each $G\in \mathbbm N$, $G\ge 3$, $\Pr\in \mathcal P_G$ and $k\in [p]$, the following conditions are satisfied:
\begin{enumerate}[(i)]
\item  $\eta \mapsto  \EP[G^{-1}\sum_{g=1}^{G}\sumi \psi_k(W_{ig},\eta)]$ is twice continuously differentiable.
\item It holds that
\begin{enumerate}[(a)]
\item $\sup_{\eta \in \mathcal H_k} \EP[G^{-1}\sum_{g=1}^{G} (\sumi \{\psi_k(W_{ig},\eta)-\psi_k(W_{ig},\eta^k)\;\})^2]\le C_0 \|\eta - \hat \eta^k\|^2_2,$ 
\item $\sup_{\eta \in \mathcal H_k}\|\partial_{\eta'}\partial_{\eta} \EP[G^{-1}\sum_{g=1}^{G} \sumi\psi_k(W_{ig},\eta)]\|_2\le D_{G}.$
\end{enumerate}
\end{enumerate}
\end{assumption}

\begin{assumption}\label{a:score_nuisance_para}
For each $G\in \mathbbm N$, $G\ge 3$, $\Pr\in \mathcal P_G$ and $k\in [p]$, the following conditions are satisfied:
\begin{enumerate}[(i)]
\item  $\hat \eta^k \in \mathcal H_k$ with probability at least $1-\Delta_G$ 
and $\sup_{\eta \in \mathcal H_k}\|\eta-\eta^k\|_2 \le \tau_G$  .
\item The collection of functions
\begin{align*}
\calF_0 =\{ \psi_k(\cdot,\eta):k\in[p], \:\eta\in \H_k \}\cup \{0\}
\end{align*}
is pointwise measurable and satisfies that for all $0<\varepsilon\le 1$,
\begin{align*}
\sup_Q\log N(\calF_0,\|\cdot\|_{Q,2},\varepsilon\|F_0\|_{Q,2})\le v_G \log(a_G/\varepsilon)
\end{align*}
where the supremum is taken over the set of all finite measures and $F_0$ is a measurable envelope of $\calF_0$ such that $\{ \EP[G^{-1}\sum_{g=1}^{G} |\sumi F_0(W_{ig})|^q]\}^{1/q}\le K_G$.
\item For all $f\in \calF_0$, we have $c_0\le \{ \EP[G^{-1}\sum_{g=1}^{G} (\sumi f(W_{ig}))^2]\}^{1/2}\le C_0$.
\item $\sqrt{G}D_{G}\tau_G^2\bigvee
\tau_G(v_G  \log a_G)^{1/2} \bigvee \{G^{-1/2+1/q}K_G  v_G \log a_G\}
\lesssim \delta_G$.
\end{enumerate}
\end{assumption}
\begin{remark}\label{rem:high_level_assumptions_1_2}
While been adapted to our cluster sampling setting, Assumptions \ref{a:score_moment}, \ref{a:score_nuisance_para} are similar to Condition 2, 3 of \cite{BCK15} and Assumption 2.1, 2.2 of \cite{BCCW18}. However, due to the additive separability of $\hat \alpha_k$, we do not need to assume Assumption 2.1(b) of \cite{BCCW18}. Also, differentiability of the orthogonal score comes directly from smoothness of logistic function. 
\end{remark}

The following result builds upon the ideas of the main results in \cite{BCK15} and \cite{BCCW18}
while allowing for cluster sampling.  Given some generic nuisance parameters estimate $\hat \eta^k$, we define the generic APE estimator for the $k$-th continuous covariate as
\begin{align}
\widehat \alpha_k=\frac{1}{G} \sumg\sumi \psi_k(W_{ig},\hat \eta^k).\label{eq:estimator_alpha_generic}
\end{align}
It is easy to verify the fact that the post-double-section estimator $\widetilde \alpha_k$, as defined in (\ref{eq:estimator_DS}), satisfies (\ref{eq:estimator_alpha_generic}) for $\hat \eta^k = (\widetilde \beta^{k\prime},\widetilde \mu^{k\prime})'$ following the first order condition of (\ref{eq:beta_check}), the definition of $\widetilde T_k$ and the definition of $\psi_k $.
\begin{theorem}[Uniform Bahadur representation]\label{theorem:bahadur_rep}
\quad\\
Suppose that we have nuisance parameter estimates $(\hat \eta^k)_{k \in [p]}$ such that Assumptions \ref{a:score_moment} and \ref{a:score_nuisance_para} are satisfied.  For the generic $(\hat \alpha_k)_{k\in[p]} $ defined based on $(\hat \eta^k)_{k \in [p]}$ following (\ref{eq:estimator_alpha_generic}), with probability at least $1-\Delta_G-(\log G)^{-1}$,
\begin{align*}
&\sup_{\Pr\in\mathcal P_G}\max_{1 \le k \le p}\Bigg|\sqrt{G}\sigma_{k}^{-1}
(\hat \alpha_k -\alpha_k)
-
\frac{1}{\sqrt{G}}\sumg \sumi \varphi_k(W_{ig},\alpha_k,\eta^k)\Bigg|\lesssim \delta_G,
\end{align*}
where $\varphi_k(W_{ig},\alpha,\eta)=- \bar\psi_k(W_{ig},\alpha,\eta)/\sigma_k$ and $\eta^k=(\beta^{0\prime},\mu^{k\prime})'$.

\end{theorem}
A proof can be found in Appendix \ref{sec:proof for theorem:bahadur_rep}. 

Let $\{\xi_g\}_{g=1}^G$ be independent standard normal random variables generated independently from data. Define the shorthand notation
$\hat \varphi_{gk}=\sumi\hat \varphi_j(W_{ig},\hat\alpha_k,\hat\eta^k)$ and $\varphi_{gk}=\sumi\varphi_j(W_{ig},\alpha_k,\eta^k)$
and let
\begin{align*}
W:=&\max_{1 \le k\le p}\frac{1}{\sqrt{G}}\sumg\xi_g\hat \varphi_{gk} \text{ and } W_0:=\max_{1 \le k\le p}\frac{1}{\sqrt{G}}\sumg\xi_g \varphi_{gk}
\end{align*}

\begin{assumption}\label{a:score_additional}
For some $q\ge 4$ and all $G \ge 3$ and $\Pr \in \mathcal P_G$, the following holds.
\begin{enumerate}[(i)]
\item There exists $B_G\ge 1$ such that for all $1\le g\le G$ and $k\in[p]$,
\begin{align*}
&\EP\left[\frac{1}{G}\sum_{g=1}^{G} \varphi_{gk}^{2}\right]\ge c_1,\quad
\EP\left[\frac{1}{G}\sum_{g=1}^{G} |\varphi_{gk}|^{2+\kappa}\right]\le B_G^\kappa,\qquad\text{for }\kappa=1,2,\\
&\EP\left[\max_{k\in[p]}\left|\varphi_{gk}\right|^q\right]\le B_G^q,
\end{align*} 
and $G^{-1 } B_G^4\log^7a_G\bigvee G^{-1+2/q}B_G^2\log^3a_G=o(1)$.
\item 
For each $k\in[p]$, with probability at least $1- \Delta_G$, it holds that
$G^{-1}\sum_{g=1}^{G}  (\hat \varphi_{gk}-
\varphi_{gk})^2\le \bar\sigma_G^2$.
\item $\bar \sigma_G   \log a_G=o(1)$, $\delta_G^2\log p=o(1)$.
\end{enumerate}
\end{assumption}
\begin{remark}\label{rem:high_level_assumptions_3}
Assumption \ref{a:score_additional} (i) is required by the high-dimensional central limit theorem of \cite{CCK13} (see their Corollary 2.1). Assumption \ref{a:score_additional} (ii) is discussed in the next remark.  Assumption \ref{a:score_additional} (iii) is a technical assumption that turns out to be mild, as shown in the sufficient conditions in Section \ref{sec:main}.
\end{remark}
\begin{remark}[Double/debiased machine learning]\label{rem:double_machine_learning}
One could potentially employ sample splitting to eliminate the dependence between the orthogonal score and nuisance parameters. This procedure is known as "double/debiased machine learning" (cf \cite{CCDDHNR18}). This would allow us to relax Assumption \ref{a:score_additional} (ii). We did not make use of sample splitting due to the following considerations.  First, we do not assume each cluster is identically distributed since it is not suitable for the sampling method used in the motivating example in Section \ref{sec:motivating_application}. Second, even if identical distribution is assumed, when we have binary outcome variable $Y_{ig}$, sample-splitting may results in subsamples with high percentages of outcomes equal to $1$ or $0$. In such case, the estimate for $\hat \eta^k$ could be very unreliable. Finally, relaxing Assumption \ref{a:score_additional} (ii) does not appear to allow us to relax any sufficient conditions presented in Secion \ref{sec:main}. Therefore, we do not consider sample splitting in this paper.
\end{remark}

\begin{corollary}[Multiplier cluster bootstrap of maxima]\label{corollary:MB}
Suppose that Assumptions \ref{a:score_moment}, \ref{a:score_nuisance_para} and \ref{a:score_additional} are satisfied, then let $c_W(a)$ be the $a$-th quantile of $W$, we have
\begin{align*}
\sup_{\Pr\in \mathcal P_G}\sup_{\alpha\in (0,1)}
\Big|\Pr_\Pr\left(\max_{1 \le k \le p}|\sqrt{G}\sigma_{k}^{-1}
(\hat \alpha_k -\alpha_k)|\le c_W(a)\right) 
-
a\Big|=o(1).
\end{align*}
\end{corollary}
A proof can be found in Section \ref{sec:proof for corollary:MB} in the Appendix. 
\begin{remark}[Uniform in DGP]\label{rem:unif_in_DGP}
Note that all the above results are valid uniformly over $\mathcal P_G$, the set of DGP's such that Assumptions \ref{a:score_moment}, \ref{a:score_nuisance_para}, \ref{a:score_additional} are satisfied.
This is due to the fact that Lemma A.1 of \cite{BCFH17} implies that it suffices to show that these results hold for any sequence $\Pr_G \in \mathcal P_G$, which is satisfied since all the bounds in this paper are established independently of DGP. 
\end{remark}

\section{Proofs for Results in Section \ref{sec:main_high_level}}

\subsection{Proof for Theorem \ref{theorem:bahadur_rep}}\label{sec:proof for theorem:bahadur_rep}
\begin{proof}
By (\ref{eq:estimator_alpha_generic}), it holds that $\hat \alpha_k= G^{-1}\sumg \sumi \psi_k(W_{ig},\hat\eta^k)$ for an $\hat \eta^k$ from Assumption \ref{a:score_nuisance_para}. The fact that $ \alpha_k=\EP\left[G^{-1}\sum_{g=1}^{G}  \sumi \psi_k(W_{ig},\eta^k)\right]$ implies
\begin{align*}
\hat \alpha_k - \alpha_k
=& \frac{1}{G}\sum_{g=1}^{G} \sumi  \psi_k(W_{ig},\hat \eta^k)- \alpha_k\\
=& \frac{1}{G}\sum_{g=1}^{G} \sumi \psi_k(W_{ig},\eta^k) - \alpha_k + \underbrace{\left(\EP\left[\frac{1}{G}\sum_{g=1}^{G} \sumi \psi_k(W_{ig},\hat\eta^k)\right]- \alpha_k\right)}_{I_k} \\
&+ \underbrace{\frac{1}{G} \sumg \sumi \Big\{  \psi_k(W_{ig},\hat \eta^k) - \psi_k(W_{ig}, \eta^k)  - \EP[\psi_k(W_{ig},\hat \eta^k) - \psi_k(W_{ig}, \eta^k)] \Big\}}_{II_k}.
\end{align*}
It suffices to show that $|I_k|$ and $|II_k|$ are of order $\op(1/\sqrt{G})$ uniformly over $k\in[p]$ and uniformly in $\mathcal P_G$.

\noindent \textbf{Step 1: Bound for $|I_k|$.}
By applying the mean-value expansion and under Assumption \ref{a:score_moment} (i), there exists a vector $\ddot\eta^k\in \mathcal H_k$ with each of its coordinates lies between those of $\eta^k$ and $\hat \eta^k$ such that
\begin{align*}
I_k=&  \EP\left[\frac{1}{G}\sum_{g=1}^{G}  \sumi \psi_k(W_{ig},\hat \eta^k)\right] -\alpha_k\\
=&  \left(\EP\left[\frac{1}{G}\sum_{g=1}^{G}  \sumi\psi_k(W_{ig}, \eta^k) \right]-\alpha_k\right)
+ 
\partial_\eta  \EP\left[\frac{1}{G}\sum_{g=1}^{G}  \sumi \psi_k(W_{ig}, \eta^k)'\right](\hat \eta^k -\eta^k) \\
&+ (\hat \eta^k -\eta^k)'\left\{\partial_{\eta'} \partial_\eta  \EP\left[\frac{1}{G}\sum_{g=1}^{G}  \sumi\psi_k(W_{ig}, \eta)\right]\Bigg|_{\eta=\ddot\eta^k}\right\} (\hat \eta^k -\eta^k)\\
=&0 + 0 
+ (\hat \eta^k -\eta^k)'\left\{\partial_{\eta'} \partial_\eta  \EP\left[\frac{1}{G}\sum_{g=1}^{G} \sumi \psi_k(W_{ig}, \eta)\right]\Bigg|_{\eta=\ddot\eta^k}\right\} (\hat \eta^k -\eta^k),
\end{align*}
where the last equality follows from existence condition and Neyman orthogonality condition defined in the end of Section \ref{sec:orthogonalization} of this Appendix.
Hence by the definition of induced matrix $\ell_2$-norm and Assumptions \ref{a:score_moment} (ii)(b) and \ref{a:score_nuisance_para} (i), one has
\begin{align*}
|I_k|\le \left\| \partial_{\eta'} \partial_\eta  \EP\left[\frac{1}{G}\sum_{g=1}^{G}  \sumi \psi_k(W_{ig}, \eta)\right]\Bigg|_{\eta=\ddot\eta^k} \right\|_2 \tau_G^2\le D_{G}\tau_G^2
\end{align*}
uniformly in $k$ with probability at least $1-\Delta_G$. This, together with Assumption \ref{a:score_nuisance_para} (iv), implies $\sqrt{G}|I_k|\le \sqrt{G} D_{G} \tau_G^2\lesssim \delta_G$ with the same probability.

\noindent \textbf{Step 2: Bound for $|II_k|$.}
Recall that our modelling assumption suggest $n_g\le \overline n$ for all $g\in[G]$. Hence for each $g\in[G]$, $W_g$ can be written as
\begin{align*}
W_{g}=
\begin{cases}
(W_{1g}',\quad\:\:\;...\;\:\:\quad,W_{\bar n g}')' &\text{ if $n_g=\bar n$},\\
(W_{1g}',...,W_{ n_g g}',0,...,0)' &\text{ if $n_g<\bar n$}.
\end{cases}
\end{align*}
Our goal is to find a uniform entropy bound for the class
\begin{align*}
\calF=\Big\{W_g\mapsto \sum_{i=1}^{\bar n}\psi_k(W_{ig},\eta)\cdot\1\{W_{ig}\ne 0 \}\; :k=1,...,p,\: \eta \in \H_k\Big\},
\end{align*}
which then allows us to apply the maximal inequality of Lemma \ref{lemma:maximal_ineq}. 
Let us define the one function class
$
\calG_j=\{W_{g}\mapsto \1\{W_{jg}\ne 0\} \}
$.
Thus each of them is a VC-subgraph class with VC index $1$ and themselves as their envelopes (denoted as $\mathfrak g_j$). Thus for any $0<\varepsilon\le 1$, it holds that
$
\sup_Q \log N(\calG_j,\|\cdot\|_{Q,2},\varepsilon\|\mathfrak g_j\|_{Q,2})\lesssim 1+ \log(1/\varepsilon).
$
Now we define for each $j\in[\bar n]$
\begin{align*}
\calF_j=\{W_{g} \mapsto \psi_k(W_{jg},\eta):k\in [p],\: \eta \in \H_k\},
\end{align*}
with envelope $F_0$ coming from Assumption \ref{a:score_nuisance_para} (ii). Observe that $\calF\subset \bar \calF:=(\calF_1\cdot \calG_1 + ... +\calF_{\bar n}\cdot \calG_{\bar n} )$, where the addition and multiplication are component-wise.
Apply Lemma \ref{lemma:entropy_algebra}(2) under Assumption \ref{a:score_nuisance_para} (ii), for each $1\le j \le \bar n$, for  all $0<\varepsilon\le 1$, it holds that
\begin{align*}
&\sup_Q \log N(\calF_j\cdot \calG_j,\|\cdot\|_{Q,2},\varepsilon\|F_0\cdot \mathfrak g_j\|_{Q,2})\\
\le & \sup_Q \log N(\calF_j,\|\cdot\|_{Q,2},\varepsilon/2\|F_0\|_{Q,2})+ \sup_Q \log N(\calG_j,\|\cdot\|_{Q,2},\varepsilon/2\|\mathfrak g_j\|_{Q,2})\\
\lesssim & v_G \log(a_G/\varepsilon)  + \log(1/\varepsilon) .
\end{align*}
Define the transformation $\phi(f_1,...,f_{\bar n})=\sum_{j=1}^{\bar n}f_{j}$. By the triangle inequality, one has $|\phi(f_1,...,f_{\bar n})-\phi(g_1,...,g_{\bar n})|\le \sum_{j=1}^{\bar n}1\cdot |f_j-g_j|$. Applying Lemma \ref{lemma:entropy_algebra}(4), for the envelope $F(W_{g})=\sum_{j=1}^{ \bar n} F_0(W_{jg}) \1\{W_{jg}\ne 0\}$ for $\calF$, we have
\begin{align*}
&\sup_Q \log N(\calF,\|\cdot\|_{Q,2},\varepsilon\|F\|_{Q,2})\\
\lesssim &\sum_{j=1}^{\bar n} \sup_Q \log N\left(\calF_j\cdot \calG_j,\|\cdot\|_{Q,2},\frac{\varepsilon}{\bar n}\|F_0\cdot \mathfrak g_j\|_{Q,2}\right)\\
\lesssim&  \bar n v_G \log(a_G/\epsilon) + \bar n \log(1/\varepsilon) .
\end{align*}
By Assumption \ref{a:score_nuisance_para}(ii),
$
\left\{\EP\left[\max_{g\in [G]}F^2(W_g)\right]\right\}^{1/2}\le \left\{\EP\left[\sum_{g=1}^G F^q(W_g)\right]\right\}^{1/q}
\le G^{1/q} K_G. 
$
Now let
\begin{align*}
\bar\calF=\left\{W_g\mapsto\sum_{j=1}^{\bar n}\left(\psi_k(W_{jg}, \eta)- \psi_k(W_{jg}, \eta^k)\,\right)\cdot\1\{W_{jg}\ne 0\}\; :k=1,...,p, \: \eta\in \mathcal H_k\right\}.
\end{align*}
Observe that since $\bar\calF \subset \calF -\calF$, it holds that $\sup_{f\in \bar\calF}|f|\le 2 \sup_{f\in \calF}|f|\le 2F$. Assumption \ref{a:score_nuisance_para} (i),(ii) implies
\begin{align*}
\sup_{f\in \bar\calF} \EP\left[\frac{1}{G}\sumg f^2(W_g)\right]
\lesssim& \sup_{\eta \in \mathcal H_k}\EP\left[\frac{1}{G} \sumg\left(\sumi \{\psi_k(W_{ig},\eta)-\psi_k(W_{ig},\eta^k) \;\}\;\right)^2\right]\\
\le& 
C_0 \|\eta - \hat \eta^k\|^2_2 \lesssim \tau_G^2.
\end{align*}
Under Assumptions \ref{a:score_moment} (ii)(a), 
 \ref{a:score_nuisance_para} (ii), apply Lemma \ref{lemma:entropy_algebra} (2) and Lemma \ref{lemma:maximal_ineq}, we have
\begin{align*}
\sqrt{G}|II_k|
\le &\sup_{f\in \bar\calF}\Big|\frac{1}{\sqrt{G}}\sum_{g=1}^G \sum_{i=1}^{n_g}[\,f(W_{ig})-\Ep f(W_{ig}) \,]\Big| \\
\lesssim & \tau_G\sqrt{v_G  \log a_G } + G^{-1/2+1/q}K_G v_G  \log a_G\lesssim\delta_G,
\end{align*}
uniformly in $k$ with probability at least $1- C(\log G)^{-1}$, where the last inequality follows from Assumption \ref{a:score_nuisance_para} (iv).
Finally, the conclusion follows that $0<\sigma_k<\infty$ uniformly from Assumptions \ref{a:score_nuisance_para} (i)(iii).
\end{proof}

\subsection{Proof for Corollary \ref{corollary:MB}}\label{sec:proof for corollary:MB}
\begin{proof}
	Throughout this proof, denote
\begin{align*}
T=&\max_{1 \le k\le p}|\sqrt{G}\sigma^{-1}_k(\hat \alpha_k - \alpha_k)|, \quad
T_0=\max_{1 \le k\le p}\frac{1}{\sqrt{G}}\sumg \varphi_{gk}.
\end{align*}
\noindent \textbf{Step 1:} 
In this step, we bound 
\begin{align*}
\rho:=\sup_{t \in \Real}|\Pr_\Pr(T_0\le t)-\Pr_\Pr(Z_0\le t)|,
\end{align*}
where $Z_0=\max_{1\le k\le p}G^{-1/2}\sum_{g=1}^n Y_{gk}$, $\{Y_g\}_{g=1}^G$ are independently distributed $p$-dimensional centered Gaussian random vector such that $G^{-1/2}\sum_{g=1}^G Y_{gk}$ has the same covariance matrix as $G^{-1/2}\sum_{g=1}^G  \varphi_{gk}$.
First, under Assumption \ref{a:score_additional} (i), invoke Proposition 2.1 of \cite{CCK17} under their Condition (E2), we obtain
\begin{align*}
\rho\lesssim \left(\frac{B_G^2\log^7a_G}{G}\right)^{1/6} + \left(\frac{B_G^2\log^3a_G}{G^{1-2/q}}\right)^{1/3}=o(1).
\end{align*}

\noindent \textbf{Step 2:} In this step, we show
\begin{align}
&\Pr_\Pr(|T-T_0|>\theta_1)<\theta_2\label{eq:unconditional_prob_bound}\\
&\Pr_\Pr(\Pr_\xi(|W-W_0|>\theta_1)> \theta_2)<\theta_2. \label{eq:conditional_prob_bound}
\end{align}
for some appropriate $\theta_1$, $\theta_2 = o(1) $, where $\Pr_\xi$ is the law of $\xi_1$. Set $\theta_1=\delta_G \vee C  \bar \sigma_G\log^{1/2} a_G \ge \delta_G$ and $\theta_2=C( \Delta_G + (\log G)^{-1}) $. By Theorem \ref{theorem:bahadur_rep} (recall $q \ge 2$ and $G \ge 3$), Equation (\ref{eq:unconditional_prob_bound}) holds with such choice of $\theta_1$ and $\theta_2$.
We now claim (\ref{eq:conditional_prob_bound}).
We first show that
\begin{align}
|W-W_0|
\le& \max_{1\le k \le p} \left|\frac{1}{\sqrt{G} }\sumg \xi_g(\hat \varphi_{gk}-\varphi_{gk})\right| =\Op\left( \bar \sigma_G \log^{1/2} p\right).\label{eq:cor_MB_est_error}
\end{align}
Denote $\Omega_1$ for the event that $\max_{k\in[p]}\{ G^{-1}\sum_{g=1}^{G}  (\varphi_{gk}-\hat 
\varphi_{gk})^2\}^{1/2}\le \bar\sigma_G$, which, following Assumption \ref{a:score_additional} (ii), satisfies $ \Pr_\Pr(\Omega_1)\ge 1- \Delta_G$.
Conditional on $(W_g)_{g=1}^G$, $\{G^{-1/2} \sumg\xi_g(\hat\varphi_{gk}-\varphi_{gk})\}_{k\in[p]}$ is zero-mean Gaussian with variance $ G^{-1}\sum_{g=1}^{G} (\hat\varphi_{gk}-\varphi_{gk})^2\le \bar \sigma_G^2$ for all $k\in [p]$ with probability at least $1-\Delta_G$ following Assumption \ref{a:score_additional} (iii). By the Gaussian concentration inequality, for every $t>0$, we have
\begin{align*}
\Pr_\xi\left(\max_{k\in [p]}\left|\frac{1}{\sqrt{G}} \sumg \xi_g(\hat\varphi_{gk}-\varphi_{gk})\right|\ge\EP\left[\max_{k\in [p]}\left|\frac{1}{\sqrt{G}} \sumg \xi_g(\hat\varphi_{gk}-\varphi_{gk})\right|\right] +C't \right)\lesssim e^{-t^2}.
\end{align*}
Also by Gaussianity, conditional on $\Omega_1$, it holds that
\begin{align*}
\Ep_\xi\left[ \max_{1\le k \le p} \left|\frac{1}{\sqrt{G}} \sumg \xi_g(\hat\varphi_{gk}-\varphi_{gk})\right|   \right] \lesssim \bar \sigma_G \log^{1/2}p.
\end{align*}
Set $t=C'\bar \sigma \log^{1/2} G$ for some constant $C'$ large enough, conditional on $\Omega_1$, we have
\begin{align*}
\Pr_\xi\left(\max_{k\in [p]}\left|\frac{1}{\sqrt{G}} \sumg \xi_g(\hat\varphi_{gk}-\varphi_{gk})\right|\ge C'\bar \sigma_G \log^{1/2}(p\vee G) \right)\lesssim \frac{1}{G}\le (\log G)^{-1}+ \Delta_G,
\end{align*}
which implies the left hand side is greater than $(\log G)^{-1}+ \Delta_G$ only if $\Omega_1^c$ is true.
Recall that $\theta_1 \ge C \bar \sigma_G \log^{1/2} a_G$ and $\theta_2= (\log G)^{-1}+ \Delta_G$ and hence
\begin{align*}
\Pr_\Pr\left(\Pr_\xi\left(\max_{1\le k \le p} \left|\frac{1}{\sqrt{G}} \sumg\xi_g(\hat\varphi_{gk}-\varphi_{gk})\right|>\theta_1 \right)>\theta_2\right)\le \Pr_\Pr(\Omega_1^c)\le \Delta_G < \theta_2.
\end{align*}
This verifies condition (\ref{eq:conditional_prob_bound}).


\noindent \textbf{Step 3:} Here we establish bootstrap validity based on the results from preceding steps. Under Assumptions \ref{a:score_moment} (ii), \ref{a:score_nuisance_para} (ii), one can apply Theorem 3.2 of 
\cite{CCK13} and obtains that for every $\vartheta>0$,
\begin{align}
\sup_{\alpha\in (0,1)}|\Pr_\Pr(T\le c_W(\alpha))-\alpha|\lesssim  \rho + \pi(\vartheta) + \Pr_{\Pr} (\Delta>\vartheta)  
+ \theta_1 \sqrt{1 \vee \log(p/\theta_1)} + \theta_2,\label{eq:rho_bound}
\end{align}
with $\pi(\vartheta):=C_2\vartheta^{1/3} (1\vee \log(p/\vartheta))^{2/3}$ and $\Delta:= \max_{1\le j,l \le p}|\frac{1}{G} \sumg([\varphi_{gj} \varphi_{gl}]-\EP[\varphi_{gj} \varphi_{gl}])|$. 
It then suffices to show that each component on the right hand side of Equation (\ref{eq:rho_bound}) goes to zero.
We consider them one by one. First, $\rho=o(1)$ follows from the first step. Second, set $\vartheta=CB_G^2(\log p)^{3/2}/\sqrt{G}$ for some constant $C$. L'H\^opital's rule implies that $\vartheta^{1/3}(\log \vartheta)^{2/3}=o(1)$ as $\vartheta=o(1)$. Then Assumption \ref{a:score_additional}(i) implies that
\begin{align*}
\pi(\vartheta)\lesssim\frac{B_G^{2/3}\log^{1/2} p}{G^{1/6}}\left(1\vee \log^{2/3} p\right)-\vartheta^{1/3}(\log \vartheta)^{2/3} \lesssim\left(\frac{B_G^{4}\log^7 p}{G}\right)^{1/6}+o(1)=o(1),
\end{align*} 
Third, we verify $\Pr_\Pr(\Delta>\vartheta)=o(1)$. Under Assumption \ref{a:score_additional} (i), a direct application of Lemma C.1. of \cite{CCK13} gives
\begin{align*}
\EP[\Delta]\lesssim \sqrt{\frac{B_G^2 \log p}{G}}+ \frac{B_G^2 \log p}{\sqrt{G}}\lesssim  \frac{B_G^2 \log p}{\sqrt{G}}.
\end{align*}
Markov's inequality yields that
\begin{align*}
\Pr_\Pr\left(|\Delta-\EP[\Delta]|>t\right)\le \frac{\EP\left[\left|\Delta-\EP[\Delta]\right|\right]}{t}\le  \frac{\EP\left[\Delta\right]}{t}\lesssim  \frac{B_G^2 \log p}{t\sqrt{G}},
\end{align*}
where we used the fact that $\Delta\ge 0$.
Setting $t=C'B_G^2(\log p)^{3/2}/\sqrt{G}$ for an appropriate positive constant $C'$ yields
$
\Pr_\Pr(\Delta>\vartheta)\lesssim (\log G)^{-1/2}
$.
Next, under Assumption \ref{a:score_additional} (iii), $\theta_1=\delta_G\vee \bar \sigma_G \log^{1/2}a_G=o(1)$ and $\theta_1 \log^{1/2} p=o(1)$. L'H\^opital's rule implies that $\theta_1\log^{1/2} \theta_1=o(1)$. Hence the term $\theta_1\sqrt{1 \vee \log(p/\theta_1)}=o(1)$.
Finally, the last term of Equation (\ref{eq:rho_bound}) follows that $\theta_2= (\log G)^{-1}+ \Delta_G=o(1)$. This concludes the proof.
\end{proof}

\section{Proofs for Results in Section \ref{sec:main}}

\subsection{Proof for Theorem \ref{theorem:main_sufficient}}\label{sec:proof for theorem:main_sufficient}
\begin{proof}
	The results are implied by Theorem \ref{theorem:bahadur_rep} and Corollary \ref{corollary:MB}. Thus we shall verify Assumptions \ref{a:score_moment}, \ref{a:score_nuisance_para} and \ref{a:score_additional}.
	Set $\Delta_G=C(\log G)^{-1}$ and $\tau_G=C'(s\log a_G/G)^{1/2}$ for some sufficiently large positive constants $C$ and $C'$.
For each $k\in [p]$, let us define $\H_k=\H_k^G$, the bounded and convex sparse subset in $\Real^p$ shrinking with $G$, as follows
\begin{align*}
\H_k=\{\eta^k\}\cup\Bigg\{(\eta^{(1)},\eta^{(0)})\in \Real^{2p}:&\; \eta^{(0)}=\eta^{(2)}+\eta^{(3)},\; \|\eta^{(1)}\|_0 \vee \|\eta^{(2)}\|_0\vee \|\eta^{(3)}\|_0\le Cs, \\
&\|\eta^{(1)}-\beta^0\|_1\vee \|\eta^{(2)}-\zeta^k\|_1\vee \|\eta^{(3)}-\theta^k\|_1\le C\sqrt{s}\tau_G, \\
&\|\eta^{(1)}-\beta^0\|_2 \vee \|\eta^{(2)}-\zeta^k\|_2\vee \|\eta^{(3)}-\theta^k\|_2\le C\tau_G \qquad\Bigg\},
\end{align*}
The rest of this proof is divided into three steps corresponding to the verification of the three assumptions.

\noindent \textbf{Step 1.} In this step, we examine Assumption \ref{a:score_moment}. Assumption \ref{a:score_moment} (i) is clear since $\Lambda$ is infinitely continuously differentiable. 
To verify Assumption \ref{a:score_moment} (ii)(a), since for all $\eta^k\in \H_k$, $\|\eta^k -\eta \|_2\lesssim 1$, a mean value expansion and the definition of the induced matrix $\ell_2$ norm yield that
\begin{align*}
& \EP\left[\frac{1}{G}\sum_{g=1}^{G} \left(\sumi \psi_k(W_{ig},\eta) - \psi_k(W_{ig},\eta^k)\right)^2\right]\\
=& 
 \EP\left[\frac{1}{G}\sum_{g=1}^{G} \left(\sumi \partial_\eta \psi_k(W_{ig},\widetilde \eta)'(\eta-\eta^k)\right)^2\right]\\
=&
 (\eta-\eta^k)' \EP\left[\frac{1}{G}\sum_{g=1}^{G} \left(\sumi \partial_\eta \psi_k(W_{ig},\widetilde \eta))\right)\left(\sumi \partial_\eta \psi_k(W_{ig},\widetilde \eta)\right)'\right](\eta-\eta^k)\\
\le& 
\|\eta - \eta^k\|_2^2 \left\|  \EP\left[\frac{1}{G}\sum_{g=1}^{G} \left(\sumi \partial_\eta \psi_k(W_{ig},\widetilde \eta))\right)\left(\sumi \partial_\eta \psi_k(W_{ig},\widetilde \eta)\right)'\right]\right\|_2 ,
\end{align*}
where each coordinate of $\widetilde\eta$ lies between the corresponding coordinate of $\eta$ and $\eta^k$. By Assumption \ref{a:parameters} and the definition of $\tau_G$, we know $\|\widetilde \eta\|_2\lesssim 1$. Notice that
\begin{align*}
\sumi\partial_\eta \psi_k(W_{ig},\widetilde\eta)=&
\sumi
\begin{bmatrix}
\partial_\beta \psi_k(W_{ig},\widetilde\eta)\\
\partial_\mu \psi_k(W_{ig},\widetilde\eta)
\end{bmatrix}
=
\sumi
\begin{bmatrix}
\widetilde\beta_k \Lambda''(X_{ig}'\widetilde\beta)X_{ig} +\Lambda'(X_{ig}'\widetilde\beta) e_k + \widetilde\mu'X_{ig} \Lambda' (X_{ig}'\widetilde\beta)X_{ig}\\
-\{Y_{ig} - \Lambda (X_{ig}'\widetilde\beta)\}X_{ig}
\end{bmatrix}\\
=&\begin{bmatrix}
A_{g}+B_{g}+C_{g}\\
D_{g}
\end{bmatrix}.
\end{align*}
Thus, the sum of cross products can be denoted by
\begin{align*}
&\left\| \EP\left[\frac{1}{G}\sum_{g=1}^{G} \left(\sumi \partial_\eta \psi_k(W_{ig},\widetilde \eta)\right)\left(\sumi \partial_\eta \psi_k(W_{ig},\widetilde \eta)\right)'\right]\right\|_2\\
=&
\Bigg\| \EP\frac{1}{G}\sum_{g=1}^{G} 
\begin{bmatrix}
(A_{g}+B_{g}+C_{g})(A_{g}+B_{g}+C_{g})' & (A_{g}+B_{g}+C_{g})D_{g}'\\
(A_{g}+B_{g}+C_{g})'D_{g} &D_{g}D_{g}'
\end{bmatrix}
\Bigg\|_2.
\end{align*}
To further bound the right-hand side, it suffices to bound the matrix $\ell_2$-norm for each of the product terms. Under Assumption \ref{a:covariates} (4) and $\|\widetilde \mu\|_2 \lesssim 1$, we have
{\small\begin{align*}
&\left\| \EP\left[\frac{1}{G}\sum_{g=1}^{G}  C_g C_g'\right]\right\|_2 
\le
 \left\| \EP\left[\frac{1}{G}\sum_{g=1}^{G} ( \widetilde \mu' U_g  )^2U_g U_g'\right]\right\|_2
\lesssim \sup_{\|\xi\|_2 = 1}\EP\left[\frac{1}{G}\sum_{g=1}^{G} ( \widetilde \mu' U_g)^2 \xi' U_g U_g'\xi\right] \\
\le& \left\{  \EP\left[\frac{1}{G}\sum_{g=1}^{G} ( \widetilde \mu' U_g)^4 \right]\right\}^{1/2} \max_{\|\xi\|_2 = 1}\left\{ \EP\left[\frac{1}{G}\sum_{g=1}^{G}  (\xi' U_{g})^4\right]\right\}^{1/2}\le  C_1, 
\text{ and }\\
&\left\| \EP\left[\frac{1}{G}\sum_{g=1}^{G}  A_g A_g'\right]\right\|_2\bigvee \left\| \EP\left[\frac{1}{G}\sum_{g=1}^{G}  B_g B_g'\right]\right\|_2
\bigvee \left\| \EP\left[\frac{1}{G}\sum_{g=1}^{G}  D_g D_g'\right]\right\|_2
\lesssim \left\|\sup_{\|\xi\|_2 = 1} \EP\left[\frac{1}{G}\sum_{g=1}^{G} (U_g'\xi)^2\right]\right\|_2 \le  C_1.
\end{align*}}
This shows Assumption \ref{a:score_moment} (ii)(a).

To verify Assumption \ref{a:score_moment} (ii)(b), note that we can write the matrix 
\begin{align*}
&\sumi\partial_{\mu'} \partial_{\mu} \psi_k(W_{ig},\widetilde\eta)=
\begin{bmatrix}
A_g & B_g\\
B_g & 0
\end{bmatrix}\\
=&\sumi
\begin{bmatrix}
\widetilde\beta_k \Lambda'''(X_{ig}'\widetilde\beta)X_{ig}X_{ig}' + \Lambda''(X_{ig}'\widetilde\beta)e_k X'_{ig} + \Lambda''(X_{ig}'\widetilde\beta)X_{ig}e_k' + \widetilde\mu'X_{ig} \Lambda'' (X_{ig}'\widetilde\beta)X_{ig}X_{ig}' & \Lambda' (X_{ig}'\widetilde\beta)X_{ig} X_{ig}'\\
 \Lambda' (X_{ig}'\widetilde\beta)X_{ig}X_{ig}'& 0
\end{bmatrix}.
\end{align*}
So for $\eta = [\beta', \mu']'$, we have
\begin{align*}
\left\|\partial_{\mu'}  \EP\left[\frac{1}{G}\sum_{g=1}^{G}  \sumi\partial_{\mu} \psi_k(W_{ig},\widetilde\eta)\right]\right\|_2 
=&\max_{0<\|\xi\|_2\le 1} \xi' \EP\left[\frac{1}{G}\sum_{g=1}^{G}  \sumi\partial_{\mu'} \partial_{\mu} \psi_k(W_{ig},\widetilde\eta)\right] \xi\\
\le&
 \max_{0<\|\xi\|_2\le 1}  \left( \beta'  \EP\left[\frac{1}{G}\sum_{g=1}^{G}  A_g \right]\beta + 2\beta' \EP\left[\frac{1}{G}\sum_{g=1}^{G}  B_g\right]\eta \right).
\end{align*}
By Cauchy-Schwarz and the definition of matrix $\ell_2$-norm, 
$$ \beta'  \EP\left[\frac{1}{G}\sum_{g=1}^{G}  B_g\right]\eta\le \|\eta\|_2 \left\|\beta'  \EP\left[\frac{1}{G}\sum_{g=1}^{G}  B_g\right]\right\|_2 
\le \|\beta\|_2\|\eta\|_2 \left\| \EP\left[\frac{1}{G}\sum_{g=1}^{G}  B_g\right]\right\|_2 .$$ Thus a bound can be obtained similarly to \ref{a:score_moment} (ii)(a).

\noindent \textbf{Step 2.} In this step, we check Assumption \ref{a:score_nuisance_para}.
Assumption \ref{a:score_nuisance_para} (i) is a direct implication of Theorems \ref{thm:LOG_rates}, \ref{thm:REGNPL_rates} and Corollary \ref{corollary:weighted_lasso_rates}.
To verify Assumption \ref{a:score_nuisance_para} (ii), note that set $\Delta_G=C(\log G)^{-1}$ and $\tau_G=(s\log a_G/G)^{1/2}$, it follows from the convergence rate results of Theorems \ref{thm:LOG_rates}, \ref{thm:theta_k_rates} and Corollary \ref{corollary:weighted_lasso_rates}.
To verify Assumption \ref{a:score_nuisance_para} (ii), recall that
\begin{align*}
\bar\psi_k (W_{ig},\alpha_k,\eta)
 =&\alpha_k- \beta_k \Lambda'(X_{ig}'\beta)+\mu'
  \{Y_{ig}-\Lambda(X_{ig}'\beta)\}X_{ig}
\end{align*}
Pointwise measurability follows from its continuity. Now, for some constant $C$ large enough, 
define classes of functions
\begin{align*}
\calG_{1iT}=&\Big\{W_g\mapsto Y_{ig}-  \Lambda(X_{ig}'\beta): \beta\in \Real^p, T=\supp(\beta),\, \|\beta-\beta^0\|_2 \le C\tau_G \Big\},\\
\calG_{2iT}=&\Big\{W_g\mapsto  \Lambda'(X_{ig}'\beta):\beta\in \Real^p,T=\supp(\beta),\, \|\beta-\beta^0\|_2 \le C\tau_G\Big\},\\
\calG_{3ikT}=&\Big\{W_g\mapsto \mu'X_{ig}: \mu\in \Real^p,T=\supp(\mu),\, \|\mu-\mu^k\|_2 \le C\tau_G \Big\},\\
\calG_{4}=&\Big\{W_g\mapsto b: |b|\le C \Big\},\\
\calG_{5i}=&\Big\{W_g\mapsto  \Lambda'(X_{ig}'\beta^0)\Big\},\\
\calG_{6ik}=&\Big\{W_g\mapsto \mu^{k\prime}X_{ig}\Big\},\\
\calG_{7i}=&\Big\{W_g\mapsto Y_{ig}-  \Lambda(X_{ig}'\beta^0) \Big\}.
\end{align*}
Then we have $\calF_0\subset \calF_0' \cup \calF_0'' \cup \{0\}$, where
\begin{align*}
\calF_0':=&  - (\calG_{4})\cdot \sum_{i=1}^{\bar n} \left(\cup_{T\subset [p], |T|\le Cs}\calG_{2iT}\right) + \sum_{i=1}^{\bar n} \left(\cup_{k\in [p]}\cup_{T\subset [p], |T|\le Cs}\calG_{3ikT}\right)\cdot \left(\cup_{T\subset [p], |T|\le Cs}\calG_{1iT}\right),\\
\calF_0'':=& - \left(\calG_{4}\right)\cdot \sum_{i=1}^{\bar n} \calG_{5i} + \sum_{i=1}^{\bar n} \left(\cup_{k\in [p]}\calG_{6ik}\right)\cdot \calG_{8i}.
\end{align*}
Note that all these classes are uniformly bounded with the exceptions of $\calG_{3ikT}$ and $\calG_{6ik}$.
To obtain envelopes for them, note that all classes are uniformly bounded except for $\calG_{3ikT}$ and $\calG_{6ik}$. To obtain an envelope for $\calG_{3ikT}$, notice for any $ikT$, $\|\mu^k\|_1 \le \sqrt{s}C_1$ since $\|\mu^k\|_2\le C_1$ following Assumption \ref{a:parameters}. Therefore, $\|\mu\|_1\le \|\mu^k \|_1 + \|\mu - \mu^k\|_1 \le \sqrt{s}C_1 + \sqrt{s} C_1 \tau_G\lesssim \sqrt{s}$.
Set envelope $\mathfrak g$ to be such that
\begin{align*}
\mathfrak g(W_g)=\max_{k\in[p]}\max_{i\in [\bar n]}\sup_{,\mu \in \Real^p: \|\mu - \mu^k\|_1\le C\sqrt{s}\tau_G} |\mu'X_{ig}|,
\end{align*} 
then for any $\mu$ in the index set, one has
\begin{align*}
|\mu'X_{ig}|\le |(\mu-\mu^k)'X_{ig}| +|\mu^{k\prime} X_{ig}| \lesssim C\sqrt{s}\tau_G\|U_g\|_\infty +|\mu^{k\prime} X_{ig}|.
\end{align*}
Since $\mu^k= \zeta^k + \theta^k$ and $\theta^k =[-\gamma^k_1,...,-\gamma^k_{k-1},1,-\gamma^k_{k+1},...,-\gamma^k_{p-1}]/\tau^2_k$ and $\tau^{-2}_k=O(1)$ following from Assumption \ref{a:covariates} (1), we have
\begin{align*}
|\mu^{k\prime} X_{ig}|\le& |\zeta^{k\prime} X_{ig}| +|\theta^{k\prime} X_{ig}| \\
\lesssim &|S^k_{ig}| + |\varepsilon^k_{ig}|  +|D^k_{ig}| +  |\gamma^{k\prime}
X^k_{ig}|\\
\lesssim& 1+ \|V_g\|_\infty + \|U_g\|_\infty, 
\end{align*}
where the last inequality is due to $ |\gamma^{k\prime}
X^k_{ig}|\le |D^k_{ig}| + |Z_{ig}^k|$ and the definition of $V_{g}$, $U_g$. Now, Assumption \ref{a:covariates} (6) implies $\sqrt{s}\tau_G=o(1)$, thus the above implies
\begin{align*}
|\mu'X_{ig}|\lesssim \|V_g\|_\infty + \|U_g\|_\infty \le 2 (\|V_g\|_\infty \vee \|U_g\|_\infty).
\end{align*}
Under Assumption \ref{a:covariates} (5)(7),
\begin{align}
 \left\{\EP\left[\frac{1}{G}\sum_{g=1}^{G} \mathfrak g^q(W_g)\right]\right\}^{1/q} 
 \lesssim& 
  \left\{\EP\left[\frac{1}{G}\sum_{g=1}^{G} (\|V_g\|_\infty \vee \|U_g\|_\infty)^q
\right]  \right\}^{1/q} \le \left\{\EP\left[\frac{1}{G}\sum_{g=1}^{G} (\|V_g\|_\infty \vee \|U_g\|_\infty)^{2q}\right]\right\}^{1/2q}\nonumber\\
  \lesssim&  M_{G,1} \vee M_{G,2}. \label{eq:suff_envelope_lq_bound}
\end{align}
Similar argument holds for $\calG_{6ik}$ as well.

To obtain a bound for the uniform covering entropy number, first let us consider $\calG_{3ikT}$ for some fixed $i,k\in[p]$ and $|T|\le Cs$. Applying Lemma 21 of \cite{Kato17}, we have that each $\calG_{3ikT}$ is a VC-subgraph class of functions with VC-index $Cs+2=O(s)$. Thus the union of these $ p\cdot {p \choose Cs}$ class of functions is a VC-type class and has uniform covering number satisfying that for any $0<\varepsilon\le 1$, there is an $A\ge 1$ such that
\begin{align*}
\sup_Q N(\cup_{k\in [p]}\cup_{T\subset [p], |T|\le Cs}\calG_{3ikT},\|\cdot\|_{Q,2},\varepsilon\|\mathfrak g\|_{Q,2})\lesssim  a_G^{Cs+2}  \left(A/\varepsilon\right)^{Cs}.
\end{align*}
Thus we have
\begin{align*}
\sup_Q\log N(\cup_{k\in [p]}\cup_{T\subset [p], |T|\le Cs}\calG_{3ikT},\|\cdot\|_{Q,2},\varepsilon\|\mathfrak g\|_{Q,2})\lesssim s \log(a_G/\varepsilon).
\end{align*}
Similar entropy calculations hold for $\calG_{1iT}$ and $\calG_{2iT}$ as well since $\Lambda$ is monotone and $\Lambda'=\Lambda \cdot (1-\Lambda)$ and thus Lemma 22 of \cite{Kato17} and Lemma \ref{lemma:entropy_algebra} can be applied. We therefore conclude that Assumption \ref{a:score_nuisance_para} (ii) holds with $v_G=s$, $F_0=C\mathfrak g$ and $K_G=C'(M_{G,1}\bigvee M_{G,2})$.

To verify Assumption \ref{a:score_nuisance_para} (iii), note that the lower bound is implied by Assumption \ref{a:covariates} (1). It then suffices to bound $ \EP\left[G^{-1}\sum_{g=1}^{G} (\mu'
 X_{ig})^2\right]\le \left\{  \EP\left[G^{-1}\sum_{g=1}^{G} (\mu'
 X_{ig})^4\right]\right\}^{1/2}\lesssim 1$ for all $\|\mu\|_2\le C$. This follows directly from Assumption \ref{a:covariates} (4).

To verify Assumption \ref{a:score_nuisance_para} (iv),
recall that we set $v_G=s$, $\tau_G=(s\log a_G/G)^{1/2}$, $K_G=C'  (M_{G,1}\vee M_{G,2})$, then by Assumption \ref{a:covariates} (6)(8), we have
\begin{align*}
\frac{s\log a_G }{G^{1/2}}\bigvee 
\frac{s  (M_{G,1}\vee M_{G,2})\log a_G}{G^{1/2-1/q}}\lesssim\frac{s  (M_{G,1}\vee M_{G,2})\log a_G}{G^{1/2-1/q}} \lesssim \check\delta_G.
\end{align*}

\noindent \textbf{Step 3.} In this step, we examine Assumption \ref{a:score_additional}.
To verify Assumption \ref{a:score_additional} (i), we need to find $B_G$ such that $B_G^4(\log (pG))^7/G =o(1)$ and for all $1\le g\le G$, $k\in[p]$ $\kappa=1,2$,
\begin{align*}
& \EP\left[\frac{1}{G}\sum_{g=1}^{G} \left|\sumi (- \beta^0_k \Lambda(X_{ig}'\beta^0) + \mu^{k\prime}X_{ig}\{Y_{ig}- \Lambda(X_{ig}'\beta^0)\})\right|^{2+\kappa}\right]\lesssim \sigma_k^{2+\kappa} B_G^\kappa,\\
&\EP\left[\max_{1\le k\le p}\left|\sumi( - \beta^0_k \Lambda(X_{ig}'\beta^0)+ \mu^{k\prime}X_{ig}\{Y_{ig}- \Lambda(X_{ig}'\beta^0)\})\right|^4 \right]\lesssim \sigma_k^4 B_G^4.
\end{align*} 
Assumption \ref{a:score_nuisance_para} (iii) implies that $1\lesssim \sigma_k \lesssim 1$. For the first term, note that it suffices to bound $\sup_{\|\mu\|_2 =1} \EP\left[G^{-1}\sum_{g=1}^{G} (U_g'\mu)^{2+b}\right]$, which is bounded under Assumption \ref{a:covariates} (4). So the entire first term is bounded by a constant. Now, for the second term, note that using (\ref{eq:suff_envelope_lq_bound}),
\begin{align*}
&\max_{g\in[G]}\EP\left(\max_{1\le k\le p}\Big|\sumi(\alpha_k - \beta^0_k \Lambda(X_{ig}'\beta^0)+ \mu^{k\prime}X_{ig}\{Y_{ig}- \Lambda(X_{ig}'\beta^0)\})\Big|\right)^4\\
\lesssim& \EP\left[ \sumg \mathfrak g^4(W_g)\right]
 \lesssim G^{2/q} (M_{G,1}\vee M_{G,2})^4.
\end{align*}
Take $B_G= C G^{1/2q} (M_{G,1}\vee M_{G,2})$ for some $C$ large enough, we have
\begin{align*}
\frac{B_G^4(\log (pG))^7}{G} \lesssim \frac{(M_{G,1}\vee M_{G,2})^4 (\log a_G)^7}{G^{1-2/q}}=o(1)
\end{align*}
under the rate condition in Assumption \ref{a:covariates} (6) and (8).

To verify Assumption \ref{a:score_additional} (ii), observe that Assumption \ref{a:covariates} (1) implies that $\sigma_k$ is bounded away from zero uniformly in $k$. Set $\Delta_G=C(\log G)^{-1}$, $\tau_G= (s\log a_G/G)^{1/2}$ and $\bar \sigma_G=(s\log^2 a_G/G)^{1/2} $, by Assumption \ref{a:score_nuisance_para} (i), with probability $1-C\Delta_G^{-1}$, we have
$\hat \eta^k \in \H_k$. Hence by Assumption \ref{a:score_moment} (ii)(a), for any $k\in[p]$, $\EP[ G^{-1}\sum_{g=1}^{G}  (\varphi_{gk}-\hat 
\varphi_{gk})^2]\le \tau_G^2$. An application of Markov's inequality shows that with probability at least $1-C\Delta_G^{-1}$, $G^{-1}\sum_{g=1}^{G}  (\varphi_{gk}-\hat 
\varphi_{gk})^2\lesssim \bar \sigma_G^2 $.
We have thus verified Assumption \ref{a:score_additional} (ii).

Finally, Assumption \ref{a:score_additional} (iii) is satisfied by setting $\tau_G= (s\log a_G/G)^{1/2}$, $\bar \sigma_G=(s\log^2 a_G/G)^{1/2}$ and $\delta_G=(\log a_G)^{-2}$ under Assumption \ref{a:covariates} (8).
\end{proof}
\subsection{Proof for Lemma \ref{lemma:variance_est}}\label{sec:proof for lemma:variance_est}
\begin{proof}
Notice that Assumption \ref{a:covariates} (4) implies that $\sigma_k \lesssim 1$. 
By the continuous mapping theorem, it suffices to bound
\begin{align*}
|\widetilde \sigma^2_k - \sigma^2_k|\le& \left| \frac{1}{G} \sumg \left(\sumi \bar \psi_k(W_{ig},\widetilde \alpha_k, \widetilde \eta^k) \right)^2 -  \EP\left[\frac{1}{G}\sum_{g=1}^{G}  \left(\sumi \bar \psi_k(W_{ig}, \alpha_k, \eta^k) \right)^2\right] \right|\\
 \lesssim& 
 |\widetilde \alpha_k^2-\alpha_k^2| 
 + 
  \sup_{\eta \in \H_k}|\widetilde \alpha_k - \alpha_k|\left|\frac{1}{G}\sumg\sumi \psi_k(W_{ig},\eta)\right| \\
& + 
 \sup_{\eta \in \H_k}\left| \frac{1}{G} \sumg \left(\sumi  \psi_k(W_{ig},  \eta) \right)^2 -  \EP\left[\frac{1}{G}\sum_{g=1}^{G}  \left(\sumi  \psi_k(W_{ig}, \eta) \right)^2\right] \right| \\
 & + 
 \sup_{\eta \in \H_k}\left| \EP\left[\frac{1}{G}\sumg \left\{ \left(\sumi  \psi_k(W_{ig},  \eta) \right)^2 -   \left(\sumi  \psi_k(W_{ig}, \eta^k) \right)^2\right\} \right]\right| \\
=&(I) +(II) +(III)+(IV) 
\end{align*}
uniformly over $k\in [p]$.
First of all, $(IV)=\op(1) $ by the Lipschitzness of $\EP\left[G^{-1}\sumg \left(\sumi \psi_k (W_{ig},\cdot) \right)^2\right]$ and Assumption \ref{a:score_moment}, which is verified in Theorem \ref{theorem:main_sufficient}. To bound $(III)$,
let the collection of functions
\begin{align*}
\calF=\left\{ W_g \mapsto  \sum_{i=1}^{\bar n}  \sum_{j=1}^{\bar n} \psi_k(W_{ig} ,\eta) \psi_k(W_{jg},\eta)\1\{\||W_{ig}|\|_\infty \wedge \||W_{jg}|\|_\infty >0\}: k\in [p] , \eta \in \H_k \right\}.
\end{align*}
Under Assumption \ref{a:covariates} (1)(5)(6)(7), using a similar argument as in Theorem \ref{theorem:main_sufficient}, we obtain
\begin{align*}
\sup_Q N(\varepsilon\|F\|_{Q,2},\calF,\|\cdot\|_{Q,2})\lesssim  \left(\frac{A}{\varepsilon}\right)^{Cs}
\end{align*}
with an envelope $F$ defined by
\begin{align*}
F(W_g)=\max_{k\in[p]}\max_{i\in [\bar n]}\sup_{\mu \in \Real^p: \|\mu - \mu^k\|_1\le C\sqrt{s}\tau_G} \bar n^2|\mu'X_{ig}|^2 + C_3
\end{align*}
for some constant $C_3$. Furthermore, under Assumption \ref{a:covariates} (5)-(8), it holds that
\begin{align*}
\EP\left[\max_{g\in[G]} F^2(W_g)\right]\lesssim& \EP\left[\max_{g\in[G]}\max_{k\in[p]}\sup_{\mu \in \Real^p: \|\mu - \mu^k\|_1\le C\sqrt{s}\tau_G} |(\mu - \mu^k)'U_{g}|^4\right] + \EP\left[\max_{g\in[G]}\max_{i\in[\bar n]}\max_{k\in[p]} (\mu^{k\prime} X_{ig})^4\right] \\
\lesssim& G^{2/q}\left( \EP\frac{1}{G}\sum_{g=1}^{G}  s^2\tau_G^4\|U_{g}\|^{2q}_\infty + (M_{G,1}\vee M_{G,2})^{2q}
\right)^{2/q}\\
\lesssim&G^{2/q} \{ s^2\tau_G^4 M_{G,2}^4  + (M_{G,1}\vee M_{G,2})^4\}\lesssim G^{2/q}  (M_{G,1}\vee M_{G,2})^4.
\end{align*}
Applying Lemma \ref{lemma:maximal_ineq} under Assumption \ref{a:covariates} (4)(5)(6)(7)(8), with probability at least $1-C(\log G)^{-1}$, we have
\begin{align*}
& \sup_{\eta \in \H_k}\left| \frac{1}{G} \sumg \left(\sumi  \psi_k(W_{ig},  \eta) \right)^2 -  \EP\left[\frac{1}{G}\sum_{g=1}^{G}  \left(\sumi  \psi_k(W_{ig}, \eta^k) \right)^2\right] \right|\\
\lesssim &\sqrt{\frac{s \log a_G}{G}} + \frac{(M_{G,1}\vee M_{G,2})^2 s \log a_G}{G^{1-1/q}} 
= o(\log^{-1} a_G).
\end{align*} 

To bound $(I)$ and $(II)$, note that Theorem \ref{theorem:bahadur_rep} suggests
\begin{align*}
\widetilde \alpha_k - \alpha_k = \frac{1}{G} \sumg\sumi \psi_k(W_{ig},\eta^k)+\op(G^{-1/2})
\end{align*}
uniformly in $k\in [p]$. We may apply Lemma \ref{lemma:maximal_ineq} to
\begin{align*}
\calF=(\calG_{4})\cdot \sum_{i=1}^{\bar n} \calG_{5i} + \sum_{i=1}^{\bar n} (\cup_{k\in [p]}\calG_{6ik})\cdot (\calG_{7i})
\end{align*}
as the components $\calG$'s are defined in the proof of Theorem \ref{theorem:main_sufficient}. An envelope can be 
\begin{align*}
F(W_g)=\max_{k\in [p]}\max_{i\in[\bar n]}\sup_{\mu\in \Real^p:\|\mu- \mu^k\|_1 \le C\sqrt{s}\tau_G}C(1+ \mu'X_{ig})
\end{align*}
for some $C$ that does not depend on $G$.
It is then implied by (\ref{eq:suff_envelope_lq_bound}) that $ \left\{\EP\left[G^{-1}\sum_{g=1}^{G} F^q\right]\right\}^{1/q}\lesssim M_{G,1}\vee M_{G,2}$ and thus $\sqrt{\EP\left[\max_{g\in [G]}F^2(W_g)\right]}\lesssim G^{1/2q} (M_{G,1}\vee M_{G,2})$. We also have
$\sup_{f\in \calF} \EP\left[G^{-1}\sum_{g=1}^{G} f^2\right]\lesssim  C+ \sup_{\xi\in\Real^p:\|\xi\|_2=1} \EP\left[G^{-1}\sum_{g=1}^{G} (U_g'\xi)^2\right]\lesssim 1$.

Applying Lemma \ref{lemma:maximal_ineq} under Assumption \ref{a:covariates} (4)(5)(6)(7)(8) leads to 
\begin{align*}
\max_{k\in [p]}|\widetilde \alpha_k - \alpha_k| \lesssim \sqrt{\frac{\log a_G}{G}}+\frac{s (M_{G,1}\vee M_{G,2})\log a_G}{G^{1-1/2q}}=o(\log^{-1} a_G)
\end{align*}
with probability at least $1-C(\log G)^{-1}$. This implies $(I)+(II)=\op(1)$.
\end{proof}

\section{Proofs for Results in Section \ref{sec:nuisance_parameters}}

\subsection{Proof for Theorem \ref{thm:LOG_rates}}\label{sec:proof for theorem:LOG_rates}
The steps of this proof are analogous to the one of Theorem 4.1 in \cite{BCCW18} with modifications to account for cluster sampling. The major difference lies in the verification of Assumption \ref{a:M-est_lasso} (2).
\begin{proof}
We will apply Lemma \ref{lemma:M-est_lasso_rates}, \ref{lemma:M-est_lasso_sparsity} and \ref{lemma:M-est_post_lasso_rates} after verifying the  required assumptions. Let
$w_{ig}= f_{ig}^2$ and 
\begin{align*}
M (Y_{ig},X_{ig},\beta)=\hat M (Y_{ig},X_{ig},\beta)=-\{Y_{ig} X_{ig}'\beta - \log(1+\exp(X'_{ig}\beta))\}.
\end{align*} 
In order to apply Lemma \ref{lemma:M-est_WL}, we verify Assumption \ref{a:M-est_WL}. Since
\begin{align*}
S_g=& \partial_\beta M(Y_{ig},X_{ig},\beta)|_{\beta = \beta^0}\\
=&-\sumi \{Y_{ig}- \Lambda(X_{ig}'\beta^0)\} X_{ig},
\end{align*}
we have  $|S_{gj}|\le \bar n \max_{i\in[n_g]}|X_{ig,j}|= U_{gj}$. 
In addition, since $\gamma\ge 1/G$, using the fact that $1-\Phi(t)\le (2\pi)^{-1/2}t^{-1} e^{t^2/2}$, we have
$
\Phi^{-1} (1-\gamma/2p) \lesssim \sqrt{\log (pG)}\lesssim \sqrt{\log a_G}
$.
Using Assumption \ref{a:covariates} (2),(3),(4), it follows that $\log^{1/2}a_G \lesssim \check \delta_G G^{1/6}$ and $\left\{ \EP\left [G^{-1}\sum_{g=1}^{G} |S_{gj}|^3\right]\right\}^{1/3}\lesssim 1$ uniformly over $j\in[p]$. Thus Assumption \ref{a:M-est_WL} (1) is satisfied. Under Assumption \ref{a:covariates} (1),(4), it holds uniformly over $j\in[p]$ that
\begin{align*}
& \EP\left[\frac{1}{G}\sum_{g=1}^{G} S_{gj}^2\right] 
\ge
 \inf_{\|\xi\|_2 =1} \EP\left[\frac{1}{G}\sum_{g=1}^{G} \left(\sumi \{Y_{ig}-\Lambda(X_{ig}'\beta^0)\} X_{ig}'\xi\right)^2 \right]
 \gtrsim 1 \text{ and }\\
&   \EP\left[\frac{1}{G}\sum_{g=1}^{G} S_{gj}^2\right]
\le 
 \EP\left[\frac{1}{G}\sum_{g=1}^{G} U_{gj}^2\right]\lesssim 1.
\end{align*}
So Assumption \ref{a:M-est_WL} (2) is verified.

To apply Lemma \ref{lemma:M-est_lasso_rates}, we verify Assumption \ref{a:M-est_lasso}. The convexity is trivial. To show Assumption \ref{a:M-est_lasso} (2) holds, note that $S_{gj}^2\le U_{gj}^2$ and Assumption \ref{a:covariates} (4)(7) implies that if we let 
\begin{align*}
\calF=&\left\{W_g\mapsto\left(-\sum_{i=1}^{\bar n} \{Y_{ig} -\Lambda(X'_{ig} \beta^0) \} X_{ig,j}\right)^2 : j\in [p] \right\},\:
\calF_j=\left\{W_g\mapsto\left(-\sum_{i=1}^{\bar n} \{Y_{ig} -\Lambda(X'_{ig} \beta^0) \} X_{ig,j}\right)^2  \right\},
\end{align*}
then each $\calF_j$ is of VC-subgraph class since it consists of a single function, and $\calF\subset \cup_{j\in [p]}\calF_j$,
and $ \calF $ has an envelope $F$ such that $ F(W_g)=\bar n^2 \max_{j\in[p]}\max_{i\in [n_g]}|X_{ig,j}|^2$. Note that under Assumption \ref{a:covariates} (4)(7)(8), one has $\max_{j\in[p]} \EP\left[ G^{-1}\sum_{g=1}^{G} U_{gj}^4\right]\lesssim 1$, $\left( \EP\left[G^{-1}\sum_{g=1}^{G} F^2(W_g)\right]\right)^{1/2}=\left( \EP\left[G^{-1}\sum_{g=1}^{G} \|U_g\|^4_\infty\right]\right)^{1/2}\le  M_{G,2}^2$, $M_{G,2}\le \left( \EP\left[G^{-1}\sum_{g=1}^{G} \|U_g\|^{2q}_\infty\right]\right)^{1/2q}\le (\check \delta_G G^{1/2-1/q} )^{1/2}$, 
and $\sqrt{\EP[(\max_{1\le g \le G}F(W_g))^2]}\le G^{1/q} M_{G,2}^2$.
Applying Corollary 2.1 of \cite{CCK13}, with probability at least $1-c(\log G)^{-1}$, it holds that
\begin{align*}
\Big|\frac{1}{G} \sumg (S_{gj}^2-\EP [S_{gj}^2])\Big|
\lesssim&\ \sqrt{\frac{\log ( a_G )}{G}} + \frac{M_{G,2}^2}{G^{1-1/q}}\log ( a_G ) =o(1),
\end{align*}
where the last equality follows from Assumption \ref{a:covariates} (9).
This implies $G^{-1} \sumg S^2_{gj}=(1-o(1)) \EP[G^{-1}\sum_{g=1}^{G}  S^2_{gj}]$ uniformly in $j\in[p]$. Similar arguments can be used to establish the statement that $G^{-1}\sumg (n_g\sumi X^2_{ig,j})=(1-o(1)) \EP[G^{-1}\sum_{g=1}^{G}  (n_g\sumi X^2_{ig,j})]$ with probability at least $1-C(\log G)^{-1}$. Now it suffices to show that
\begin{align}
(1-o(1)) \EP\left[\frac{1}{G}\sum_{g=1}^{G}  S_{gj}^2\right] \le \hat \Psi_{j}^2 \lesssim 1 \label{eq:I.3}
\end{align}
with probability $1-c(\log G)^{-1}$ uniformly over $j\in[p]$. The case of $\bar m =0$ follows from the calculations that
\begin{align*}
\hat l^2_{j,0}=\frac{1}{4}\frac{1}{G} \sumg\left(n_g\sumi X^2_{ig,j}\right)\lesssim\frac{1-o(1)}{4}\EP\left[\frac{1}{G}\sumg\sumi X^2_{ig,j}\right]
\le 
\frac{1-o(1)}{4} \EP\left[\frac{1}{G}\sum_{g=1}^{G}  U_{gj}^2\right]
 \lesssim 1
\end{align*} 
with probability $1-c(\log G)^{-1}$ under Assumption \ref{a:covariates} (4) and 
\begin{align*}
&\frac{1}{4} \EP\left[\frac{1}{G}\sum_{g=1}^{G} \sumi X^2_{ig,j}\right]
\ge 
 \EP\left[\frac{1}{G}\sum_{g=1}^{G} \sumi f_{ig}^2 X^2_{ig,j}\right]
 = 
  \EP\left[\frac{1}{G}\sum_{g=1}^{G} \sumi \{Y_{ig}- \Lambda(X_{ig}'\beta^0)\}^2 X_{ig,j}^2\right]\\
\gtrsim&  \EP\left[\frac{1}{G}\sum_{g=1}^{G} \left(\sumi \{Y_{ig}- \Lambda(X_{ig}'\beta^0)\}X_{ig,j}\right)^2\right]= \EP\left[\frac{1}{G}\sum_{g=1}^{G}  S_{gj}^2\right],
\end{align*}
where the Cauchy-Schwarz inequalty,
the law of iterated expectations and the fact that $f_{ig}^2\le \|\Lambda(1-\Lambda)\|_\infty\le 1/4$ are used.

To show (\ref{eq:I.3}) with $m \ge 1$, suppose that (\ref{eq:I.3}) holds for $\bar m-1$, we can complete the proof and has $\|f_{ig} X_{ig} (\hat \beta - \beta^0)\|_G\lesssim (s\log a_G/G)^{1/2}$ 
with probability $1-C(\log G)^{-1}$. For $m=\bar m$, denote $\Lambda_{ig}=\Lambda(X_{ig}'\beta^0)$ and $\widetilde\Lambda_{ig}=\Lambda(X_{ig}'\widetilde\beta)$, use the fact that for positive $a,b$, $|\sqrt{a}-\sqrt{b}|\le \sqrt{|a-b|}$, we have
\begin{align*}
|\hat l_j - l_j|\le& \left( \Big| \frac{1}{G} \sumg \hat S_{gj}^2 -\frac{1}{G} \sumg S_{gj}^2  \Big|\right)^{1/2}.
\end{align*}
In addition, it holds uniformly over $j \in[p]$ that with probability at least $1-C(\log G)^{-1}$,
{\small
\begin{align*}
&\Big| \frac{1}{G} \sumg \hat S_{gj}^2 -\frac{1}{G} \sumg S_{gj}^2  \Big|\\
\le& \Big|\frac{1}{G} \sumg \left(\sumi \{\Lambda_{ig}- \widetilde \Lambda_{ig}\}X_{ig,j}\right)^2\Big| + 2 \Big|\frac{1}{G} \sumg \left(\sumi \{Y_{ig} - \Lambda_{ig}\}X_{ig,j}\right) \left(\sumi \{\Lambda_{ig}- \widetilde \Lambda_{ig}\}X_{ig,j}\right)\Big|\\
\le& \Big|\frac{1}{G} \sumg \left(\sumi X_{ig,j}^2\right)\left(\sumi \{\Lambda_{ig}- \widetilde \Lambda_{ig}\}^2\right)\Big| + 2 \Big|\frac{1}{G} \sumg \left(\sumi \{Y_{ig} - \Lambda_{ig}\}X_{ig,j}\right) \left(\sumi \{\Lambda_{ig}- \widetilde \Lambda_{ig}\}X_{ig,j}\right)\Big|\\
\lesssim & \max_{1\le g \le G}\|U_{g}\|_\infty^2\Big|\frac{1}{G} \sumg \sumi \{(f_{ig}X_{ig}'(\beta^0- \widetilde \beta) \}^2\Big|
+  \max_{1\le g \le G}\|U_{g}\|_\infty\Big|\frac{1}{G} \sumg \sumi \{(f_{ig}X_{ig}'(\beta^0- \widetilde \beta) \}^2\Big|^{1/2}\\
\lesssim&  \frac{M_{G,2} s\log a_G}{G^{1-1/q}} + \frac{ M_{G,2} (s \log a_G)^{1/2}}{G^{1/2-1/2q}}=o(1),
\end{align*}}
where the second inequality follows Cauchy-Schwarz inequality and the third follows Assumption \ref{a:covariates} (7) and the last holds
following $|\Lambda(t + \Delta t)-\Lambda(t)|\lesssim \Lambda'(t)\Delta t$ for $|\Delta t|\le 1$ as in inequality (I6) in \cite{BCCW18}, the
rates from $m=1$, Assumption \ref{a:covariates} (8), and the fact $\Lambda'$ is Lipschitz. This verifies Assumption \ref{a:M-est_lasso} (2).

We now apply Lemma \ref{lemma:M-est_WL} and obtain that
with some $c'>c$ and $\gamma=\gamma_G\in[1/G,1/\log G]$, one has
\begin{align*}
\Pr_\Pr\left( \frac{\lambda}{G} \ge c    \Big\|\hat \Psi^{-1}\frac{1}{G} \sumg S_g\Big\|_\infty\right)\ge 1-\gamma -o(\gamma).
\end{align*}

Assumption \ref{a:M-est_lasso} (1) is trivial since we have $\hat M(y,x,\beta)=M(y,x,\beta)$ in this case. Assumption \ref{a:M-est_lasso} (3) holds for any $A$ and $C_G\lesssim (s\log a_G/G)^{1/2}$ following Lemma \ref{lemma:minoration}.

Now, let us define 
\begin{align*}
\bar q_A=\inf_{\delta \in A} \frac{(\frac{1}{G} \sumg\sumi w_{ig}|X_{ig}'\delta|^2)^{3/2}}{\frac{1}{G} \sumg\sumi w_{ig}|X_{ig}'\delta|^3}.
\end{align*}
To apply Lemma \ref{lemma:M-est_lasso_rates}, we need to verify the condition 
\begin{align*}
\bar q_A=\bar q_{A_1} \wedge \bar q_{A_2} \ge (L+\frac{1}{c})\|\hat \Psi_0\|_\infty \frac{\lambda \sqrt{s}}{G \bar \kappa_{2 \widetilde c}} + 6\widetilde c C_G
\end{align*}
for $A=\Delta_{2\widetilde c}\cup \{\delta\in \Real^p : \|\delta\|_1  \le \frac{3G}{\lambda} \frac{c \|\hat \Psi_0^{-1}\|_\infty}{\ell c-1} C_G \|\sqrt{w_{ig}  }X_{ig}'\delta\|_G \}=A_1 \cup A_2$. 

Note that under Assumptions \ref{a:covariates} (6)(7)(8) and \ref{a:sparse_eigenvalues}, we have
\begin{align*}
\bar q_{A_1} \ge& \inf_{\delta \in A_1 } \frac{(\frac{1}{G} \sumg\sumi w_{ig}|X_{ig}'\delta|^2)^{1/2}}{\max_{1\le g \le G}\|U_g\|_\infty \|\delta\|_1}
\gtrsim_\Pr \inf_{\delta \in A_1 } \frac{(\frac{1}{G} \sumg\sumi w_{ig}|X_{ig}'\delta|^2)^{1/2}}{ G^{1/2q} M_{G,2} \|\delta\|_1}\\
\ge&  \inf_{\delta \in A_1 } \frac{(\frac{1}{G} \sumg\sumi w_{ig}|X_{ig}'\delta|^2)^{1/2}}{ G^{1/2q} M_{G,2}  (1+2 \widetilde c)\sqrt{s}\|\delta_T\|_2}
\gtrsim  \frac{\bar \kappa_{2 \widetilde c}}{ G^{1/2q} M_{G,2}  (1+2 \widetilde c)\sqrt{s}}\\
\gtrsim& \frac{1}{\check \delta^{1/2} G^{1/4}} \gtrsim \sqrt{\frac{s\log a_G}{\check \delta G}}.
\end{align*}
Next, using Assumptions \ref{a:covariates} (7)(8), since $\lambda\lesssim \sqrt{G\log a_G}$ and $C_G\lesssim (s\log a_G/G)^{1/2}$, some calculations yield
\begin{align*}
\bar q_{A_2} \ge& \inf_{\delta \in A_2 } \frac{(\frac{1}{G} \sumg\sumi w_{ig}|X_{ig}'\delta|^2)^{1/2}}{\max_{1\le g \le G}\|U_g\|_\infty \|\delta\|_1}
\gtrsim_\Pr \inf_{\delta \in A_2 } \frac{(\frac{1}{G} \sumg\sumi w_{ig}|X_{ig}'\delta|^2)^{1/2}}{ G^{1/2q} M_{G,2} \|\delta\|_1}\\
\ge & \frac{\lambda}{3G C_G} \frac{\ell c -1}{c} \frac{\|\hat \Psi_0^{-1}\|_\infty^{-1}}{G^{1/2q} M_{G,2}}\gtrsim_\Pr \frac{\lambda}{C_G G^{1+1/2q} M_{G,2}}\\
\gtrsim&_\Pr \frac{1}{G^{1/2q} M_{G,2} \sqrt{s}} \ge \frac{1}{\check \delta^{1/2} G^{1/4}} \gtrsim \sqrt{\frac{s\log a_G}{\check\delta G}}.
\end{align*} 
Furthermore, we have
\begin{align*}
(L+\frac{1}{c})\|\hat \Psi_0\|_\infty \frac{\lambda \sqrt{s}}{G \bar \kappa_{2 \widetilde c}} + 6\widetilde c C_G\lesssim \sqrt{\frac{s\log a_G}{G}}
\end{align*}
since $\|\hat \Psi_0\|_\infty\lesssim 1$ with probability $1-C(\log G)^{-1}$. So all conditions required by Lemma \ref{lemma:M-est_lasso_rates} are satisfied. An application of the Lemma leads to
\begin{align*}
\|\sqrt{w_{ig}}X_{ig}'(\hat \beta -\beta^0)\|_G \lesssim \sqrt{\frac{s\log a_G}{G}} \text{ and } \|\hat \beta -\beta^0\|_1 \lesssim \sqrt{\frac{s^2\log a_G}{G}}.
\end{align*}

Now, to apply Lemma \ref{lemma:M-est_lasso_sparsity}, we need to verify condition (\ref{eq:M-est_L6}). First, using Assumption \ref{a:covariates} (7)(8), we have
\begin{align*}
\max_{1\le g \le G}\max_{i\in[n_g]} |X_{ig}'(\hat \beta - \beta^0)| \lesssim&_\Pr G^{1/2q} M_{G,2} \|\hat \beta - \beta^0\|_1
\lesssim \sqrt{\frac{M^2_{G,2} s^2 \log a_G}{G^{1-1/q}}} \lesssim \check \delta_G=o(1).
\end{align*}
Also, following equation (I.6) of \cite{BCCW18}, one has $|\Lambda(t+\Delta t)-\Lambda(t)|\lesssim \Lambda'(t) |\Delta t|$ uniformly over $t$ and $\Delta t$ with $|\Delta t|\le 1$.
It holds uniformly over $ig$ that 
\begin{align*}
[\partial_\beta \hat M(Y_{ig},X_{ig},\hat \beta) -\partial_\beta \hat M(Y_{ig},X_{ig},\beta^0) ]\}'\delta\lesssim& |\Lambda(X_{ig}'\hat \beta)-\Lambda(X_{ig}' \beta^0)|\cdot|X_{ig}'\delta|\\
\lesssim & \Lambda'(X'_{ig}\beta^0) \cdot |X_{ig}'(\hat \beta - \beta^0)|\cdot|X_{ig}'\delta|.
\end{align*}
Since $|\Lambda'(X_{ig}'\beta^0)|\lesssim w_{ig}\le \sqrt{w_{ig}}$, with probability at least $1-C(\log G)^{-1}$, we have
\begin{align*}
&\Big|\frac{1}{G} \sumg\sumi[\partial_\beta \hat M(Y_{ig},X_{ig},\hat \beta) -\partial_\beta \hat M(Y_{ig},X_{ig},\beta^0) ]\}'\delta\Big|\\
\le & C\|\sqrt{w_{ig}}X'_{ig} (\hat \beta - \beta^0)\|_G \| X_{ig}'\delta \|_G \le L_G  \| X_{ig}'\delta \|_G
\end{align*}
for some $L_G \lesssim (s\log a_G/G)^{1/2}$. Thus condition (\ref{eq:M-est_L6}) is satisfied. In addition, Lemma \ref{lemma:M-est_lasso_sparsity} implies $\|\hat \beta\|_0\lesssim s$.

Finally, to establish the convergence rates for $\widetilde \beta$, we apply Lemma \ref{lemma:M-est_post_lasso_rates}. 
We verify condition (\ref{eq:M-est_L7}) on $\bar q_A$ for $A=\{\delta\in \Real^p:\|\delta\|_0\le Cs\}$ for a constant $\hat s+ s \le Cs$ with probability $1- o(1)$. Note it holds that
\begin{align*}
\bar q_A =&
\inf_{\delta \in A}\frac{(\frac{1}{G} \sumg\sumi w_{ig}|X_{ig}'\delta|^2)^{1/2}}{\max_{1\le g \le G} \|U_{g}\|_\infty \|\delta\|_1} \\
\ge& \inf_{\|\delta\|_0 \le Cs}\frac{(\frac{1}{G} \sumg\sumi w_{ig}|U_{g}'\delta|^2)^{1/2}}{ \max_{1\le g \le G} \|U_{g}\|_\infty \sqrt{Cs}\|\delta\|_2} 
\gtrsim_\Pr\inf_{\|\delta\|_0 \le Cs}\frac{\sqrt{\semax{Cs}}}{ \sqrt{s}G^{1/2q} M_{G,2}} \gtrsim \frac{\log^{1/4} a_G}{\check \delta_G G^{1/4}} 
\end{align*}
under Assumptions \ref{a:covariates} (7)(8) and \ref{a:sparse_eigenvalues}. On the other hand, it follows from (\ref{eq:M-est_L8}) that with probability $1-C(\log G)^{-1}$,
\begin{align*}
\frac{1}{G} \sumg\sumi\partial_\beta \hat M(Y_{ig},X_{ig},\widetilde\beta)- \frac{1}{G} \sumg\sumi\partial_\beta \hat M(Y_{ig},X_{ig},\beta^0)\lesssim \frac{s \log a_G}{G} 
\end{align*}
since $\lambda/G \lesssim (\log a_G/G)^{1/2}$, $\|\hat \beta - \beta^0\|_1\lesssim (s\log a_G/G)^{1/2}$ and $\|\hat \Psi_0\|_\infty \lesssim 1$
with probability $1-C(\log G)^{-1}$. Also $C_G\lesssim (s^2\log a_G/G)^{1/2}$,
\begin{align*}
\Big\|\frac{1}{G} \sumg S_g\Big\|_\infty \le \Big\|\hat \Psi_0\Big\|_\infty \Big\| \hat \Psi_0^{-1} \frac{1}{G} \sumg S_g\Big\|_\infty \lesssim \frac{\lambda}{G}
\end{align*}
with probability $1-C(\log G)^{-1}$. So right-hand side of (\ref{eq:M-est_L7}) is bounded by $(s\log a_G/G)^{1/2}$. So by  Lemma \ref{lemma:M-est_post_lasso_rates}, we have the desired results.

Finally, since $s\ge 1$, we can without loss of generality assume the $k$-th coordinate is always in the support of $\hat\beta$ and this does not affect the rate of convergence in post-lasso (see Comment D.1. of \cite{BCK15}). Also, since all $k\in[p]$ share the same regularized event, the convergence rate holds uniformly for all $k\in[p]$.
\end{proof}
\subsection{Proof for Theorem \ref{thm:REGNPL_rates}}\label{sec:proof for thm:REGNPL_rates}
The proof follows analogously of the Proof of Theorem 4.2 in \cite{BCCW18} with modifications to account for cluster sampling. The major difference lies in the verification of our Assumption \ref{a:M-est_lasso} (2).
\begin{proof}
Let $\bar r^j_{ig}=X^j_{ig} (\gamma^j - \bar \gamma^j)$, $w_{ig}=\hat f_{ig}^2$ and
\begin{align}
M (D^j_{ig},X^j_{ig},\gamma)=& f_{ig}^2 (D_{ig}^j - X^j_{ig}\gamma -\bar r^j_{ig} )^2,\nonumber\\
\widehat M (D^j_{ig},X^j_{ig},\gamma)=& \hat f_{ig}^2 (D_{ig}^j - X^j_{ig}\gamma )^2.  \label{eq:REGNPL_loss_func}
\end{align}
Then, the sparse approximation $\bar\gamma^j$ can be identified by
\begin{align*}
\bar\gamma^j=\underset{\gamma \in \Real^{p-1}}{\text{argmin}}\: \EP\left[\frac{1}{G}\sum_{g=1}^{G} \sumi M (D^j_{ig},X^j_{ig},\gamma)\right].
\end{align*}
We will first show that the regularized events (\ref{eq:M-est_regularized_event_L5}) holds uniformly over $j\in [p]$. Subsequently, we apply Lemmas \ref{lemma:M-est_lasso_rates}, \ref{lemma:M-est_lasso_sparsity} and \ref{lemma:M-est_post_lasso_rates} to bound different norms of $(\widetilde \gamma^j -\bar \gamma^j)$. Then bounds for $(\widehat\gamma^j - \gamma^j)$ follow from Assumption \ref{a:sparsity}.

 
First, we verify Assumption \ref{a:M-est_WL}. For Assumption \ref{a:M-est_WL} (1), note  that
\begin{align*}
S_g^j =& \sumi\partial_\gamma M(D^j_{ig},X^j_{ig},\bar\gamma^j)=2\sumi  f^2_{ig} (D^j_{ig} - X^j_{ig}\bar\gamma^j - \bar r^j_{ig}) (X^j_{ig})'
= 2 \sumi f^2_{ig} Z_{ig}^j (X^j_{ig})'.
\end{align*}
where $a^j=\beta^0$.
Since $\Phi^{-1} (1-\gamma/2p)\le \sqrt{\log (1/t)}$ for all $t\in(0,1/2)$, along with Assumption \ref{a:covariates} (3), we have
\begin{align*}
\left\{ \EP\left[\frac{1}{G}\sum_{g=1}^{G} |S^j_{gk}|^3\right]\right\}^{1/3} \Phi^{-1} (1-\gamma/2p) \lesssim \left\{ \EP\left[\frac{1}{G}\sum_{g=1}^{G} |V^j_{g} U_{gk}|^3\right]\right\}^{1/3} \log^{1/2} a_G\le \check \delta_G G^{1/6}
\end{align*}
uniformly in $j\in[p]$ and $k\in[p]\setminus \{j\}$. This shows Assumption \ref{a:M-est_WL} (1).

To show Assumption \ref{a:M-est_WL} (2), notice that Assumption \ref{a:covariates} (2) implies $ \EP\left[ G^{-1}\sum_{g=1}^{G} |S^j_{gk}|^2\right]\gtrsim 1$ and 
\begin{align*}
 \EP\left[\frac{1}{G}\sum_{g=1}^{G} (S^j_{gk})^2\right]\le  \EP\left[\frac{1}{G}\sum_{g=1}^{G} (V^j_g U_{gk})^2\right]\le  \EP\left[\frac{1}{G}\sum_{g=1}^{G} (|V^j_g|^4  + |U_{gk}|^4)\right]\lesssim 1
\end{align*}
uniformly over $j\in[p]$ and $k\in[p]\setminus \{j\}$ by Assumption \ref{a:covariates} (4).

The convexity requirement is trivially satisfied.
To show Assumptions \ref{a:M-est_lasso} (1), we first claim that with probability $1-C(\log G)^{-1}$,
\begin{align}
\max_{j\in [p]}\|(\hat f^2_{ig} - f^2_{ig})  Z^j_{ig}/\hat f_{ig}\|_G\lesssim (s\log a_G/G)^{1/2}. \label{eq:M-est_f_approxi_rate}
\end{align}
Now, since by Theorem \ref{thm:LOG_rates} and Assumption \ref{a:covariates} (7)(8), one has
\begin{align*}
\max_{i,g}|X'_{ig} (\hat\beta - \beta^0)|\le \max_{i,g}\|X_{ig}\|_\infty \|\hat\beta - \beta^0\|_1\lesssim_\Pr G^{1/2q} M_{G,2} (s^2\log a_G/G)^{1/2} \le\check \delta_G=o(1)
\end{align*} 
with probability $1-C(\log G)^{-1}$, we then have with probability $1-C(\log G)^{-1}$
\begin{align}
|\hat f_{ig}^2 - f^2_{ig}|\le |\Lambda(X_{ig}'\widetilde \beta) -\Lambda(X_{ig}'\beta^0)|\lesssim \Lambda'(X_{ig}'\beta^0)|X'_{ig} (\hat\beta - \beta^0)|\le f_{ig}^2/2\le 1 \label{eq:f2_approx}
\end{align}
uniformly over all $i,g$. Note we have used the fact that for $|\widetilde t - t |\le 1$, $|\Lambda( t )-\Lambda(\widetilde{t})|\lesssim \Lambda'(t)|t-\widetilde{t}|$. Also, some calculations give that for $G$ large enough, let $\widetilde t_{ig}=X_{ig}\widetilde \beta$, $t_{ig}=X_{ig}\beta^0$, then it holds that
\begin{align*}
|\hat f^2_{ig} - f^2_{ig}|=& |\Lambda(\widetilde t_{ig})-\Lambda^2(\widetilde t_{ig}) -\Lambda( t_{ig})+\Lambda^2( t_{ig})|\\
\le& |\Lambda(\widetilde t_{ig})- \Lambda(t_{ig})|+  |\Lambda^2(\widetilde t_{ig})- \Lambda^2(t_{ig})| \\
\le& |\Lambda(\widetilde t_{ig})- \Lambda(t_{ig})| + \Lambda(\widetilde t_{ig})|\Lambda(\widetilde t_{ig})- \Lambda(t_{ig})|+ \Lambda(t_{ig}) |\Lambda(\widetilde t_{ig})- \Lambda(t_{ig})|\\
\lesssim& |\Lambda(\widetilde t_{ig})- \Lambda(t_{ig})|
\lesssim\Lambda'(\widetilde t_{ig})|\widetilde t_{ig}- t_{ig}|=f^2_{ig}|\widetilde t_{ig}- t_{ig}|.
\end{align*}
Thus, with probability at least $1-C(\log G)^{-1}$, one has
\begin{align*}
\max_{j\in [p]} \frac{1}{G} \sumg \sumi (\hat f^2_{ig} - f^2_{ig})^2  (Z^j_{ig}/\hat f_{ig})^2\lesssim& \max_{j\in [p]}  \frac{1}{G} \sumg \sumi (\Lambda'(X_{ig}'\widetilde\beta))^2|X'_{ig} (\beta^0 - \widetilde  \beta)|^2  (Z^j_{ig}/\hat f_{ig})^2\\
\lesssim& \max_{j\in [p]}  \frac{1}{G} \sumg \sumi |X'_{ig} (\beta^0 - \widetilde  \beta)|^2  (Z^j_{ig})^2 \\
\le&  \max_{j\in [p]} \sup_{\|\delta\|_0\le Cs, \|\delta\|_2=1} \frac{s\log a_G}{G}\frac{1}{G} \sumg \sumi (X_{ig}'\delta)^2  (Z^j_{ig})^2\\
\le& \frac{s\log a_G}{G} O(1),
\end{align*}
where the last inequality follows from Assumption \ref{a:sparse_eigenvalues}.
Therefore, with probability $1-C(\log G)^{-1}$, we have
\begin{align*}
\max_{j\in [p]}\|(\hat f^2_{ig} - f^2_{ig})  Z^j_{ig}/\hat f_{ig}\|_G \lesssim \max_{j\in [p]}\|(\hat f^2_{ig} - f^2_{ig})  Z^j_{ig}/ f_{ig}\|_G.
\end{align*}
Recall that $\bar r^j_{ig}=X^j_{ig} (\gamma^j - \bar \gamma^j)$. Assumption \ref{a:M-est_lasso} (1) can be examined by noting that it holds uniformly in $j\in[p]$ that
\begin{align*}
&\left|\left[\frac{1}{G} \sumg\sumi\left(\partial_\gamma  \hat M(Y_{ig},X_{ig},\bar\gamma^j) - \partial_\gamma  M(Y_{ig},X_{ig},\bar\gamma^j)\right)\right]'\delta\right| \\
\le& \Big|\frac{2}{G} \sumg\sumi (\hat f_{ig}^2 \bar r_{ig}^j  +(\hat f^2_{ig} - f^2_{ig} )Z_{ig}^j)X^j_{ig}\delta \Big|\\
\le& 2\left(\|\hat f_{ig}\bar r^j_{ig}\|_G + \|(\hat f_{ig}^2 -f_{ig}^2)Z^j_{ig}/\hat f_{ig}\|_G \right)\|\sqrt{w_{ig} } X^j_{ig}\delta\|_G\\
\le& C_G \|\sqrt{w_{ig}} X_{ig}' \delta\|_G.
\end{align*}
We now verify the condition $C_G\lesssim (s\log a_G/G)^{1/2}$ in Assumption \ref{a:M-est_lasso} (1). Notice that one has
\begin{align}
\|\hat f_{ig}\bar r^j_{ig}\|_G\le \|\bar r^j_{ig}\|_G\lesssim (s\log a_G/G)^{1/2}\label{eq:M-est_r_rate}
\end{align}
with probability at least $1-C(\log G)^{-1}$ following the same arguments as in Lemma J1 of \cite{BCCW18} under Assumption \ref{a:parameters}, \ref{a:sparsity}, \ref{a:covariates}, \ref{a:sparse_eigenvalues}. To see this, let
\begin{align*}
&\calG=\{W_{g}\mapsto \sum_{i=1}^{\bar n} X^j_{ig} (\gamma^j - \bar \gamma^j):j\in [p]\},\\
&\calG_{ijT}=\{W_g\mapsto  X^j_{ig} (\gamma^j -  \gamma^j_T):j\in [p]\}.
\end{align*}
Note $\bar \gamma^j = \gamma^j_T$ for some $T$ by Assumption \ref{a:parameters}. Thus one has $\calG\subset\cup_{j\in[p],T\le s}\sum_{i=1}^{\bar n}\calG_{ijT}$. So for $\calG^2$, we have an envelope $\mathfrak g(\textbf{w})=\|u(\textbf{w})\|_\infty \max_{j\in[p]}\|\bar\gamma^j - \gamma^j\|_1^2$ with
\begin{align*}
\left\{ \EP\left[\frac{1}{G}\sum_{g=1}^{G} \max_{1\le g\le G} \mathfrak g^2(W_g) \right]\right\}^{1/2} \lesssim \frac{s^2 M_{G,2}\log a_G}{G^{1-1/q}}.
\end{align*}
In addition, for all finite discrete measures $Q$ and $0<\epsilon\le 1$, it holds that
\begin{align*}
\sup_{Q}\log N(\epsilon\|G\|_{Q,2},\calG^2,\|\cdot\|_{Q,2})\lesssim s \log(a_G/\epsilon). 
\end{align*}
Thus by applying Lemma \ref{lemma:maximal_ineq}, one has with probability at least $1-C(\log G)^{-1}$,
\begin{align*}
\max_{j\in [p]}\left|\frac{1}{G} \sumg\sumi  (\bar r^2_{ig} - \EP[\bar r^2_{ig}] )\right|\lesssim \frac{s\log a_G}{G}.
\end{align*}
Finally, $ \EP\left[G^{-1}\sum_{g=1}^{G} \sumi \bar r^2_{ig}\right]\lesssim \sup_{\|\xi\|_2=1} \EP\left[\frac{1}{G}\sum_{g=1}^{G} (U_g'\xi)^2\|\gamma^j-\bar \gamma^j\|_2^2\right]\lesssim s\log a_G/G$ by Assumption \ref{a:parameters}. This shows (\ref{eq:M-est_r_rate}) and thus Assumption \ref{a:M-est_lasso} (1).

Note that Assumption \ref{a:M-est_lasso} (3) holds with $\check \Delta_G=0$ and $\bar q_A=\infty$ for any $A$ since
\begin{align*}
&\frac{1}{G} \sumg \sumi \widehat M_j(D^j_{ig},X^j_{ig},\bar\gamma^j+ \delta) - \frac{1}{G} \sumg\sumi \widehat M(Y_{ig},X_{ig},\bar\gamma^j) \\
& -2\frac{1}{G} \sumg \sumi \hat f_{ig}^2 (D^j_{ig}-X^j \bar\gamma^j)X^j_{ig} \delta = \frac{1}{G} \sumg\sumi (\hat f_{ig} X^j_{ig}\delta)^2 
\end{align*}
and $\frac{1}{G} \sumg \sumi (\hat f_{ig} X^j_{ig}\delta)^2 = \|\sqrt{w_{ig}}X_{ig}^j\delta\|^2_G$.

To check Assumption \ref{a:M-est_lasso} (2), note that under Assumption \ref{a:covariates} (5)(6)(7)(8), one has
\begin{align*}
&\EP\left[\max_{1\le g \le G} \max_{j \in[p]}\Big\|\sumi f^2_{ig}Z^j_{ig}X^j_{ig}\Big\|_\infty^2\right]\\
\le&
 \EP\left[\max_{1\le g \le G} \max_{j \in[p]} |V^j_g|^2 \|U_g\|_\infty^2 \right]\\
\lesssim& G^{2/q} (M_{G,1}+M_{G,2}).
\end{align*}
Thus, an application of Lemma \ref{lemma:maximal_ineq} gives 
\begin{align*}
\max_{k \in [p]}\max_{j \in [p]\setminus \{k\}}\left|\frac{1}{G} \sumg\left[\left(\sumi S_{gk}^j\right)^2- \EP\left(\sumi S_{gk}^j\right)^2\right]\right| 
\lesssim&_\Pr\:
G^{-(1/2 - 1/q)} (M_{G,1}^2 + M_{G,2}^2) \log a_G\\
\le& \:\check \delta_G \log a_G = o(1). 
\end{align*}
where the last equality follows the rate assumption in statement of the Theorem.
Therefore, since $l_{j0k}=\{\frac{1}{G} \sumg(\sumi S^j_{gk})^2\}^{1/2}$, we have $1\lesssim \hat \Psi^\gamma_{j0k}\lesssim 1$
with probability at least $1-C(\log G)^{-1}$ uniformly over $j\in[p]$ and $k\in[p]\setminus \{j\}$. 

For $m=0$,
with probability $1-C(\log G)^{-1}$, we have
\begin{align*}
\hat l_{jk,0}\gtrsim 2\left\{\frac{1}{G} \sumg\left(\sumi \hat f_{ig}^2 Z^j_{ig} X^j_{ig,k}\right)^2\right\}^{1/2} 
\gtrsim 
 2\left\{\frac{1}{G} \sumg\left(\sumi  f_{ig}^2 Z^j_{ig} X^j_{ig,k}\right)^2\right\}^{1/2} \gtrsim 1 
\end{align*}
uniformly over $j\in[p]$ and $k\in[p]\setminus \{j\}$.
This follows from the fact that $|\hat f_{ig}^2-f_{ig}^2|\le f_{ig}^2$ with probability $1-C(\log G)^{-1}$. To obtain an upperbound, note under Assumption \ref{a:covariates} (4)(7) and the fact that $\hat f_{ig}\le 1$, one has
\begin{align*}
\hat l_{jk,0} \lesssim  2\max_{ g \in[G]}\max_{i \in [n_g]}|\hat f_{ig}X_{ig,k}|\left\{\frac{1}{G} \sumg\left(\sumi \hat f_{ig} D^j_{ig}\right)^2\right\}^{1/2}\lesssim_\Pr G^{1/2q} M_{G,2}. 
\end{align*}
Thus for $m=0$, Assumption \ref{a:M-est_lasso} (2) holds with $L\lesssim G^{1/2q} M_{G,2} \log^{1/2} a_G$ and $\ell \gtrsim 1$.
For $m\ge 1$, suppose that the statement holds for $m=\bar m-1$, we can complete the proof and obtain the bound
\begin{align*}
\max_{j\in[p]}\|\hat f_{ig} (\widetilde \gamma^j - \bar \gamma^j)\|_G\lesssim (L+1)(s^2 \log a_G/G)^{1/2}
\end{align*} 
for $L\lesssim G^{1/2q} M_{G,2} \log^{1/2} a_G$. In addition, under Assumption \ref{a:sparsity} and \ref{a:covariates} (7), it holds uniformly over $j\in[p]$ that
\begin{align*}
\|\hat f_{ig} X^j_{ig} (\bar \gamma^j -\gamma^j)\|_G \le & \max_{1 \le g \le G} \|U_g\|_\infty \cdot \|\bar \gamma^j - \gamma^j\|_1\\
\lesssim&_\Pr G^{1/2q} M_{G,2} (s^2 \log a_G/G)^{1/2}.
\end{align*}
Thus by the triangle inequality, we have
\begin{align*}
\max_{j\in[p]}\|\hat f_{ig} X^j_{ig} (\widetilde \gamma^j -\gamma^j)\|_G \lesssim (L+1)  (s^2 \log a_G/G)^{1/2}
\end{align*}
for $L\lesssim G^{1/2q} M_{G,2} \log^{1/2} a_G$. 
Using the fact that for positive $a,b$, $|\sqrt{a}-\sqrt{b}|\le \sqrt{|a-b|}$, we have, for $m=\bar m$, it holds uniformly over $j\in[p]$, $k\in[p-1]$ that
\begin{align*}
|\hat l_{jk,m}- l_{j0k}|=&2\left|\left\{\frac{1}{G} \sumg\left(\sumi \hat f^2_{ig} (D^j_{ig}-X^j_{ig}\widetilde\gamma^j)X^j_{ig,k} \right)^2 \right\}^{1/2}
- 
 \left\{\frac{1}{G} \sumg\left(\sumi f^2_{ig} (D^j_{ig}-X^j_{ig}\gamma^j) X^j_{ig,k}\right)^2 \right\}^{1/2} \right|\\
\le&  
2\left\{\left|\frac{1}{G} \sumg\left(\sumi \hat f^2_{ig}  (D^j_{ig}-X^j_{ig}\widetilde\gamma^j) X^j_{ig,k} \right)^2 
-
 \frac{1}{G} \sumg\left(\sumi f^2_{ig} (D^j_{ig}-X^j_{ig}\gamma^j) X^j_{ig,k} \right)^2 \right| \right\}^{1/2}\\
=& 
2\left\{\left|\frac{1}{G} \sumg\left(\sumi \hat f^2_{ig} X^j_{ig,k} \hat Z^j_{ig}\right)^2 - \frac{1}{G} \sumg\left(\sumi f^2_{ig} X^j_{ig,k} Z_{ig}^j \right)^2 \right| \right\}^{1/2},
\end{align*}
where $\hat Z^j_{ig}=D^j_{ig}-X^j_{ig}\widetilde \gamma^j$.
To bound the right-hand side, note by adding and subtracting terms and the triangle inequality, 
\begin{align*}
&\Big|\frac{1}{G} \sumg\left(\sumi \hat f^2_{ig} X^j_{ig,k} \hat Z^j_{ig}\right)^2 - \frac{1}{G} \sumg\left(\sumi f^2_{ig} X^j_{ig,k} Z_{ig}^j \right)^2 \Big| \\
\le 
&\Big|\frac{1}{G} \sumg\left(\sumi  \hat f^2_{ig} X^j_{ig,k} X^j_{ig} (\gamma^j - \widetilde \gamma^j)\right)^2\Big| + 2\Big|\frac{1}{G} \sumg\left(\sumi \hat f^2_{ig} X^j_{ig,k} X^j_{ig} (\gamma^j - \widetilde \gamma^j)\right) \left(\sumi  \hat f^2_{ig} X^j_{ig,k}  Z^j_{ig}\right)\Big|\\
&+ \Big|\frac{1}{G} \sumg\left(\sumi (\hat f^2_{ig} - f^2_{ig}) X^j_{ig,k}  Z^j_{ig}\right)^2  \Big| + 2 \Big|\frac{1}{G} \sumg\left(\sumi (\hat f^2_{ig} - f^2_{ig}) X^j_{ig,k}  Z^j_{ig}\right) \left(\sumi  f^2_{ig} X^j_{ig,k}  Z^j_{ig}\right) \Big|
=o(1)
\end{align*}
uniformly over $j\in[p]$, $k\in[p-1]$
with probability at least $1-C(\log G)^{-1}$.
The inequality holds following the Cauchy-Schwarz inequality. Then under Assumption \ref{a:covariates} (7)(8), with probability at least $1-C(\log G)^{-1}$, one has
\begin{align*}
\frac{1}{G} \sumg\left(\sumi (\hat f^2_{ig} - f^2_{ig}) X^j_{ig,k}  Z^j_{ig}\right)^2 \le& \frac{1}{G} \sumg\left(\sumi (\hat f^2_{ig} - f^2_{ig})^2 ( Z^j_{ig})^2/\hat f^2_{ig}\right) \left(\sumi \hat f^2_{ig} ( X^j_{ig,k})^2 \right)\\
\le&  \max_{1\le g \le G}\|U_{gk}\|_\infty^2 \|(\hat f^2_{ig} - f^2_{ig})  Z^j_{ig}/\hat f_{ig}\|_G^2\le \frac{ M^2_{G,2} s\log a_G}{G^{1-1/q}}=o(1) 
\end{align*}
uniformly over $j\in[p]$, $k\in[p-1]$. Here, we have used $\|(\hat f^2_{ig} - f^2_{ig})  Z^j_{ig}/\hat f_{ig}\|_G\lesssim (s\log a_G/G)^{1/2}$ with probability at least $1-C(\log G)^{-1}$ by equation (\ref{eq:M-est_f_approxi_rate}). Similar arguments show that by Assumption \ref{a:covariates} (8), with probability at least $1-C(\log G)^{-1}$, we have
\begin{align*}
\frac{1}{G} \sumg\left(\sumi  \hat f^2_{ig} X^j_{ig,k} X^j_{ig} (\gamma^j - \widetilde \gamma^j)\right)^2\le&  \frac{1}{G} \sumg\left(\sumi  (\gamma^j - \widetilde \gamma^j)' \hat f^2_{ig} (X^j_{ig})'X^j_{ig} (\gamma^j - \widetilde \gamma^j)\right) \left(\sumi  \hat f^2_{ig} (X^j_{ig,k})^2\right)\\
\le& \max_{1\le g \le G}\|U_{gk}\|_\infty^2 \|\hat f_{ig } X_{ig}^j(\widetilde \gamma^j-\gamma^j)\|_G^2\\
\le& L\frac{ M_{G,2} s\log a_G}{G^{1-1/q}} \le \frac{ M_{G,2}^3 s\log^{3/2} a_G}{G^{1-3/2q}}=o(1)
\end{align*}
uniformly over $j\in[p]$, $k\in[p-1]$. Furthermore, by the Cauchy-Schwarz inequality, we have\small
\begin{align*}
&\frac{1}{G} \sumg\left(\sumi (\hat f^2_{ig} - f^2_{ig}) X^j_{ig,k}  Z^j_{ig}\right) \left(\sumi  f^2_{ig} X^j_{ig,k}  Z^j_{ig}\right) \\
\le& 
\left\{
\frac{1}{G} \sumg\left(\sumi (\hat f^2_{ig} - f^2_{ig}) X^j_{ig,k}  Z^j_{ig}\right)^2 \frac{1}{G} \sumg\left(\sumi  f^2_{ig} X^j_{ig,k}  Z^j_{ig}\right)^2
\right\}^{1/2}
\end{align*}
as well as
\begin{align*}
&\frac{1}{G} \sumg\left(\sumi \hat f^2_{ig} X^j_{ig,k} X^j_{ig} (\gamma^j - \widetilde \gamma^j)\right) \left(\sumi  \hat f^2_{ig} X^j_{ig,k}  Z^j_{ig}\right)\\
\le & 
\left\{
\frac{1}{G} \sumg\left(\sumi \hat f^2_{ig} X^j_{ig,k} X^j_{ig} (\gamma^j - \widetilde \gamma^j)\right)^2\frac{1}{G} \sumg \left(\sumi  \hat f^2_{ig} X^j_{ig,k}  Z^j_{ig}\right)^2
\right\}^{1/2}.
\end{align*}
\normalsize
From the preceding results, it suffices to show the claim uniformly over $j\in[p]$ and $k\in[p-1]$,
\begin{align*}
\frac{ M_{G,2}^3 s\log^{3/2} a_G}{G^{1-3/2q}}\frac{1}{G} \sumg \left(\sumi  \hat f^2_{ig} X^j_{ig,k}  Z^j_{ig}\right)^2=o_p(1).
\end{align*}
Under Assumption \ref{a:covariates} (4)(7), since $\hat f_{ig}^2\le 1$, the Cauchy-Schwarz inequality gives
\begin{align*}
\max_{j\in[p]}\max_{k\in [p]\setminus \{j\}}\frac{1}{G} \sumg \left(\sumi  \hat f^2_{ig} X^j_{ig,k}  Z^j_{ig}\right)^2\lesssim& \max_{k\in[p]}\frac{1}{G} \sumg U_{gk}^4 + \max_{k\in[p]}\frac{1}{G} \sumg(V^j_g)^4\lesssim_\Pr 1
\end{align*}
by Assumption \ref{a:covariates} (4). The same bound holds even if $\hat f_{ig}$ is used in place of $f_{ig}$. The claim then follows from Assumption \ref{a:covariates} (9). Thus for $m=\bar m$, the result holds for some $L,\ell, \check \Delta_G$ with $L \lesssim 1$, $\ell \gtrsim 1$ and $\check \Delta_G=o(1)$. This verifies Assumption \ref{a:M-est_lasso} (2).

Note that $\|\hat\Psi_0\|_\infty\lesssim 1$ and $\|\hat\Psi_0^{-1}\|_\infty\lesssim 1$ with probability $1-C(\log G)^{-1}$ following the preceding arguments. By Lemma \ref{lemma:M-est_WL}, (\ref{eq:M-est_regularized_event_L5}) holds with probability $1-C(\log G)^{-1}$. Furthermore, following Assumption \ref{a:sparse_eigenvalues} and the fact $|\hat f_{ig}^2 -f^2_{ig}|\le f_{ig}^2/2$ with probability $1-C(\log G)^{-1}$, we have, for some $\ell_G \to \infty$, it holds that  
\begin{align*}
1\lesssim \min_{\|\delta\|_0 \le \ell_G s} \frac{\|f_{ig} X_{ig}'\delta\|_G^2}{\|\delta\|^2_2} \le \max_{\|\delta\|_0 \le \ell_G s} \frac{\| X_{ig}'\delta\|_G^2}{\|\delta\|^2_2}\lesssim 1.
\end{align*} 
Thus, by Lemma \ref{lemma:M-est_lasso_rates}, one has
\begin{align*}
\|\hat f_{ig} X^j_{ig} (\hat \gamma^j - \bar \gamma^j)\|_G \lesssim (s\log a_G/G)^{1/2} \text{ and } \|\hat \gamma^j - \bar \gamma^j\|_1 \lesssim (s^2\log a_G/G)^{1/2}
\end{align*}
with probability $1-C(\log G)^{-1}$ uniformly over $j\in [p]$.

By the Cauchy-Schwarz inequality and the fact that $\hat f_{ig} \le 1$, we have
\begin{align*}
\Big| \Big\{\frac{1}{G} \sumg \sumi [\partial_\gamma \widehat M(Y_{ig},X_{ig},\hat \gamma^j) -\partial_\gamma \widehat M(Y_{ig},X_{ig},\bar \gamma^j) ]\Big\}'\delta \Big|\le 
\|\hat f_{ig}X^j_{ig} (\hat \gamma^j - \bar \gamma^j)\|_G \|\hat f_{ig} X^j_{ig} \delta\|_G\le
 L_G \| X'_{ig} \delta \|_G
\end{align*}
with probability $1-C(\log G)^{-1}$ uniformly over $j\in[p]$ for some $L_G \lesssim  (s\log a_G/G)^{1/2}$. Since Assumption \ref{a:sparse_eigenvalues} implies that there is a $\ell_G\to \infty$ such that $\semax{\ell_G s}\lesssim 1$ with probability $1-C(\log G)^{-1}$, it follows Lemma \ref{lemma:M-est_lasso_sparsity} that $\|\hat\gamma^j\|_0\lesssim s$ with probability $1-C(\log G)^{-1}$ uniformly over $j\in[p]$.

Note that condition (\ref{eq:M-est_L7}) holds with $\bar q_A=\infty$. Also, with probability $1-C(\log G)^{-1}$, it holds that
\begin{align*}
\frac{1}{G} \sumg\sumi  \widehat M(Y_{ig},X_{ig},\hat \gamma^j) - \frac{1}{G} \sumg\sumi  \widehat M(Y_{ig},X_{ig},\bar\gamma^j)\lesssim s \log a_G/G
\end{align*}  
since $\lambda/G\lesssim (s\log a_G/G)^{1/2}$, $\max_{j \in [p]}\|\hat \gamma^j - \bar \gamma^j\|_1\lesssim (s^2 \log a_G)^{1/2}$ and $\max_{j \in [p]}\|\hat \Psi_{j0}\|_\infty\lesssim 1$ with probability $1-C(\log G)^{-1}$. Finally, one has $C_G\lesssim (s^2\log a_G/G)^{1/2}$,
\begin{align*}
\Big\|\frac{1}{G} \sumg S^j_g\Big\|_\infty \le \|\hat \Psi_0\|_\infty 
\Big\| \hat \Psi_0^{-1} \frac{1}{G} \sumg S^j_g\Big\|_\infty \lesssim \frac{\lambda}{G}
\end{align*}
with probability $1-C(\log G)^{-1}$. This concludes the proof. 
\end{proof}
\subsection{Proof for Corollary \ref{corollary:weighted_lasso_rates}}\label{sec:proof for corollary:weighted_lasso_rates}
\begin{proof}
Define $\ddot r_{ig}^k=X_{ig}'(\bar\zeta^k - \zeta^k)$ and the lost functions be
\begin{align*}
 M (S^k_{ig},X_{ig},\zeta)=& f_{ig}^2 ( S_{ig}^k - X'_{ig}\zeta-\ddot r^k_{ig})^2,\\
 \widehat M (S^k_{ig},X'_{ig}, \zeta)=& \hat f_{ig}^2 (\hat S_{ig}^k - X_{ig}\zeta)^2.
\end{align*}
The sparse approximation $\bar \zeta^k$ is identified by
\begin{align*}
\bar\zeta^k= \underset{\zeta\in \Real^{p}}{\argmin} \EP\left[\frac{1}{G}\sum_{g=1}^{G} \sumi M (S^k_{ig},X_{ig},\zeta)\right].
\end{align*}
Then the proof follows the same steps in the proof for Theorem \ref{thm:REGNPL_rates} as long as one can verify that Assumption \ref{eq:M-est_lasso} (1) is still satisfied with $\hat S^k_{ig}$ in place of $S^k_{ig}$. Thus, it suffices to show
\begin{align*}
\left|\left[\frac{1}{G} \sumg\sumi  \{\hat f_{ig}^2(\hat S^k_{ig} - S^k_{ig}) X_{ig} \}\right]'\delta \right|
=&\left|\left[\frac{1}{G} \sumg\sumi  \{\hat f_{ig}^2(\hat S^k_{ig} - X_{ig}'{\zeta^k}) X_{ig} - \hat f_{ig}^2(S^k_{ig} - X_{ig}'{\zeta^k}) X_{ig} \}\right]'\delta  \right|\\ \lesssim& \|\widetilde \beta - \beta^0\|_2 \left\{ \frac{1}{G} \sumg\sumi  ( \hat f_{ig} X_{ig}' \delta )^2\right\}^{1/2}.
\end{align*}
Observe that the left-hand side equals
\begin{align*}
\left[\frac{1}{G} \sumg\sumi  \{ \hat f_{ig}^2( \hat S^k_{ig}  - \widetilde S^k_{ig} ) X_{ig} +  \hat f^2_{ig} ( \widetilde S^k_{ig}  - S^k_{ig} ) X_{ig}\} \right]'\delta= (i) + (ii).
\end{align*}
Notice that
\begin{align*}
|(ii)|\le 2\|\Lambda\|_\infty|\widetilde \beta^k_k - \beta^0|\Big|\left[\frac{1}{G} \sumg\sumi \hat f_{ig}^2 X_{ig}    \right]'\delta\Big| 
\lesssim
 |\widetilde \beta^k_k - \beta^0_k|\max_{i,g}|f_{ig}| \left\{ \frac{1}{G} \sumg\sumi  ( \hat f_{ig} X_{ig}' \delta )^2\right\}^{1/2}.
\end{align*}
A mean value expansion and an application of H\"older's inequality give that with probability at least $1-C(\log G)^{-1}$,
\begin{align*}
|(i)| \le& 2 |\widetilde \beta_k^k| \|\Lambda'\|_\infty\frac{1}{G} \sumg\left[\sumi  \{ \hat f_{ig}^2 X_{ig}' (\widetilde \beta- \beta^0)X_{ig} \right]'\delta\\
\lesssim& 
\left\{ \frac{1}{G} \sumg\sumi  \left( \hat f_{ig} X_{ig}' (\widetilde \beta- \beta^0) \right)^2\right\}^{1/2} \left\{ \frac{1}{G} \sumg\sumi  \left( \hat f_{ig} X_{ig}' \delta \right)^2\right\}^{1/2}\\
\le& (s\log a_G/G)^{1/2}\|\sqrt{w_{ig}}X_{ig}'\delta\|_G\\
\le& C_G\|\sqrt{w_{ig}}X_{ig}'\delta\|_G.
\end{align*}
This concludes the proof.
\end{proof}

\subsection{Proof for Lemma \ref{lemma:Theta_rates}}\label{sec:proof for lemma:Theta_rates}
\begin{proof}
Throughout the proof, we denote $\|v\|^2_G=v'v/G$ and $(u,v)_G=u'v/G$ for $u,v\in \Real^n$. 
For each $j=1,...,p$, denote $D^j=\{D^j_{ig}:1\le i \le n_g, 1\le g \le G\}$, an $n\times 1$ vector, $\textbf{X}^j=\{X^j_{ig}:1\le i \le n_g, 1\le g \le G\}$, a $n \times (p-1)$ matrix.
We also make use of the notations $\hat F=\diag\{\hat f_{ig}: i\in [n_g], g\in [G]\}$ and $ F=\diag\{ f_{ig}: i\in [n_g], g\in [G]\}$. 

\noindent \textbf{Step 1.} We first derive the identity
\begin{align}
\hat \tau^2_{j}= D^{j\prime}\hat F^2(D^{j}-\textbf{X}^{j}\widetilde \gamma^{j})/G.\label{eq:Theta_tau_sq}
\end{align}
The first order condition of nodewise post-lasso gives
\begin{align}
-\textbf{X}^{j\prime}_{\hat T^j}\hat F^2 (D^{j}-\textbf{X}^{j}\widetilde \gamma^{j})/G=0 \label{eq:nodewise_KKT}
\end{align}
where $\hat T^j=\supp(\widetilde \gamma^j)$.

Multiplying both sides by $\widetilde \gamma'_{j}$, we have 
\begin{align}
-\widetilde \gamma^{j\prime}\textbf{X}^{j\prime}\hat F^2 D^{j}/G
+\widetilde \gamma^{j\prime}\textbf{X}^{j\prime}\hat F^2\textbf{X}^{j}\widetilde \gamma^{j}/G=0.\label{eq:Theta_KKT}
\end{align}
Using its definition, some calculations yield that
\begin{align*}
\hat \tau^2_{j}=D^{j\prime}\hat F^2 D^{j}/G-2 \hat \gamma^{j\prime}\textbf{X}^{j\prime} \hat F^2 D^{j}  /G+   \hat \gamma^{j\prime}\textbf{X}^{j\prime} \hat F^2 \textbf{X}^{j} \hat \gamma^{j}/G.
\end{align*}
Subtracting (\ref{eq:Theta_KKT}) from this gives (\ref{eq:nodewise_KKT}).

\noindent \textbf{Step 2.}
Applying Theorem \ref{thm:REGNPL_rates},
we have the convergence rates
\begin{align*}
\|\widetilde \gamma^{j}-\gamma^{j}\|_1\lesssim s\sqrt{\frac{\log a_G}{G}}\qquad
 \text{ and }
 \qquad
 \|\hat f_{ig} X^{j\prime}_{ig} (\widetilde \gamma^j -\gamma^j)\|_G\vee\|\widetilde \gamma^{j}-\gamma^{j}\|_2\lesssim \sqrt{\frac{s\log a_G}{G}} 
\end{align*}
uniformly in $j$ with probability $1-C(\log G)^{-1}$.

\noindent \textbf{Step 3.}
Since
$
F D^j = F \textbf{X}^{j} \gamma^j + F Z^j,
$
by Step 1, one has
\begin{align*}
\hat \tau_j^2 =& D^{j\prime}\hat F^2 ( D^{j} - \textbf{X}^{j}\widetilde \gamma^j)/G \\
=&D^{j\prime} (\hat F^2 -F^2)(D^j-\textbf{X}^j \widetilde \gamma^j) /G  + D^{j\prime} F^2 (D^j-\textbf{X}^j \widetilde \gamma^j) /G .
\end{align*}
Note we only need to consider bounding $ D^{j\prime} F^2 (D^j-\textbf{X}^j \widetilde \gamma^j) /G $ term since the first term is of smaller order following the fact that $|\hat f_{ig} - f_{ig}|\lesssim f_{ig}$ holds with probability at least $1-C(\log G)^{-1}$ by (\ref{eq:f2_approx}) in the proof of Theorem \ref{thm:REGNPL_rates}. Now, decompose it into
\begin{align*}
D^{j\prime} F^2 (D^j-\textbf{X}^j \widetilde \gamma^j) /G
=& D^{j\prime} F^2 \textbf{X}^j (\gamma^j -  \widetilde \gamma^j)/G  + \gamma^j \textbf{X}^j F^2 Z^j /G + Z^{j\prime } F^2 Z^j/G \\
=& (I)_j + (II)_j +(III)_j.
\end{align*}

First, we bound $(I)_j$. Under Assumption \ref{a:sparsity}, \ref{a:covariates} (4) and Cauchy-Schwarz inequality, it holds uniformly that
\begin{align*}
\max_{j \in [p]}|(I)_j|\le& \max_{j \in [p]} (D^{j\prime}F, F \textbf{X}^j(\widetilde \gamma^j - \gamma^j))_G\\
\le & 
\max_{k \in [p]}\left\{\frac{1}{G} \sumg \sumi X_{ig,k}^2\right\}^{1/2} \max_{j \in [p]} \|f_{ig} X^{j\prime}_{ig} (\widetilde \gamma^j - \gamma^j)\|_G\\
\lesssim&_\Pr \: 
\Op(1)\cdot  \sqrt{\frac{s\log a_G}{G}}
\end{align*}
with probability $1-C(\log G)^{-1}$.

We now bound $(II)_j$. The property of projection implies $ \EP\frac{1}{G}\sum_{g=1}^{G} \sumi f_{ig}^2 X^j_{ig} Z^j_{ig}=0$,
\begin{align*}
\max_{j \in [p]} |(II)_j|\le& \max_{j \in [p]}\|\gamma^j\|_1\| \textbf{X}^j F^2 Z^j /G\|_\infty\\
\le &\max_{j \in [p]}\|\gamma^j\|_1\max_{j,k}\left|\frac{1}{G} \sumg \sumi f^2_{ig} X_{ig,k}^j Z_{ig}^j\right|\\
\le & C_1 \sqrt{s}  \max_{j,k}\left|\frac{1}{G} \sumg \sumi \left(f_{ig}^2 X_{ig,k}^j Z_{ig}^j-\EP f_{ig}^2 X_{ig,k}^j Z_{ig}^j\right)\right|.
\end{align*} 
For each $j,k \in[p]$, denote the classes of functions
\begin{align*}
&\calG=\Big\{ W_g\mapsto \sum^{\bar n}_{i=1} \Lambda'(X_{ig}'\beta^0) X^j_{ig,k} Z^j_{ig}\1\{W_{ig}\ne 0\}: j,k\in[p]  \Big\},\\
&\calG_{j,k}=\Big\{W_g\mapsto \sum^{\bar n}_{i=1} \Lambda'(X_{ig}'\beta^0) X^j_{ig,k} Z^j_{ig}\1\{W_{ig}\ne 0\}  \Big\}.
\end{align*}
Then each $\calG_{j,k}$ contains only one function and thus is a VC-subgraph class with VC index equals unity with itself as an envelope. Also $\calG \subset \cup_{j,k\in [p]}\calG_{j,k}$. Since $|f_{ig}|\le 1$, a measurable envelope for $\calG$ is $H(W_g)=\max_{j,k}|U_{gk}V^j_g|$.

Some calculations and Assumption \ref{a:covariates} (5)(6)(7)(8) give
\begin{align*}
\EP\left[\max_g |H(W_g)|^2\right]\lesssim&   \EP \left[\max_g\|U_{g}\|_\infty^4\right] +  \EP \left[\max_g\max_{j\in[p]}|V_g^j|^4\right] \lesssim G^{2/q} (M^4_{G,1}+M^4_{G,2}).
\end{align*}
The fact that $\sqrt{a+b}\le\sqrt{a}+\sqrt{b}$ for $a$, $b>0$ suggests $\left\{\EP\left[\max_g |H(W_g)|^2\right]\right\}^{1/2}\lesssim  G^{1/q} (M^2_{G,1}+M^2_{G,2})$. Similarly, under Assumption \ref{a:covariates} (4), we have $\sup_{g \in \calG} \EP\left[G^{-1}\sum_{g=1}^{G} g^2(W_g)\right]\lesssim 1$.
Applying Lemma \ref{lemma:entropy_algebra} (1) and (2), we have for any $0<\epsilon\le 1$,
\begin{align*}
  N\left(\epsilon\Vert H\Vert _{Q,2},\calG, \| \cdot \|_{Q,2}\right)\lesssim p^2 \max_{j,k} N(\epsilon\|G_{j,k}\|_{Q,2},\calG_{j,k},\|\cdot\|_{Q,2})\lesssim p^2 \left(\frac{1}{\epsilon}\right).
\end{align*}
Thus one has
$
\sup_Q \log N\left(\epsilon\Vert H\Vert _{Q,2},\calG, \| \cdot \|_{Q,2}\right)\lesssim \log p \lesssim \log a_G.
$
Applying Lemma \ref{lemma:maximal_ineq}, we have with probability at least $1-C(\log G)^{-1}$,
\begin{align*}
 \max_{j,k\in [p]}\left|\frac{1}{G} \sumg \sumi \left(f_{ig}^2 X_{ig,k}^j Z_{ig}^j- \EP f_{ig}^2 X_{ig,k}^j Z_{ig}^j\right)\right| \lesssim \sqrt{\frac{\log a_G}{G}} + \frac{ (M_{G,1}^2\vee M_{G,2}^2)\log  a_G}{G^{1-1/q}}.
\end{align*}
Therefore, under Assumption \ref{a:covariates} (6)(8), 
$$\max_{j\in[p]}|(II)_j|\lesssim  \sqrt{\frac{s\log a_G}{G}} + \frac{ \sqrt{s}\log  a_G(M_{G,1}^2\vee M_{G,2}^2)}{G^{1-1/q}}\lesssim \sqrt{\frac{s\log a_G}{G}}.$$ 
 
Now, we show $|(III)_j-\tau_j^2|=\op(1)$. Under Assumption \ref{a:covariates} (4)(5)(6), using Lemma \ref{lemma:entropy_algebra} (1) and (2), a similar argument leads to that with probability at least $1-C(\log G)^{-1}$,
\begin{align*}
\max_{j \in [p]}| Z^{j\prime}F^2 Z^j/G-\tau^2_j|\lesssim \sqrt{\frac{\log a_G}{G}} + \frac{M^2_{G,1}\log a_G }{G^{1-1/q}} \lesssim \sqrt{\frac{s\log a_G}{G}}.
\end{align*}
Therefore, we conclude that with probability at least $1-C(\log G)^{-1}$, one has 
\begin{align*}
\max_{j \in [p]}|\hat \tau_j^2 -\tau_j^2 |\lesssim \sqrt{\frac{s\log a_G}{G}}.
\end{align*}

\noindent \textbf{Step 4.}
By invoking Assumption \ref{a:covariates} (1), we have for any $G$, one has $\tau^2_{j}=1/\Theta_{j,j}\ge 1/\Lambda_{\max} (\Theta) =\Lambda_{\min} (\Sigma)=\min_{\|\xi\|_2=1} \EP[G^{-1}\sum_{g=1}^{G} \sumi (f_{ig} X_{ig}'\xi)^2]=c_1>0$. 
This implies that with probability at least $1-C(\log G)^{-1}$, one has
\begin{align*}
\max_{j \in [p]}|1/\hat\tau^2_{j}-1/\tau^2_{j}|\lesssim \sqrt{\frac{s\log a_G}{G}}. 
\end{align*}

\noindent \textbf{Step 5.}
We now conclude the proof by deriving a bound for $\max_{j \in [p]}\|\hat \Theta_{j}-\Theta_{j}\|_2$.
By (\ref{eq:surrogate_inverse_pop}), Assumption \ref{a:sparsity} and use preceding steps, we have
\begin{align*}
\max_{j \in [p]}\|\hat \Theta_{j}-\Theta_{j}\|_2=&\max_{j \in [p]}\|\hat C_{j}/\hat \tau^2_{j}-C_{j}/\tau^2_{j}\|_2\\
\le &\max_{j \in [p]}\|\widetilde \gamma^{j}-\gamma^{j}\|_2/\hat \tau^2_{j} +\max_{j \in [p]} (\|\bar\gamma^{j}\|_2 + \|\gamma^j -\bar\gamma^j\|_2 )  |1/\hat \tau^2_{j}-1/\tau^2_{j}|\\
\lesssim &   \sqrt{\frac{s\log a_G}{G}}\cdot \Op(1) + \Op(1) \cdot \sqrt{\frac{ s\log a_G}{G}}\lesssim   \sqrt{\frac{s\log a_G}{G}}
\end{align*}
with probability at least $1-C(\log G)^{-1}$. Similar arguments give
$
\max_{j \in [p]}\|\hat \Theta_{j}-\Theta_{j}\|_1 \lesssim (s^2\log a_G/G)^{1/2}
$
with probability at least $1-C(\log G)^{-1}$.
\end{proof}
\subsection{Proof for Theorem \ref{thm:theta_k_rates}}\label{sec:proof for thm:theta_k_rates}
\begin{proof}
Since $\|\Lambda''\|_\infty \lesssim 1$, one has
\begin{align*}
\max_{k\in[p]}\|\hat \theta^k - \theta^k \|_2 
\le& 
\max_{k\in[p]}\|\hat \Theta_k - \Theta_k\|_2  \left|\frac{1}{G} \sumg \sumi \Lambda'(X_{ig}'\widetilde\beta)\right|+\max_{k\in[p]}\|\Theta_k\|_2\left|\frac{1}{G} \sumg \sumi \left(\Lambda'(X_{ig}'\widetilde\beta)- \Lambda'(X_{ig}'\beta^0)\right)\right|.
\end{align*}
Assumptions \ref{a:parameters} and \ref{a:covariates} (2) imply $\max_{k\in[p]}\|\Theta_k\|_2\le C_1$. Furthermore, using equation (I.6) of \cite{BCCW18} and the fact $\Lambda'=\Lambda\cdot (1-\Lambda)$, suppose that $|X'_{ig} (\widetilde \beta - \beta^0)|\le 1$, it holds that
\begin{align*}
\left|\frac{1}{G} \sumg\sumi \left(\Lambda'(X_{ig}'\widetilde\beta)- \Lambda'(X_{ig}'\beta^0)\right)\right|
=&
\left|\frac{1}{G} \sumg\sumi \left(\widetilde\Lambda_{ig} (1-\widetilde\Lambda_{ig})-\Lambda_{ig} (1-\widetilde\Lambda_{ig})+\Lambda_{ig} (1-\widetilde\Lambda_{ig})-\Lambda_{ig} (1-\Lambda_{ig})\right)\right|\\
\lesssim& (\|\Lambda\|_\infty + \|1-\Lambda\|_\infty)\max_{i,g}|f_{ig}|
 \left|\frac{1}{G} \sumg\sumi f_{ig} X_{ig}'(\widetilde \beta - \beta^0)\right| \\
\lesssim& 
O(1)\cdot \|f_{ig} X_{ig}'(\widetilde \beta - \beta^0)\|_G\\
\lesssim&
\sqrt{\frac{s\log a_G}{G}}
\end{align*}
with probability $1-C(\log G)^{-1/2}$, where $\Lambda_{ig}=\Lambda'(X_{ig}'\beta)$, $\widetilde\Lambda_{ig}=\Lambda'(X_{ig}'\widetilde\beta)$ and the second inequality is due to an application of Cauchy-Schwarz inequality.
The condition $|X'_{ig} (\widetilde \beta - \beta^0)|\le 1$ holds asymptotically with probability $1-o(1)$ since
\begin{align*}
\max_{i,g}\|X_{ig}\|_\infty \|\widetilde \beta -\beta^0\|_1
\lesssim&_\Pr \frac{ M_{G,2} s(\log a_G)^{1/2}}{G^{1/2-1/2q}}=o(1)
\end{align*} 
with probability $1-C(\log G)^{-1}$ under Assumption \ref{a:covariates} (7)(8) and Theorem \ref{thm:LOG_rates}. Furthermore, $$
\max_{k\in[p]}\|\hat \Theta_k - \Theta_k\|_2 \lesssim_\Pr \sqrt{\frac{s\log a_G}{G}}
$$
following Theorem \ref{thm:REGNPL_rates} and $\hat\tau^{-2}=O(\tau^{-2})=O(1)$.
So 
$$\max_{k\in[p]}\|\widetilde \theta^k - \theta^k \|_2\le \sqrt{\frac{s\log a_G}{G}}.$$
The bound $\max_{k\in[p]}\|\widetilde \theta^k - \theta^k \|_1\le s\sqrt{\frac{\log a_G}{G}}$ with probability at least $1-C(\log G)^{-1}$ can be established following similar arguments and the fact that $\max_{k \in[p]}\|\Theta_k\|_1\le \sqrt{s}C_1$.
\end{proof}





\section{Additional Theoretical Results}

\subsection{Properties of $\tau^2_{j}$} \label{sec:properies_tau_2}
\text{}\\
In this Section, we derives some important properties of $\tau^2_{j}$, which is based on the work of \cite{Kock16}, a panel data generalization of the nodewise lasso in \cite{vdGBRD14}. Denote $\Sigma= \EP\left[\frac{1}{G}\sum_{g=1}^{G} \sumi f^2_{ig}X_{ig}X'_{ig}\right]$. Let $\Sigma_{-j,-j}$ be the $(p-1)\times(p-1)$ submatrix of $\Sigma$
with the $j$-th column and row removed. $\Sigma_{j,-j}$ represents the $j$-th row of $\Sigma$ with its $j$-th element removed and $\Sigma_{-j,j}$ is defined analogously. From the inverse formula of a partitioned matrix, we have
\begin{align*}
\Theta_{j,j}&=(\Sigma_{j,j}-\Sigma_{j,-j}\Sigma_{-j,-j}^{-1}\Sigma_{-j,j})^{-1},\\
\Theta_{j,-j}&=(\Sigma_{j,j}-\Sigma_{j,-j}\Sigma_{-j,-j}^{-1}\Sigma_{j,-j})\Sigma_{j,-j}\Sigma_{-j,-j}^{-1}=-\Theta_{j,j}\Sigma_{j,-j}\Sigma_{-j,-j}^{-1}.
\end{align*}
Now, by solving (\ref{eq:population_gamma}), we have
\begin{align*}
\gamma^{j}=&\left\{ \EP\left[\frac{1}{G}\sum_{g=1}^{G} \sumi f^2_{ig}X_{ig}^{j\prime} X^j_{ig}\right] \right\}^{-1}\cdot \EP\left[\frac{1}{G}\sum_{g=1}^{G} \sumi f^2_{ig} X^{j\prime}_{ig} D^j_{ig} \right]\\
=&\Sigma_{-j,-j}^{-1}\Sigma_{j,-j}'
\end{align*}
Combining with above, we have
\begin{align}
\Theta_{j,-j}=-\Theta_{j,j}\gamma^{j\prime}.\label{eq:theta_gamma_relation}
\end{align}
Furthermore, using $D^{j}=\textbf X^j\gamma^{j}+Z^{j}$ and $\EP[Z^{j\prime}F^2 X^j ]=0$, we have
\begin{align*}
\Sigma_{j,j}=& \EP [D^{j\prime}F^2 D^j]\\
=&\gamma^{j\prime} \EP [\textbf X^{j\prime}F^2 \textbf X^j]\gamma^{j}+ \EP[ Z^{j\prime}F^2 Z^{j}]+2 \EP [Z^{j\prime}F^2\textbf X^j] \gamma^{j}\\
=&\Sigma_{j,-j}\Sigma_{-j,-j}^{-1}\Sigma_{-j,j}'+\tau^2_{j}+0.
\end{align*}
Therefore we have
\begin{align}
\tau^2_{j}=\Sigma_{j,j}-\Sigma_{j,-j}\Sigma_{-j,-j}^{-1}\Sigma_{-j,j}'=1/\Theta_{j,j}. \label{eq:tau2_theta_relation}
\end{align}
Now define
\begin{align*}
C=
\begin{bmatrix}
1 & - \gamma^1_{1 }& \dots & -\gamma^1_{p-1 }\\
- \gamma^2_{ 1 }& 1 & \dots &-\gamma^2_{p -1}\\
\vdots &\vdots &\ddots&\vdots\\
-\gamma^p_{ 1 }&- \gamma^p_{ 2 }&\dots& 1
\end{bmatrix}
\end{align*}
and $T^2=\text{diag}\{\tau^2_{1},...,\tau^2_{p}\}$, using (\ref{eq:theta_gamma_relation}) and (\ref{eq:tau2_theta_relation}), we have 
\begin{align}
\Theta=T^{-2}C. \label{eq:surrogate_inverse_pop}
\end{align}

\subsection{Results for Nuisance Parameters Estimation}\label{sec:nuisance}

\text{}\\
The following results adapt lemmas in Appendix L of \cite{BCCW18} to cluster sampling. Their proofs follow closely those of Lemma L1-L4 of \cite{BCCW18} while we only consider an increasing finite index set for simplicity.

\subsubsection{$\ell$-1 Penalized M-Estimation with Clustered Data}\label{sec:M-estimation}
Consider a data generating process with an outcome variable $Y^k_{ig}$ and $p$-dimensional covariates $X^k_{ig}$, both indexed by $k\in\mathcal U_G$ for some $\mathcal U_G \subset [p]$. We maintain the cluster sampling setting as before. The parameter of interest 
\begin{align}
\mu^k \in  \underset{\mu\in \Real^p}{\argmin}  \EP\left[\frac{1}{G}\sum_{g=1}^{G} \sumi M_k(Y^k_{ig},X^k_{ig},\mu)\right]. \label{eq:M-est_lasso_pop}
\end{align}
Define the lasso and post-lasso estimators 
\begin{align}
&\hat \mu^k \in \underset{\mu\in \Real^p}{\argmin} \frac{1}{G} \sumg\sumi \widehat M(Y^k_{ig},X^k_{ig},\mu) + \frac{\lambda}{G}\|\hat \Psi_k \mu\|_1, \label{eq:M-est_lasso}\\
&\widetilde \mu^k \in \underset{\mu\in \supp(\hat \mu_k)}{\argmin} \frac{1}{G} \sumg \sumi \widehat M_k(Y^k_{ig},X^k_{ig},\mu).  \label{eq:M-est_post-lasso}
\end{align}
For each $k\in\mathcal U_G$, denote the \textit{ideal} penalty loadings $\hat \Psi_{k0}=\diag(\{l_{k0j}:j\in [p]\})$, where
\begin{align*}
l_{k0j}=\left\{\frac{1}{G} \sumg \left(\sumi\partial_{\mu_j} M_k(Y^k_{ig},X^k_{ig},\mu^k)\right)^2 \right\}^{1/2}=\left\{\frac{1}{G} \sumg  (S^k_{gj})^2 \right\}^{1/2},
\end{align*}
where $S^k_{gj}=\sumi\partial_{\mu_j} M_k(Y^k_{ig},X^k_{ig},\mu^k)$. We also denote the feasible penalty loadings by $\hat \Psi_k=\diag(\{l_{kj}:j\in [p]\})$ for some $l_{kj}$
\begin{align*}
l_{kj}=\left\{\frac{1}{G} \sumg \left(\sumi\partial_{\mu_j} \widehat M_k(Y^k_{ig},X^k_{ig},\hat \mu^k)\right)^2 \right\}^{1/2}=\left\{\frac{1}{G} \sumg  (\hat S^k_{gj})^2 \right\}^{1/2},
\end{align*}
where $\hat S^k_{gj}=\sumi\partial_{\mu_j} \widehat  M_k(Y^k_{ig},X^k_{ig},\hat \mu^k)$. Also write $S^k_g=(\{S^k_{gj}:j\in[p]\})$ and  $\hat S^k_g=(\{\hat S^k_{gj}:j\in[p]\})$. Denote $T_k=\supp(\mu^k)$ and $\hat T_k=\supp(\hat \mu^k)$. We assume $\lambda$ is chosen such that with high probability, 
\begin{align}
\frac{\lambda}{G}\ge c\max_{k\in\mathcal U_G}\left\|\hat \Psi_{0}^{-1} \sumi\partial_{\mu} M(Y^k_{ig},X^k_{ig},\mu^k)\right\|_\infty, \label{eq:M-est_regularized_event_L5}
\end{align}
for a fixed constant $c>1$. This will be shown to happen under some sufficient conditions in Section \ref{sec:concentration_regularized_event}. Let $L\ge \ell >1/c$ be some fixed constants and let
\begin{align*}
\widetilde c=\frac{Lc+1}{\ell c-1}\max_{k\in\mathcal U_G} \|\hat \Psi_{k0}\|_\infty \|\hat \Psi^{-1}_{k0}\|_\infty.
\end{align*}
Denote $s_k=\|\mu^k\|_0$ and let $\widetilde \Delta_G$ be a sequence of positive constants converging to zero, let $\widetilde C_G $ be a sequence of random variables and $w_{ig}=w(X_{ig})$ be some weights such that $0\le w_{ig}\le 1$ almost surely. Finally, let $A_k$ be a random subset of $\Real^p$ and $\bar q_{A_k}$ a random variable depends possibly on $A_k$.
\begin{assumption}\label{a:M-est_lasso}
Suppose that $\max_{k\in \mathcal U_G}\|\mu^k\|_0=s$ and for each $k\in[p]$ $\mu \mapsto  \widehat M_k(y,x,\mu)$ is convex almost surely and with probability at least $1-\widetilde \Delta_G$ for all $\delta\in \Real^p$, it holds that for all $k\in \mathcal U_G$,
\begin{enumerate}
\item $\Big|\Big\{\frac{1}{G} \sumg\sumi[\partial_\mu  \widehat  M_k(Y^k_{ig},X^k_{ig},\mu^k) - \partial_\mu M_k(Y^k_{ig},X^k_{ig},\mu^k)]\Big\}'\delta\Big| \le C_G \|\sqrt{w_{ig}} X^{k\prime}_{ig} \delta\|_G$ for all $\delta\in \Real^p$;
\item $\ell \hat \Psi_{k0} \le \hat \Psi_k \le L \hat \Psi_{k0}$;
\item for all $\delta\in A_k$,
\begin{align*}
&\frac{1}{G} \sumg\sumi  \widehat  M_k(Y^k_{ig},X^k_{ig},\mu^k+ \delta) - \frac{1}{G} \sumg\sumi  \widehat  M_k(Y^k_{ig},X^k_{ig},\mu^k) - \frac{1}{G} \sumg\sumi[ \partial_\mu  \widehat  M_k(Y^k_{ig},X^k_{ig},\mu^k+ \delta)]'\delta \\
+& 2 C_G  \|\sqrt{w_{ig}} X^{k\prime}_{ig} \delta\|_G 
\ge \{  \|\sqrt{w_{ig}} X^{k\prime}_{ig}  \delta\|_G^2\} \wedge \{\bar q_{A_k}  \|\sqrt{w_{ig}} X^{k\prime}_{ig}  \delta\|_G\}.
\end{align*}
\end{enumerate}
\end{assumption}

Define the restricted eigenvalue
\begin{align*}
\bar \kappa_{2 \widetilde c}=\min_{k\in \mathcal U_G}\inf_{\delta \in \Delta_{2 \widetilde c,k}}\frac{\|\sqrt{w_{ig} }X'_{ig}\delta\|_G }{\|\delta_{T_k}\|_2},
\end{align*}
where $\Delta_{2\widetilde c,k}=\{\delta\in\Real^p: \|\delta_{T^c_k}\|_1 \le 2 \widetilde c \|\delta_{T_k}\|_1\}$. In addition, define the minimum and maximum sparse eigenvalues
\begin{align*}
\semin{m,k}=\min_{1 \le \|\delta\|_0\le m}  \frac{\|\sqrt{w_{ig} }X^{k\prime}_{ig}\delta\|^2_G }{\|\delta\|_2^2}\: \text{ and } \:\semax{m,k}=\max_{1 \le \|\delta\|_0\le m}  \frac{\|\sqrt{w_{ig} }X^{k\prime}_{ig}\delta\|_G^2 }{\|\delta\|^2_2}.
\end{align*}
Boundedness of minimum and maximum sparse eigenvalues with probability goes to $1$ implies that restricted eigenvalue is bounded away from $0$ with probability goes to $1$. For its proof, see Lemma 4.1 of \cite{BRT09}.

\begin{lemma}\label{lemma:M-est_lasso_rates}
Suppose that Assumption \ref{a:M-est_lasso} holds with 
\begin{align*}
A_k=\Delta_{2\widetilde c,k}\cup \{\delta\in \Real^p : \|\delta\|_1  \le \frac{3G}{\lambda} \frac{c \|\hat \Psi_{k0}^{-1}\|_\infty}{\ell c-1} C_G \|\sqrt{w_{ig}  }X_{ig}^{k\prime}\delta\|_G \},
\end{align*}
and $\bar q_{A_k} \ge (L+\frac{1}{c})\|\hat \Psi_{k0}\|_\infty \frac{\lambda \sqrt{s}}{G \bar \kappa_{2 \widetilde c}} + 6\widetilde c C_G$. In addition, suppose that $\lambda$ satisfies condition \ref{eq:M-est_regularized_event_L5} with probability at least $1-\widetilde \Delta_G$. Then, with probability at least $1- 2 \widetilde \Delta_G$, we have
\begin{align*}
\|\sqrt{w_{ig}}X_{ig}^{k\prime} (\hat \mu^k - \mu^k)\|_G \le& \left(L+ \frac{1}{c}\right)\|\hat \Psi_{k0} \|_\infty \frac{\lambda \sqrt{s}}{G \ktc} +6 \widetilde c C_G,\\
\|\hat \mu^k - \mu^k\|_1\le& \left(\frac{(1+2\widetilde c)\sqrt{s}}{\ktc}+ \frac{3G}{\lambda} \frac{c\|\hat\Psi_{k0}^{-1}\|_\infty}{\ell c - 1}C_G\right) \left( \left(L+ \frac{1}{c}\right)\|\hat \Psi_{k0}^{-1} \|_\infty \frac{\lambda \sqrt{s}}{G\ktc } + 6 \widetilde c C_G\right)
\end{align*}
uniform for $k\in \mathcal U_G$.

\end{lemma}

\begin{lemma}\label{lemma:M-est_lasso_sparsity}
In addition to conditions of Lemma \ref{lemma:M-est_lasso_rates}, suppose that with probability $1-\widetilde \Delta_G$, for some random variable $L_G$ such that for all $\delta\in \Real^p$, it holds that
\begin{align}
\Big| \Big\{\frac{1}{G} \sumg\sumi [\partial_\mu  \widehat M_k(Y_{ig}^k,X_{ig}^k,\hat \mu^k)  -\partial_\mu  \widehat  M_k(Y_{ig}^k,X_{ig}^k,\mu^k) ]\Big\}'\delta \Big|\le L_G \| X_{ig}^{k\prime} \delta \|_G. \label{eq:M-est_L6}
\end{align}
Then with probability $1-3 \widetilde \Delta_G$, we have for all $k\in \mathcal U_G$,
\begin{align*}
\hat s_k \le \min_{m\in \mathcal M_k} \semax{m,k} L^2_k,
\end{align*}
where $\mathcal M_k=\{m\in \NN:m \ge 2 \semax{m,k}L^2_k\}$ and $L_k=\frac{c\|\hPsi^{-1}\|_\infty}{c\ell -1}\frac{G}{\lambda}\{C_G + L_G\}$.

\end{lemma}

\begin{lemma}\label{lemma:M-est_post_lasso_rates}
Suppose that Assumption \ref{a:M-est_lasso} holds with $A_k=\{\delta\in\Real^p: \|\delta\|_0\le \hat s_k + s_k\}$ and
{\small 
\begin{align}
\bar q_{A_k}>
  2 \max\Bigg\{& \left( \frac{1}{G} \sumg \sumi [ \widehat  M_k(Y_{ig}^k,X_{ig}^k,\widetilde \mu^k) -  \widehat  M_k(Y_{ig}^k,X_{ig}^k, \mu^k) ]\right)^{1/2}_+, \nonumber\\
  &\left(
\frac{\sqrt{\hat s_k + s_k}\|\frac{1}{G} \sumg\sumi\partial_\mu M_k(Y_{ig}^k,X_{ig}^k,\mu^k)\|_\infty}{\sqrt{\semin{\hat s_k + s_k}}} +3 C_G
 \right) \Bigg\}.
 \label{eq:M-est_L7}
\end{align} }
Then with probability at least $1-\widetilde \Delta_G$, 
\begin{align*}
\|\sqrt{w_{ig}} X_{ig}^{k\prime} (\widetilde \mu^k - \mu^k)\|_G \le& \Big\{ \frac{1}{G} \sumg \sumi  [ \widehat  M_k(Y_{ig}^k,X_{ig}^k,\widetilde \mu^k) -  \widehat  M_k(Y_{ig}^k,X_{ig}^k, \mu^k)] \Big\}^{1/2}_+ \\
&+
\frac{\sqrt{\hat s_k + s_k}\|\frac{1}{G} \sumg\sumi\partial_\mu M_k(Y_{ig}^k,X_{ig}^k,\mu^k)\|_\infty}{\sqrt{\semin{\hat s_k + s_k}}}+ 3 C_G
\end{align*}
uniform for $k\in \mathcal U_G$.
In addition, with probability at least $1-\widetilde \Delta_G$, one has
\begin{align}
\frac{1}{G} \sumg\sumi\partial_\mu  \widehat M_k(Y_{ig}^k,X_{ig}^k,\widetilde\mu^k)- \frac{1}{G} \sumg\sumi\partial_\mu  \widehat M_k(Y_{ig}^k,X_{ig}^k,\mu^k)\le L\frac{\lambda}{G}\|\hat \mu^k - \mu^k\|_1\|\hPsi\|_\infty. \label{eq:M-est_L8}
\end{align}
Therefore, with probability at least $1-\widetilde \Delta_G$, we have
\begin{align*}
\|\widetilde \mu^k - \mu^k\|_1 \le \frac{\sqrt{\hat s_k + s_k}}{\sqrt{\semax{\hat s_k + s_k}} \min_{i,g} w_{ig}^2} 
\left(L\frac{\lambda}{G}\|\hat \mu^k - \mu^k\|_1\|\hPsi\|_\infty + \frac{\lambda\sqrt{\hat s_k + s_k}}{c G\sqrt{\semin{\hat s_k + s_k}}}+ 3 C_G \right)
\end{align*}
uniform for $k\in \mathcal U_G$.

\end{lemma}

\subsubsection{Concentration for Regularized Events}\label{sec:concentration_regularized_event}
We now provide sufficient conditions for \ref{eq:M-est_regularized_event_L5}. Denote $|\mathcal U_G|=\widetilde p$.
\begin{assumption}\label{a:M-est_WL}
Suppose that the following holds for each $G$,
\begin{enumerate}
\item $\max_{k\in \mathcal U_G} \max_{j\in [p]} ( \EP\frac{1}{G}\sum_{g=1}^{G} |S^k_{gj}|^3)^{1/3} \Phi^{-1} (1-\gamma/2p)\le \widetilde \varphi_G G^{1/6}$ for $j \in[\widetilde p]$.
\item $\underline C\le ( \EP\frac{1}{G}\sum_{g=1}^{G} |S^k_{gj}|^2)^{1/2}\le \overline C$ for all $k\in \mathcal U_G$ for $j \in[\widetilde p]$.
\end{enumerate}
\end{assumption} 
Let
\begin{align}\label{eq:lambda}
\lambda=c' \sqrt{G}\Phi^{-1} (1-\gamma/2p \widetilde p),
\end{align}
where $\gamma=\gamma_G =o(1)$.
\begin{lemma}\label{lemma:M-est_WL}
Suppose that \ref{a:M-est_WL} holds and $\lambda$ satisfies (\ref{eq:lambda}) with some $c'>c$ and $\gamma=\gamma_G\in[1/G,1/\log G]$. Then
\begin{align*}
\Pr_\Pr\left( \frac{\lambda}{G} \ge c  \max_{k\in \mathcal U_G}  \left\|\hat \Psi^{-1}_k\frac{1}{G} \sumg S_g^k\right\|_\infty\right)\ge 1-\gamma -o(\gamma).
\end{align*}

\end{lemma}

\section{Proof for Additional Technical Results } \label{sec:proofs_M-estimation}
\subsection{Proof for Lemma \ref{lemma:M-est_lasso_rates} }\label{sec:proof_M_est_lasso_rates}
\begin{proof}
Denote $ \delta^k =\hat\mu^k - \mu^k$. Assume the events of Assumption \ref{a:M-est_lasso} and (\ref{eq:M-est_regularized_event_L5}) holds. This happens with probability at least $1-2 \widetilde \Delta_G$.
By definition of $\hat \mu$,
\begin{align}
\frac{1}{G} \sumg\sumi \widehat M_k(Y_{ig}^k,X_{ig}^k,\hat \mu^k)  - \frac{1}{G} \sumg\sumi  \widehat  M_k(Y_{ig}^k,X_{ig}^k, \mu^k)\le&
 \frac{\lambda}{G}\|\hat \Psi \mu^k\|_1  -  \frac{\lambda}{G}\|\hat \Psi \hat \mu^k\|_1 \nonumber\\
 \le& L\frac{\lambda}{G}\|\hat \Psi_{k0} \delta_{k,T_k}\|_1  -  \ell\frac{\lambda}{G}\|\hat \Psi_{k0} \delta_{k,T^c_k}\|_1. \label{eq:M1}
\end{align}
Furthermore, Assumption \ref{a:M-est_lasso} (a) and the convexity of $M$ in $\mu$ as well as condition (\ref{eq:M-est_regularized_event_L5}) suggest
\begin{align}
\frac{1}{G} \sumg\sumi  \widehat  M_k(Y_{ig}^k,X_{ig}^k,\hat \mu^k)   - \frac{1}{G} \sumg\sumi \widehat  M_k(Y_{ig}^k,X_{ig}^k, \mu^k)\nonumber\\ 
\ge \frac{1}{G} \sumg\sumi[ \partial_\mu  \widehat  M_k(Y_{ig}^k,X_{ig}^k, \mu^k)]'\delta_{k} \ge -\frac{\lambda}{G}\frac{1}{c}  - C_G\|\sqrt{w_{ig}}X_{ig}^{k\prime}\delta_{k} \|_G. \label{eq:M2}
\end{align}
Combining (\ref{eq:M1}) and (\ref{eq:M2}) gives 
\begin{align}
\frac{\lambda}{G}\frac{\ell c -1}{c} \|\hPsi \delta_{k,T^c_k}\|_1 \le \frac{\lambda}{G}\frac{Lc+1}{c} \|\hPsi \delta_{k,T_k}\|_1 +  C_G\|\sqrt{w_{ig}}X_{ig}^{k\prime}\delta_{k} \|_G. \label{eq:M4}
\end{align}
Thus 
\begin{align*}
\|\delta_{k,T^c_k}\|_1 \le \widetilde c \|\delta_{k,T_k}\|_1 + \frac{G}{\lambda}\frac{c\|\hPsi^{-1}\|_\infty}{\ell c-1} C_G\|\sqrt{w_{ig}}X_{ig}^{k\prime}\delta_{k} \|_G.
\end{align*}
Consider the case that $\delta\not\in \Delta_{2\widetilde c,k}$, then since $\widetilde c\ge 1$,
\begin{align*}
\|\delta_{k,T_k}\|_1 \le   \frac{G}{\lambda}\frac{c\|\hPsi^{-1}\|_\infty}{\ell c-1} C_G\|\sqrt{w_{ig}}X_{ig}^{k\prime}\delta_k \|_G.
\end{align*}
Also from above,
\begin{align*}
\|\delta_{k,T^c_k}\|_1 \le \frac{1}{2} \|\delta_{k,T^c_k}\|_1 + \frac{G}{\lambda}\frac{c\|\hPsi^{-1}\|_\infty}{\ell c-1} C_G\|\sqrt{w_{ig}}X_{ig}^{k\prime}\delta_k \|_G,
\end{align*}
and thus
\begin{align*}
\|\delta_{k,T^c_k}\|_1 \le  \frac{2G}{\lambda}\frac{c\|\hPsi^{-1}\|_\infty}{\ell c-1} C_G\|\sqrt{w_{ig}}X_{ig}^{k\prime}\delta_k \|_G.
\end{align*}
Adding them up, one has
\begin{align*}
\|\delta_k\|_1 \le  \frac{3G}{\lambda}\frac{c\|\hPsi^{-1}\|_\infty}{\ell c-1} C_G\|\sqrt{w_{ig}}X_{ig}^{k\prime}\delta_k \|_G:=I_k.
\end{align*}
Now suppose that $\delta\in \Delta_{2\widetilde c,k}$, the definition of $\ktc$ gives
\begin{align*}
\|\delta_{k,T_k}\|_1 \le \sqrt{s} \|\delta_{k,T_k}\|_2 \le \frac{\sqrt{s}}{\ktc} \|\sqrt{w_{ig}}X_{ig}^{k\prime}\delta_k \|_G=II_k.
\end{align*}
So by combining two cases, we have
\begin{align}
\|\delta_{k,T_k}\|_1 \le I_k+ II_k \label{eq:M6}.
\end{align}

Recall that 
\begin{align*}
A_k=\left\{\delta\in \Real^p : \|\delta\|_1  \le \frac{3G}{\lambda} \frac{c \|\hat \Psi_{k0}^{-1} \|_\infty}{\ell c-1} C_G \|\sqrt{w_{ig}  }X_{ig}^{k\prime}\delta\|_G \right\}.
\end{align*}
By invoking Assumption \ref{a:M-est_lasso} (3), we have
\begin{align*}
&\{\|\sqrt{w_{ig}  }X_{ig}^{k\prime}\delta_k\|_G^2 \}\wedge \{\bar q_{A_k} \|\sqrt{w_{ig}  }X_{ig}^{k\prime}\delta_k\|_G \}\}\\
\le& \frac{1}{G} \sumg\sumi \widehat  M_k(Y_{ig}^k,X_{ig}^k,\mu^k+ \delta_k) - \frac{1}{G} \sumg\sumi  \widehat M_k(Y_{ig}^k,X_{ig}^k,\mu^k) - \frac{1}{G} \sumg\sumi [\partial_\mu \widehat  M_k(Y_{ig}^k,X_{ig}^k,\mu^k+ \delta_k)]'\delta_k \\
&+ 2 C_G  \|\sqrt{w_{ig}} X_{ig}^{k\prime} \delta_k\|_G \\
\le &\left(L + \frac{1}{c}\right)\frac{\lambda}{G}\|\hPsi \delta_{k,T_k}\|_1 + 3 C_G\|\sqrt{w_{ig}} X_{ig}^{k\prime} \delta_k\|_G \\
\le& \left(L + \frac{1}{c}\right)\frac{\lambda}{G}\|\hPsi\|_\infty(I_k+II_k) + 3 C_G\|\sqrt{w_{ig}} X_{ig}^{k\prime} \delta_k\|_G \\
\le& \left\{\left(L + \frac{1}{c}\right)\|\hPsi\|_\infty \frac{\lambda\sqrt{s}}{G\ktc} + 6\widetilde c C_G\right\}\|\sqrt{w_{ig}} X_{ig}^{k\prime} \delta_k\|_G.
\end{align*}
The definition of $A$ implies that the minimum on the left-hand side must be achieved by the quadratic term and thus
\begin{align*}
\|\sqrt{w_{ig}  }X_{ig}^{k\prime}\delta_k\|_G  \le \left\{\left(L + \frac{1}{c}\right)\|\hPsi\|_\infty \frac{\lambda\sqrt{s}}{G\ktc} + 6\widetilde c C_G\right\}.
\end{align*}
Finally,
\begin{align*}
\|\delta_k\|_1 \le (1+2 \widetilde c) II_k +  I_k\le \left(\frac{(1+2 \widetilde c)\sqrt{s}}{\ktc} + \frac{3G}{\lambda} \frac{c\|\hPsi^{-1}\|_\infty}{\ell c -1}C_G\right)
\end{align*}
uniform for $k\in \mathcal U_G$.
\end{proof}
\subsection{Proof for Lemma \ref{lemma:M-est_lasso_sparsity}}\label{sec:proof_for_lemma:M_est_lasso_sparsity}
\begin{proof}
Let $ S_G^k= \frac{1}{G} \sumg\sumi M_k(Y_{ig}^k,X_{ig}^k,\mu^k)$. 
Assume the events of Assumption \ref{a:M-est_lasso}, conditions (\ref{eq:M-est_regularized_event_L5}) and (\ref{eq:M-est_L6}) holds. This happens with probability at least $1-3 \widetilde \Delta_G$.

By definition of $\hat \mu^k$, for all $j\in \hat T_k$,
\begin{align*}
\left|\hat \Psi^{-1}_k \frac{1}{G} \sumg\sumi \partial_{\mu_j}  \widehat  M_k(Y_{ig}^k,X_{ig}^k,\hat \mu^k)   \right|=\frac{\lambda}{G}.
\end{align*}
Therefore, using Assumption \ref{a:M-est_lasso} (1),(2), and inequalities (\ref{eq:M-est_regularized_event_L5}),(\ref{eq:M-est_L6}),
\begin{align*}
&\frac{\lambda}{G}\sqrt{s_k}=\left\| \left(\hat \Psi^{-1}_k \frac{1}{G} \sumg\sumi \partial_\mu  \widehat  M_k(Y_{ig}^k,X_{ig}^k,\hat \mu^k)   \right)_{\hat T_k} \right\|_2\\
\le & \|\left(\hat \Psi^{-1}_k  S_G^k\right)_{\hat T_k}\|_2 + \left\|\left(\hat \Psi^{-1}_k\{ \frac{1}{G} \sumg\sumi \partial_\mu  \widehat  M_k(Y_{ig}^k,X_{ig}^k,\mu^k)   -S_G^k\}\right)_{\hat T_k}\right\|_2\\
&+\left\|\left(\hat \Psi^{-1}_k \frac{1}{G} \sumg\sumi\{ \partial_\mu  \widehat  M_k(Y_{ig}^k,X_{ig}^k,\hat \mu^k)  -\partial_\mu  \widehat  M_k(Y_{ig}^k,X_{ig}^k,\mu^k) \}\right)_{\hat T_k}\right\|_2\\
\le & \sqrt{s}\|\hat \Psi^{-1}_k \hPsi\|_\infty \|S_G^k\|_\infty + \|\hat \Psi^{-1}_k\|_\infty C_G \sup_{\|\delta\|_2=1,\|\delta\|_0\le \hat s_k} \|\sqrt{w_{ig}}X_{ig}^{k\prime}\delta\|_G\\
&
+\|\hat \Psi^{-1}_k\|_\infty  \sup_{\|\delta\|_2=1,\|\delta\|_0\le \hat s_k} \Big|\frac{1}{G} \sumg\sumi[ \partial_\mu  \widehat  M_k(Y_{ig}^k,X_{ig}^k,\hat \mu^k)  -\partial_\mu   \widehat  M_k(Y_{ig}^k,X_{ig}^k,\mu^k) ]'\delta\Big|\\
\le &\frac{\lambda}{c\ell G}\sqrt{s_k} +\frac{\|\hPsi^{-1}\|_\infty}{\ell}\{C_G + L_G\} \sup_{\|\delta\|_2=1,\|\delta\|_0\le \hat s_k}  \|X_{ig}^{k\prime}\delta\|_G.
\end{align*}
Note that $\sup_{\|\delta\|_2=1,\|\delta\|_0\le \hat s_k}  \|X_{ig}^{k\prime}\delta\|_G=\semax{\hat s_k,k}$,
\begin{align*}
\hat s_k \le \semax{\hat s_k}L^2_k.
\end{align*}
The rest follows from the sublinearity of maximum sparse eigenvalue and minimizing over $M\in \mathcal M_k$.
\end{proof}

\subsection{Proof for Lemma \ref{lemma:M-est_post_lasso_rates}}\label{sec:proof_for_M_est_post_lasso_rates}
\begin{proof}
First, note that by definition of $\widetilde \mu^k$ and $\hat \mu^k$
\begin{align*}
&\EG\sumi\partial_\mu \hat M_k(Y_{ig}^k,X_{ig}^k,\widetilde\mu^k)- \frac{1}{G} \sumg\sumi\partial_\mu \hat M_k(Y_{ig}^k,X_{ig}^k,\mu^k)\\
\le & \frac{1}{G} \sumg\sumi\partial_\mu\hat M_k(Y_{ig}^k,X_{ig}^k,\hat \mu^k)- \frac{1}{G} \sumg\sumi\partial_\mu \hat M_k(Y_{ig}^k,X_{ig}^k,\mu^k)\\
\le& L\frac{\lambda}{G}\|\hat \mu^k - \mu^k\|_1\|\hPsi\|_\infty
\end{align*}
with probability at least $1-\widetilde \Delta_G$.

To show the first claim, let us suppose the events of Assumption \ref{a:M-est_lasso} holds with probability $1-\widetilde \Delta_G$. Denote $\delta_k=\widetilde \mu^k - \mu^k$ and $S_G^k= \frac{1}{G} \sumg\sumi M_k(Y_{ig}^k,X_{ig}^k,\mu^k)$ and $t_k=\|\sqrt{w_{ig}}X_{ig}^{k\prime}\delta_k\|_G$. Assumption \ref{a:M-est_lasso} (3) gives
\begin{align*}
t^2_k\wedge\{\bar q_{A_k} t_k\} \le& \frac{1}{G} \sumg\sumi  \widehat  M_k(Y_{ig}^k,X_{ig}^k,\widetilde\mu^k)-\frac{1}{G} \sumg\sumi  \widehat  M_k(Y_{ig}^k,X_{ig}^k,\mu^k)\\
&
-\frac{1}{G} \sumg\sumi[ \partial_\mu  \widehat  M_k(Y_{ig}^k,X_{ig}^k,\mu^k)]'\delta_k +2 C_G t_k\\
\le& \frac{1}{G} \sumg\sumi  \widehat  M_k(Y_{ig}^k,X_{ig}^k,\widetilde\mu^k)-\frac{1}{G} \sumg\sumi  \widehat  M_k(Y_{ig}^k,X_{ig}^k,\mu^k)\\
&+ \|S_G^k\|_\infty \|\delta_k\|_1 +  3 C_G t_k\\
\le& \frac{1}{G} \sumg\sumi  \widehat  M_k(Y_{ig}^k,X_{ig}^k,\widetilde\mu^k)-\frac{1}{G} \sumg\sumi  \widehat M_k(Y_{ig}^k,X_{ig}^k,\mu^k)\\
&+ \left( \frac{\sqrt{\hat s_k + s_k}\|S_G^k\|_\infty  }{\sqrt{\semin{\hat s_k + s_k ,k }}} + 3 C_G \right)t_k.
\end{align*}
where the last inequality follows from
\begin{align*}
\|\delta_k\|_1 \le \sqrt{\hat s_k + s_k}\|\delta_k\|_2 \le \frac{\sqrt{\hat s_k + s_k}  }{\sqrt{\semin{\hat s_k + s_k ,k }}} \|\sqrt{w_{ig}}X_{ig}^{k\prime}\delta_k\|_G.
\end{align*}
We then consider two cases. First, suppose $t^2_k>\bar q_{A_k} t_k$, by definition of $\bar q_{A_k}$
\begin{align*}
\bar q_{A_k} t_k \le \frac{\bar q_{A_k}}{2}\Big\{\frac{1}{G} \sumg\sumi  \widehat  M_k(Y_{ig}^k,X_{ig}^k,\widetilde\mu^k)-\frac{1}{G} \sumg\sumi  \widehat  M_k(Y_{ig}^k,X_{ig}^k,\mu^k)\Big\}^{1/2}_+ 
+ \frac{\bar q_{A_k}}{2}t_k,
\end{align*}
and thus $t_k\le \{\frac{1}{G} \sumg\sumi  \widehat  M_k(Y_{ig}^k,X_{ig}^k,\widetilde\mu^k)-\frac{1}{G} \sumg\sumi  \widehat  M_k(Y_{ig}^k,X_{ig}^k,\mu^k)\}^{1/2}_+ $. Now suppose $t^2_k\le \bar q_{A_k}  t_k$, then
\begin{align*}
t^2_k \le& \Big\{\frac{1}{G} \sumg\sumi  \widehat  M_k(Y_{ig}^k,X_{ig}^k,\widetilde\mu^k)-\frac{1}{G} \sumg\sumi  \widehat  M_k(Y_{ig}^k,X_{ig}^k,\mu^k)\Big\}+ \left( \frac{\sqrt{\hat s_k + s_k}\|S_G^k\|_\infty  }{\sqrt{\semin{\hat s_k + s_k ,k }}} + 3 C_G \right)t_k.
\end{align*}
Since for any positive numbers $a,b,c$, $a^2\le b+ac$ implies $a\le \sqrt{b}+c$, one has
\begin{align*}
t_k\le \Big\{\frac{1}{G} \sumg\sumi  \widehat  M_k(Y_{ig}^k,X_{ig}^k,\widetilde\mu^k)-\frac{1}{G} \sumg\sumi  \widehat M_k(Y_{ig}^k,X_{ig}^k,\mu^k)\Big\}^{1/2}_+ + \left( \frac{\sqrt{\hat s_k + s_k}\|S_G^k\|_\infty  }{\sqrt{\semin{\hat s_k + s_k ,k }}} + 3 C_G \right).
\end{align*}
\end{proof}
\subsection{Proof for Lemma \ref{a:M-est_WL}}\label{sec:proof_for_M_est_WL}
\begin{proof}
By Assumption \ref{a:M-est_WL}, we have for $\ell_G=c''/\widetilde \varphi_G$, $c''$ a constant depends only on $\underline C$, $\overline C$,
\begin{align*}
0\le \Phi^{-1} (1-\gamma/2p) \le \frac{G^{1/6} ( \EP\frac{1}{G}\sum_{g=1}^{G} |S^k_g|^2)^{1/2}/( \EP\frac{1}{G}\sum_{g=1}^{G} |S^k_g|^3)^{1/3} }{\ell_G} -1.
\end{align*}
for all $k\in \mathcal U_G$. Applying inequalty for self-normalized sums (Lemma 5 in \cite{BCCH12}), we have
\begin{align*}
\Pr_\Pr\left( \frac{\lambda}{G} \ge c  \max_{k\in \mathcal U_G}  \left\|\hat \Psi^{-1}_k\frac{1}{G} \sumg S_g^k\right\|_\infty\right)
\ge& \Pr_\Pr\left( \Phi^{-1} \left(1-\frac{\gamma}{2p \widetilde p}\right)
 \ge
  \max_{k\in \mathcal U_G}\max_{j\in[p]} {\small \frac{\left| \sqrt{G} \frac{1}{G} \sumg S^k_{gj}\right|}{\sqrt{\frac{1}{G} \sumg (S^{k}_{gj})^2}}}\right)\\
\ge& 1-\gamma -o(\gamma).
\end{align*}
\end{proof}

\section{Technical Lemmas}\label{sec:technical_lemmas}
For completeness, we collect some of the technical results used in our proofs in this Section. They are either direct restated from other papers or their straightforward modifications.

\subsection{A Maximal Inequality}\label{sec:max_ineq}
Define $\|F\|_{P_n,2}=\left(n^{-1}\sum_{i=1}^n f^2(X_i)\right)^{1/2}$. 
\begin{lemma}[Maximal inequality]\label{lemma:maximal_ineq}
 Given $X_1,...,X_n$ independent (but not necessarily identically distributed) $\Ss$-valued random variables. Suppose $0< \EP\left[n^{-1}\sum_{i=1}^{n}  F^2(X_i)\right] <\infty$, and let $\sigma^2>0$ be any positive constant such that $\sup_{f\in \calF} \EP\left[n^{-1}\sum_{i=1}^n  f^2(X_i)\right]\le \sigma^2 \le \EP\left[n^{-1}\sum_{i=1}^n  F^2(X_i)\right]$. Let $\delta=\sigma/\left(  \EP\left[n^{-1}\sum_{i=1}^n  F^2(X_i)\right]\right)^{1/2}$. Define $B=\sqrt{\EP[\max_{1\le i\le n}F^2(X_i)]}$. Then we have
\begin{align*}
\EP\left[\left\| \frac{1}{\sqrt{n}}\sum_{i=1}^n (f(X_i)-\EP [f(X_i)]) \right\|_\calF\right]
\le
 C \left\{ J(\delta,\calF,F)\left( \EP\left[\frac{1}{n}\sum_{i=1}^{n}  F^2(X_i)\right]\right)^{1/2} + \frac{BJ^2(\delta,\calF,F)}{\delta^2\sqrt{n}} \right\},
\end{align*}
where $C>0$ is a universal constant. In addition, suppose that $\calF$ is a VC type class with characteristics $(A,v)$, $A\ge e$ and $v\ge 1$,
then we have
\begin{align*}
\EP\left[\left\| \frac{1}{\sqrt{n}}\sum_{i=1}^n (f(X_i)-\EP [f(X_i)]) \right\|_\calF\right]
\le C\left\{\sigma\sqrt{v\log(A\vee n)}+ \frac{B}{\sqrt{n}}v\log (A\vee n)\right\}.
\end{align*}
\begin{proof}
The first result follows directly from Lemma B.1 in \cite{cattaneo2022}.
For the second result, notice that Lemma B.2 in \cite{cattaneo2022} implies that
\begin{align*}
\EP\left[\left\| \frac{1}{\sqrt{n}}\sum_{i=1}^n (f(X_i)-\EP [f(X_i)]) \right\|_\calF\right]
\le C\left\{\sigma\sqrt{v\log\overline A}+ \frac{B}{\sqrt{n}}v\log \overline A)\right\},
\end{align*}
where $\overline A:=\left\{A\cdot\left(\EP\left[n^{-1}\sum_i F^2(X_i)\right]\right)^{1/2}/\sigma\right\}$. The result follows from replacing the $\sigma$ on the right hand side by $\sigma'=\sigma\vee\left\{n^{-1/2} \left(\EP\left[n^{-1}\sum_i F^2(X_i)\right]\right)^{1/2}\right\}$. Note that $\sigma'$ also satisfies the requirements for $\sigma$.
\end{proof}
\end{lemma}

\subsection{Additional Technical Lemmas}\label{sec:additional_lemmas}
\text{}\\

The following is a restate of Lemma K.1 in \cite{BCFH17}. 
\begin{lemma}\text{  }\label{lemma:entropy_algebra} \\
Let $\calF$ denote a class of measurable functions $f:\mathcal W \to \Real$ with a measurable envelope $F$.\\
(1) Let $\calF$ be a VC subgraph class with a finite VC index $k$ or any
other class whose entropy is bounded above by that of such a VC subgraph class, then
the uniform entropy numbers of $\calF$ obey
\begin{equation*}
 \sup_{Q} \log  N(\epsilon \|F\|_{Q,2}, \calF,  \| \cdot \|_{Q,2}) \lesssim 1+ k \log (1/\epsilon)\vee 0
\newline
\end{equation*}
(2) For any measurable classes of functions $\calF$ and $\calF^{\prime
} $ mapping $\mathcal{W}$ to $\Bbb{R}$,
\begin{align*}
&\log N(\epsilon \Vert F+F^{\prime }\Vert _{Q,2},\calF+\calF^{\prime
}, \| \cdot \|_{Q,2})\\
&\qquad \leq \log   N\left(\mbox{$ \frac{\epsilon }{2}$}\Vert F\Vert _{Q,2},\calF, \| \cdot \|_{Q,2}\right)
+ \log N\left( \mbox{$ \frac{\epsilon }{2}$}\Vert F^{\prime }\Vert _{Q,2},\calF^{\prime
}, \| \cdot \|_{Q,2}\right), \\
&\log  N(\epsilon \Vert F\cdot F^{\prime }\Vert _{Q,2},\calF\cdot \calF^{\prime
}, \| \cdot \|_{Q,2})\\
&\qquad \leq \log   N\left( \mbox{$ \frac{\epsilon }{2}$}\Vert F\Vert _{Q,2},\calF, \| \cdot \|_{Q,2}\right)
+ \log N\left( \mbox{$ \frac{\epsilon }{2}$}\Vert F^{\prime }\Vert _{Q,2},\calF^{\prime
}, \| \cdot \|_{Q,2}\right), \\
& N(\epsilon \Vert F\vee F^{\prime }\Vert _{Q,2},\calF\cup \calF^{\prime
}, \| \cdot \|_{Q,2})\\
&\qquad \leq   N\left(\epsilon\Vert F\Vert _{Q,2},\calF, \| \cdot \|_{Q,2}\right)
+ N\left( \epsilon\Vert F^{\prime }\Vert _{Q,2},\calF^{\prime
}, \| \cdot \|_{Q,2}\right).
\end{align*}
(3)  For any measurable class of functions $%
\mathcal{F}$ and a fixed function $f$ mapping $\mathcal{W}$ to $\Bbb{R}$,
\begin{equation*}
 \log \sup_{Q} N(\epsilon \Vert |f|\cdot F\Vert _{Q,2},f\cdot\calF, \| \cdot \|_{Q,2})\leq \log \sup_{Q} N\left(
\epsilon /2\Vert F\Vert _{Q,2},\calF, \| \cdot \|_{Q,2}\right)
\end{equation*}
(4)  Given measurable classes $\calF_j$ and envelopes $F_j$, $j=1,\ldots,k$, mapping $\mathcal{W}$ to $\Real$, a mapping $\phi\colon\Bbb{R}^k\to\Bbb{R}$ such that for $f_j,g_j\in\calF_j$, the following Lipschitz condition holds:
$ |\phi(f_1,\ldots,f_k) - \phi(g_1,\ldots,g_k) | \leq \sum_{j=1}^k L_j(x)|f_j(x)-g_j(x)|$ for $L_j(x)\geq 0$, and some fixed functions $\bar f_j \in \calF_j$,  the class of functions $\mathcal{L}=\{\phi(f_1,\ldots,f_k)-\phi(\bar f_1,\ldots,\bar f_k)\colon f_j \in\mathcal{F}_j, j=1,\ldots,k\}$ satisfies
\begin{align*}
&\log \sup_Q N\left(\epsilon\Big\|\sum_{j=1}^kL_jF_j\Big\|_{Q,2},\mathcal{L}, \| \cdot \|_{Q,2}\right)\\
&\qquad \leq \sum_{j=1}^k\log  \sup_Q  N\left(
\mbox{$\frac{\epsilon}{k}$}\|F_j\|_{Q,2},\calF_j, \| \cdot \|_{Q,2}\right).
\end{align*}
\end{lemma}

The following generalizes Lemma 9 of \cite{BCW16} to allow for cluster sampling. The proof follows closely to the orginal. Denote $M=\frac{1}{G} \sumg\sumi f_{ig}^2 X_{ig}X_{ig}'$.
\begin{lemma}[Minoration Lemma]\label{lemma:minoration}\text{}\\
Suppose that for each $G$, $L(\beta)=-\frac{1}{G} \sumg\sumi \{Y_{ig} X_{ig}'\beta - \log(1+\exp(X_{ig}^{\prime}\beta))\}$. For any $\delta\in A \subset \Real^p$,
\begin{align*}
L( \beta^0 + \delta ) - L(\beta^0) -\nabla L(\beta^0)' \delta 
\ge 
\frac{1}{3G} \delta' M \delta \wedge \frac{1}{3G}\bar q_A \sqrt{\delta' M \delta}
\end{align*}
\begin{proof}
The proof is divided into two steps.

\textbf{Step 1. (Minoration)}
Write $F(\delta)=L( \beta^0 + \delta ) - L(\beta^0) -\nabla L(\beta^0)' \delta$. Define
\begin{align*}
r_A =:\sup\Big\{r\in \Real: F(\delta)\ge \frac{1}{3G} \delta' M \delta \text{ for all } \delta\in A, \sqrt{\delta' M \delta}\le r \Big\}
\end{align*}
So for any $\delta\in A$, if $\sqrt{\delta' M \delta}\le r_A$, then by construction of $r_A$,
\begin{align*}
F(\delta) \ge \frac{1}{3G} \delta' M \delta.
\end{align*}
Otherwise if $\sqrt{\delta' M \delta}> r_A$, by convexity of $t\mapsto F(t \delta)$ and the fact that $\frac{r_A}{\sqrt{\delta' M \delta}}< 1$,
\begin{align*}
F(\delta)\ge \frac{\sqrt{\delta' M \delta}}{r_A}F\left(\frac{r_A}{\sqrt{\delta' M \delta}}\delta\right)
\end{align*}
Now, let $\bar \delta=\frac{r_A}{\sqrt{\delta' M \delta}} \delta$, then $\sqrt{\bar \delta'M \bar \delta}\le r_A$ and thus
\begin{align*}
F(\delta)\ge \frac{\sqrt{\delta' M \delta}}{r_A}F(\bar \delta) \ge  \frac{\sqrt{\delta' M \delta}}{r_A} \frac{1}{3G}  r_A^2 \ge     \frac{1}{3G}  \bar q_A \sqrt{\delta' M \delta}.
\end{align*}
where the last inequality follows from  $r_A \ge \bar q_A $ that is shown in the next step. Combining these two cases, we have
\begin{align*}
F(\delta)\ge \frac{1}{3G} \delta' M \delta \wedge \frac{1}{3G}  \bar q_A \sqrt{\delta' M \delta}.
\end{align*}

\textbf{Step 2.} We now prove $r_A \ge \bar q_A $. Define $f_{ig} (t)=\log \{1+ \exp(X_{ig}^{\prime} \beta^0)\}$, then 
\begin{align*}
F(\delta)=\frac{1}{G} \sumg \sumi[f_{ig} (1)-f_{ig} (0)-1\cdot f'_{ig} (0)].
\end{align*}
By Lemma 7 and 8 of \cite{BCW16}, we have
\begin{align*}
f_{ig} (1)-f_{ig} (0)-1\cdot f'_{ig} (0) \ge f_{ig}^2 \Big\{\frac{|X_{ig}'\delta|^2}{2} - \frac{|X_{ig}'\delta|^3}{6}\Big\}.
\end{align*}
Summing over $i$, we have
\begin{align*}
F(\delta) \ge \frac{1}{2}  \frac{1}{G} \sumg\sumi f_{ig}^2 |X_{ig}'\delta|^2
- \frac{1}{6}  \frac{1}{G} \sumg\sumi f_{ig}^2 |X_{ig}'\delta|^3.
\end{align*}
Now, for any $\delta \in A$ such that $\sqrt{\delta' M \delta}\le \bar q_A$, the definition of $\bar q_A$ gives
\begin{align*}
\sqrt{\delta' M \delta}\le \bar q_A \le \frac{(\delta'M\delta)^{3/2}}{\frac{1}{G} \sumg\sumi f_{ig}^2 |X_{ig}'\delta|^3}
\end{align*}
This implies $\frac{1}{G} \sumg\sumi f_{ig}^2 |X_{ig}'\delta|^3\le \frac{1}{G} \sumg\sumi f_{ig}^2 |X_{ig}'\delta|^2$ and thus
\begin{align*}
F(\delta) \ge \frac{1}{2}  \frac{1}{G} \sumg\sumi f_{ig}^2 |X_{ig}'\delta|^2
- \frac{1}{6}  \frac{1}{G} \sumg\sumi f_{ig}^2 |X_{ig}'\delta|^3 \ge \frac{1}{3} \frac{1}{G} \sumg\sumi f_{ig}^2 |X_{ig}'\delta|^2= \frac{1}{3G} \delta'M\delta.
\end{align*}
The definition of $r_A $ then suggests $  r_A \ge \bar q_A$.
\end{proof}
\end{lemma}
\bibliographystyle{ecta}
\bibliography{biblio}

\begin{thebibliography}{43}
\newcommand{\enquote}[1]{``#1''}
\expandafter\ifx\csname natexlab\endcsname\relax\def\natexlab#1{#1}\fi

\bibitem[\protect\citeauthoryear{Athey, Imbens, and Wager}{Athey
  et~al.}{2018}]{AIW18}
\textsc{Athey, S., G.~W. Imbens, and S.~Wager} (2018): \enquote{Approximate
  residual balancing: debiased inference of average treatment effects in high
  dimensions,} \emph{Journal of the Royal Statistical Society}.

\bibitem[\protect\citeauthoryear{Belloni, Chen, Chernozhukov, and
  Hansen}{Belloni et~al.}{2012}]{BCCH12}
\textsc{Belloni, A., D.~Chen, V.~Chernozhukov, and C.~Hansen} (2012):
  \enquote{Sparse models and methods for optimal instruments with an
  application to eminent domain,} \emph{Econometrica}, 80, 2369--2429.

\bibitem[\protect\citeauthoryear{Belloni, Chernozhukov, Chetverikov, and
  Wei}{Belloni et~al.}{2018}]{BCCW18}
\textsc{Belloni, A., V.~Chernozhukov, D.~Chetverikov, and Y.~Wei} (2018):
  \enquote{Uniformly valid post-regularization confidence regions for many
  functional parameters in {Z}-estimation framework,} \emph{Annals of
  Statistics}, 46, 3643--3675.

\bibitem[\protect\citeauthoryear{Belloni, Chernozhukov, Fern{\'a}ndez-Val, and
  Hansen}{Belloni et~al.}{2017}]{BCFH17}
\textsc{Belloni, A., V.~Chernozhukov, I.~Fern{\'a}ndez-Val, and C.~Hansen}
  (2017): \enquote{Program evaluation and causal inference with
  high-dimensional data,} \emph{Econometrica}, 85, 233--298.

\bibitem[\protect\citeauthoryear{Belloni, Chernozhukov, and Hansen}{Belloni
  et~al.}{2014}]{BCH14}
\textsc{Belloni, A., V.~Chernozhukov, and C.~Hansen} (2014): \enquote{Inference
  on treatment effects after selection among high-dimensional controls,}
  \emph{The Review of Economic Studies}, 81, 608--650.

\bibitem[\protect\citeauthoryear{Belloni, Chernozhukov, Hansen, and
  Kozbur}{Belloni et~al.}{2016{\natexlab{a}}}]{BCHK16}
\textsc{Belloni, A., V.~Chernozhukov, C.~Hansen, and D.~Kozbur}
  (2016{\natexlab{a}}): \enquote{Inference in high-dimensional panel models
  with an application to gun control,} \emph{Journal of Business \& Economic
  Statistics}, 34, 590--605.

\bibitem[\protect\citeauthoryear{Belloni, Chernozhukov, and Kato}{Belloni
  et~al.}{2015}]{BCK15}
\textsc{Belloni, A., V.~Chernozhukov, and K.~Kato} (2015): \enquote{Uniform
  post-selection inference for least absolute deviation regression and other
  Z-estimation problems,} \emph{Biometrika}, 102, 77--94.

\bibitem[\protect\citeauthoryear{Belloni, Chernozhukov, and Wei}{Belloni
  et~al.}{2016{\natexlab{b}}}]{BCW16}
\textsc{Belloni, A., V.~Chernozhukov, and Y.~Wei} (2016{\natexlab{b}}):
  \enquote{Post-selection inference for generalized linear models with many
  controls,} \emph{Journal of Business \& Economic Statistics}, 34, 606--619.

\bibitem[\protect\citeauthoryear{Bickel, Ritov, Tsybakov et~al.}{Bickel
  et~al.}{2009}]{BRT09}
\textsc{Bickel, P.~J., Y.~Ritov, A.~B. Tsybakov, et~al.} (2009):
  \enquote{Simultaneous analysis of Lasso and Dantzig selector,} \emph{The
  Annals of statistics}, 37, 1705--1732.

\bibitem[\protect\citeauthoryear{Cameron and Miller}{Cameron and
  Miller}{2015}]{CM15}
\textsc{Cameron, C.~A. and D.~L. Miller} (2015): \enquote{A practitioner’s
  guide to cluster-robust inference,} \emph{Journal of Human Resources}, 50,
  317--372.

\bibitem[\protect\citeauthoryear{Caner}{Caner}{2017}]{Caner17}
\textsc{Caner, M.} (2017): \enquote{Delta Theorem in the Age of High
  Dimensions,} \emph{arXiv preprint arXiv:1701.05911}.

\bibitem[\protect\citeauthoryear{Caner and Kock}{Caner and Kock}{2018}]{CK18}
\textsc{Caner, M. and A.~B. Kock} (2018): \enquote{Asymptotically honest
  confidence regions for high dimensional parameters by the desparsified
  conservative lasso,} \emph{Journal of Econometrics}, 203, 143--168.

\bibitem[\protect\citeauthoryear{Cattaneo, Feng, and Underwood}{Cattaneo
  et~al.}{2022}]{cattaneo2022}
\textsc{Cattaneo, M., Y.~Feng, and W.~Underwood} (2022): \enquote{Uniform
  Inference for Kernel Density Estimators with Dyadic Data,} \emph{preprint}.

\bibitem[\protect\citeauthoryear{Chamberlain}{Chamberlain}{1984}]{chamberlain84}
\textsc{Chamberlain, G.} (1984): \enquote{Panel data,} \emph{Handbook of
  econometrics}, 2, 1247--1318.

\bibitem[\protect\citeauthoryear{Chernozhukov, Chetverikov, Demirer, Duflo,
  Hansen, Newey, and Robins}{Chernozhukov
  et~al.}{2018{\natexlab{a}}}]{CCDDHNR18}
\textsc{Chernozhukov, V., D.~Chetverikov, M.~Demirer, E.~Duflo, C.~Hansen,
  W.~Newey, and J.~Robins} (2018{\natexlab{a}}): \enquote{Double/debiased
  machine learning for treatment and structural parameters,} \emph{The
  Econometrics Journal}, 21, C1--C68.

\bibitem[\protect\citeauthoryear{Chernozhukov, Chetverikov, and
  Kato}{Chernozhukov et~al.}{2013}]{CCK13}
\textsc{Chernozhukov, V., D.~Chetverikov, and K.~Kato} (2013):
  \enquote{Gaussian approximations and multiplier bootstrap for maxima of sums
  of high-dimensional random vectors,} \emph{Annals of Statistics}, 41,
  2786--2819.

\bibitem[\protect\citeauthoryear{Chernozhukov, Chetverikov, and
  Kato}{Chernozhukov et~al.}{2017}]{CCK17}
---\hspace{-.1pt}---\hspace{-.1pt}--- (2017): \enquote{Central limit theorems
  and bootstrap in high dimensions,} \emph{Annals of Probability}, 45,
  2309--2352.

\bibitem[\protect\citeauthoryear{Chernozhukov, Newey, and Singh}{Chernozhukov
  et~al.}{2018{\natexlab{b}}}]{CNS18}
\textsc{Chernozhukov, V., W.~Newey, and R.~Singh} (2018{\natexlab{b}}):
  \enquote{Learning L2 Continuous Regression Functionals via Regularized Riesz
  Representers. arXiv e-prints, page,} \emph{arXiv preprint arXiv:1809.05224}.

\bibitem[\protect\citeauthoryear{Djogbenou, MacKinnon, and Nielsen}{Djogbenou
  et~al.}{2019}]{DMN18}
\textsc{Djogbenou, A.~A., J.~G. MacKinnon, and M.~{\O}. Nielsen} (2019):
  \enquote{Asymptotic theory and wild bootstrap inference with clustered
  errors,} \emph{Journal of Econometrics}, 212, 393--412.

\bibitem[\protect\citeauthoryear{Farrell}{Farrell}{2015}]{Farrell15}
\textsc{Farrell, M.~H.} (2015): \enquote{Robust inference on average treatment
  effects with possibly more covariates than observations,} \emph{Journal of
  Econometrics}, 189, 1--23.

\bibitem[\protect\citeauthoryear{Gentzkow, Kelly, and Taddy}{Gentzkow
  et~al.}{2019{\natexlab{a}}}]{GKT19}
\textsc{Gentzkow, M., B.~Kelly, and M.~Taddy} (2019{\natexlab{a}}):
  \enquote{Text as data,} \emph{Journal of Economic Literature}, 57, 535--74.

\bibitem[\protect\citeauthoryear{Gentzkow, Shapiro, and Taddy}{Gentzkow
  et~al.}{2019{\natexlab{b}}}]{GST19}
\textsc{Gentzkow, M., J.~M. Shapiro, and M.~Taddy} (2019{\natexlab{b}}):
  \enquote{Measuring group differences in high-dimensional choices: method and
  application to congressional speech,} \emph{Econometrica}, 87, 1307--1340.

\bibitem[\protect\citeauthoryear{Gin{\'e} and Nickl}{Gin{\'e} and
  Nickl}{2016}]{GN16}
\textsc{Gin{\'e}, E. and R.~Nickl} (2016): \emph{Mathematical foundations of
  infinite-dimensional statistical models}, vol.~40, Cambridge University
  Press.

\bibitem[\protect\citeauthoryear{Ginsberg, Mohebbi, Patel, Brammer, Smolinski,
  and Brilliant}{Ginsberg et~al.}{2009}]{G_etal09}
\textsc{Ginsberg, J., M.~H. Mohebbi, R.~S. Patel, L.~Brammer, M.~S. Smolinski,
  and L.~Brilliant} (2009): \enquote{Detecting influenza epidemics using search
  engine query data,} \emph{Nature}, 457, 1012--1014.

\bibitem[\protect\citeauthoryear{Hagemann}{Hagemann}{2017}]{Hagemann17}
\textsc{Hagemann, A.} (2017): \enquote{Cluster-robust bootstrap inference in
  quantile regression models,} \emph{Journal of the American Statistical
  Association}, 112, 446--456.

\bibitem[\protect\citeauthoryear{Hirshberg and Wager}{Hirshberg and
  Wager}{2017}]{HS17}
\textsc{Hirshberg, D.~A. and S.~Wager} (2017): \enquote{Balancing out
  regression error: efficient treatment effect estimation without smooth
  propensities,} \emph{arXiv preprint arXiv:1712.00038}.

\bibitem[\protect\citeauthoryear{Hirshberg and Wager}{Hirshberg and
  Wager}{2018}]{HW18}
---\hspace{-.1pt}---\hspace{-.1pt}--- (2018): \enquote{Debiased inference of
  average partial effects in single-index models,} \emph{arXiv preprint
  arXiv:1811.02547}.

\bibitem[\protect\citeauthoryear{Javanmard and Montanari}{Javanmard and
  Montanari}{2014}]{JM14}
\textsc{Javanmard, A. and A.~Montanari} (2014): \enquote{Confidence intervals
  and hypothesis testing for high-dimensional regression,} \emph{The Journal of
  Machine Learning Research}, 15, 2869--2909.

\bibitem[\protect\citeauthoryear{Jegadeesh and Wu}{Jegadeesh and
  Wu}{2013}]{JW13}
\textsc{Jegadeesh, N. and D.~Wu} (2013): \enquote{Word power: A new approach
  for content analysis,} \emph{Journal of financial economics}, 110, 712--729.

\bibitem[\protect\citeauthoryear{Kato}{Kato}{2017}]{Kato17}
\textsc{Kato, K.} (2017): \enquote{Lecture notes on empirical process theory,}
  Tech. rep., technical report.

\bibitem[\protect\citeauthoryear{Kline and Santos}{Kline and
  Santos}{2012}]{KS12}
\textsc{Kline, P. and A.~Santos} (2012): \enquote{A score based approach to
  wild bootstrap inference,} \emph{Journal of Econometric Methods}, 1, 23--41.

\bibitem[\protect\citeauthoryear{Kock}{Kock}{2016}]{Kock16}
\textsc{Kock, A.~B.} (2016): \enquote{Oracle inequalities, variable selection
  and uniform inference in high-dimensional correlated random effects panel
  data models,} \emph{Journal of Econometrics}, 195, 71--85.

\bibitem[\protect\citeauthoryear{Kock and Tang}{Kock and Tang}{2015}]{KT18}
\textsc{Kock, A.~B. and H.~Tang} (2015): \enquote{Uniform inference in
  high-dimensional dynamic panel data models,} \emph{arXiv preprint
  arXiv:1501.00478}.

\bibitem[\protect\citeauthoryear{MacKinnon and Webb}{MacKinnon and
  Webb}{2017}]{MW17}
\textsc{MacKinnon, J.~G. and M.~D. Webb} (2017): \enquote{Wild bootstrap
  inference for wildly different cluster sizes,} \emph{Journal of Applied
  Econometrics}, 32, 233--254.

\bibitem[\protect\citeauthoryear{P{\"o}tscher and Leeb}{P{\"o}tscher and
  Leeb}{2009}]{PL09}
\textsc{P{\"o}tscher, B.~M. and H.~Leeb} (2009): \enquote{On the distribution
  of penalized maximum likelihood estimators: The LASSO, SCAD, and
  thresholding,} \emph{Journal of Multivariate Analysis}, 100, 2065--2082.

\bibitem[\protect\citeauthoryear{van~de Geer, B{\"u}hlmann, Ritov, and
  Dezeure}{van~de Geer et~al.}{2014}]{vdGBRD14}
\textsc{van~de Geer, S., P.~B{\"u}hlmann, Y.~Ritov, and R.~Dezeure} (2014):
  \enquote{On asymptotically optimal confidence regions and tests for
  high-dimensional models,} \emph{The Annals of Statistics}, 42, 1166--1202.

\bibitem[\protect\citeauthoryear{van~der Vaart and Wellner}{van~der Vaart and
  Wellner}{1996}]{vdVW96}
\textsc{van~der Vaart, A.~W. and J.~A. Wellner} (1996): \emph{Weak Convergence
  and Empirical Processes}, Springer.

\bibitem[\protect\citeauthoryear{Wooldridge and Zhu}{Wooldridge and
  Zhu}{2017}]{WZ17}
\textsc{Wooldridge, J. and Y.~Zhu} (2017): \enquote{Inference in approximately
  sparse correlated random effects probit models,} \emph{Journal of Business
  and Economic Statistics, Forthcoming}.

\bibitem[\protect\citeauthoryear{Wooldridge}{Wooldridge}{2005}]{wooldridge05}
\textsc{Wooldridge, J.~M.} (2005): \enquote{Unobserved heterogeneity and
  estimation of average partial effects,} \emph{Identification and inference
  for econometric models: Essays in honor of Thomas Rothenberg}, 27--55.

\bibitem[\protect\citeauthoryear{Wooldridge}{Wooldridge}{2010}]{Wooldridge10}
---\hspace{-.1pt}---\hspace{-.1pt}--- (2010): \emph{Econometric analysis of
  cross section and panel data}, MIT press.

\bibitem[\protect\citeauthoryear{Wooldridge}{Wooldridge}{2019}]{wooldridge18}
---\hspace{-.1pt}---\hspace{-.1pt}--- (2019): \enquote{Correlated random
  effects models with unbalanced panels,} \emph{Journal of Econometrics}, 211,
  137--150.

\bibitem[\protect\citeauthoryear{Wu}{Wu}{2018}]{Wu18}
\textsc{Wu, A.~H.} (2018): \enquote{Gendered language on the economics job
  market rumors forum,} in \emph{AEA Papers and Proceedings}, vol. 108,
  175--79.

\bibitem[\protect\citeauthoryear{Zhang and Zhang}{Zhang and Zhang}{2014}]{ZZ14}
\textsc{Zhang, C.-H. and S.~S. Zhang} (2014): \enquote{Confidence intervals for
  low dimensional parameters in high dimensional linear models,} \emph{Journal
  of the Royal Statistical Society: Series B: Statistical Methodology},
  217--242.

\end{thebibliography}
\end{document}